# A Survey of Software Engineering Practices in Turkey (Technical report-Extended version)

Vahid Garousi[1,2], Ahmet Coşkunçay[3], Aysu Betin-Can[3], Onur Demirörs[3]

1: System and Software Quality Engineering Research Group (SySoQual)
Department of Software Engineering
Atilim University
Ankara, Turkey
vahid.garousi@atilim.edu.tr

2: Software Quality Engineering Research Group (SoftQual)
Department of Electrical and Computer Engineering
University of Calgary, Calgary, Alberta, Canada

3: Informatics Institute
Middle East Technical University
Ankara, Turkey
{cahmet, betincan, demirors}@metu.edu.tr

**Abstract**

**Context:** Understanding the types of software engineering practices and techniques used in the industry is important. There is a wide spectrum in terms of the types and maturity of software engineering practices conducted in each software team and company. To characterize the type of software engineering practices conducted in software firms, a variety of surveys have been conducted in different countries and regions. Turkey has a vibrant software industry and it is important to characterize and understand the state of software engineering practices in this industry.

**Objective:** Our objective is to characterize and grasp a high-level view on type of software engineering practices in the Turkish software industry. Among the software engineering practices that we have surveyed in this study are the followings: software requirements, design, development, testing, maintenance, configuration management, release planning and support practices. The current survey is the most comprehensive of its type ever conducted in the context of Turkish software industry.

**Method:** To achieve the above objective, we systematically designed an online survey with 46 questions based on our past experience in the Canadian and Turkish contexts and using the Software Engineering Body of Knowledge (SWEBOK). 202 practicing software engineers from the Turkish software industry participated in the survey. We analyze and report in this paper the results of the questions. Whenever possible, we also compare the trends and results of our survey with the results of a similar 2010 survey conducted in the Canadian software industry.

**Results:** The survey results reveal important and interesting findings about software engineering practices in Turkey. Among the findings are the followings: (1) The software industries of the military and defense domains are quite strong in Turkey, especially in Ankara, and many SE practitioners work for those firms. (2) In terms of types of software developed in the companies, the top categories are: business applications software, safety-critical and mission-critical software, and web applications, (3) 54% of the participants reported not using any software size measurement methods, whereas 33% mentioned that they have measured lines of code (LOC) and 15% used use-case points, (4) development phase is the phase where teams spend the most effort on (with an average of 31%), (5) After the development phase, software testing, requirements, design and maintenance phases come next and have similar average values (14%, 12%, 12% and 11% respectively), (6) respondents experience the most challenge in the requirements phase, (7) Waterfall, as a rather old but still widely used lifecycle model, is the model that more than half of the respondents (53%) use. The next most preferred lifecycle models are incremental and Agile/lean development models with usage rates of 38% and 34%, respectively, (8) Around 45% of the respondents from larger companies operate in the military and defense sector. Majority of the respondents from smaller companies are in engineering, manufacturing, IT and telecommunication sectors, and (9) The Waterfall and Agile methodologies have slight negative correlations, denoting that if one is used in a firm, the other will less likely to be used.

**Conclusion:** The results of our survey will be of interest to software engineering professionals both in Turkey and world-wide. It will also benefit researchers in observing the latest trends in software engineering industry identifying the areas of strength and weakness, which would then hopefully encourage further industry-academia collaborations in those areas.

**Table of Contents**

## 1 Introduction

Software Engineering (SE) has become a mature field. The term SE first appeared in the late 1960s and was introduced by Bauer to describe ways to develop, manage and maintain software so that the resulting products are reliable, correct, efficient and flexible [1].

Similar to many other engineering fields, the extent to which the SE academia and industry collaborate with each other is rather limited in many countries. To analyze and characterize the state of SE practices and encourage further academic-industrial collaborations, various authors have conducted and reported surveys on the topic since 1980's, e.g., [2-18]. For example, Zelkowitz et al. performed an in-depth survey [3] of 30 companies in which they characterized the state of practice in the SE industry in the USA and Japan in 1984. Their survey revealed that— at that time — practice was around 10 years behind the SE research.

Going on a quote from Strachey at the 1969 NATO conference on SE where he states that "there is a need for a greater mutual understanding between the communities of software development practice and software research" [1], we can only observe that this gap has existed for a very long time. In this respect, literature suggests two distinct ways for reducing the gap that exists between industry and academia: education and technology transfer. For example, Lethbridge et al. [19] argue that we should better understand the dimensions of the industrial practice so that we can focus education appropriately. Focusing on education can also have direct benefits on practice, firstly, because academia is currently educating the next generation of software engineers, ready to take recent knowledge from academia to industry, and secondly, because we should also remember to continue educating existing practitioners to continually increase the level of professionalism in SE.

Turkey has a vibrant software industry. As of 2011, there were about 1,600 software development firms in Turkey [20]. Similar to other countries, it is important to continually characterize and understand the state of practices in the Turkish software industry. The IT sector worldwide faces enormous challenges in delivering quality products on time and within budget [21, 22]. Ongoing challenges of the software industry in delivering projects on time and on budget lead us to question the SE methods and practices used in the industry. Indeed, following proper and systematic SE practices by all the software firms across the globe including Turkey is a major keystone in determining the success or failure of software projects.

The goal of the survey reported in this paper is to characterize the SE practices in Turkey for the purpose of identifying the trends, and also to provide a view on the latest SE techniques, tools and metrics used by practitioners (professionals) in Turkey and the challenges faced by them, to be able to benefit both SE professionals and also researchers both in Turkey and world-wide, for observing the latest trends in the SE industry and identifying the areas of strength and weakness and encouraging more academia-industry collaborations.

Building on top of our track-record on conducting and reporting similar surveys in SE in Canada [23, 24] and Turkey [25], and also our international industrial connections, e.g., [26-29], we planned designed, and conducted a survey in year 2013 across Turkey which received responses from 202 participants. We report and analyze the results in this article.

The remainder of this article is structured as follows. Background and a survey of the related work are presented in Section 2. We describe the design of the survey goal, design and its execution aspects in Section 3. In Section 4, we present and analyze the survey's results. Section 5 summarizes the findings and discusses the lessons learned. Finally, in Section 6, we draw conclusions, and suggest areas for further research.

## 2 Background and Related Work

We discuss in this section the followings:

- A brief review on the state of the software industry in Turkey
- Related work:
    - Surveys on SE practices in Turkey
    - Surveys on SE practices world-wide

- Surveys on sub-areas of SE world-wide
- Articles pointing out (posing) industrial "open problems" for the research community to work on

## 2.1 State of the Software Industry in Turkey

To provide a review on the state of the Turkish software industry, we provide a brief summary of the existing reports and articles surveying this topic, i.e., [20, 30-32]. Table 1 shows a summary of these reports, sorted by year of publication. Only one of these reports is in English, while the rest are in Turkish. We discuss below a summary which we have adapted from these sources. Interestingly, all the five reports are recent, published between 2009 and 2012.

We should note that we have separated sources which have explicitly conducted surveys on SE practices in Turkey from the articles only expressing the state of the software industry in Turkey, without getting into technical details of SE practices (e.g., how companies conduct requirements engineering). The former group of papers is discussed in this section, while the latter group is discussed in Section 2.2.1.

**Table 1-A summary of reports showing the state of the Turkish software industry (sorted by year of publication)**

| Reference | Year | Subject | Topics covered | Authors/Organization | Language |
|---|---|---|---|---|---|
| [30] | 2009 | Software: the new strength of the economy | • Importance of software sector and economy created by software<br>• Government's role in supporting software sector – state of the world<br>• Government's role in supporting software sector – state of Turkey and evaluation<br>• Strategies and suggestions for supporting software sector | Turkish Software Industry Association (acronym in Turkish: YASAD) | Turkish |
| [31] | 2010 | Software industry in Turkey | • Distribution of software types in Turkey<br>• IT resources in Turkey<br>• Fast growing ICT companies<br>• ICT, Telecom and IT market in Turkey<br>• Distribution of software companies in sectors<br>• Turkey's software export strength | Doğan Ufuk Güneş, YASAD | English |
| [20] | 2011 | Turkish software sector and value added by this sector | • Turkish software sector<br>• Value added by software<br>• Success stories<br>• National strategies for software-intensive services<br>• Government's role in supporting the software sector | Gülara Tirpançeker, YASAD | Turkish |
| [32] | 2012 | The software sector in Turkey | • Turkish Software Industry: overview, potential opportunities and threats<br>• Support and incentives for the software industry<br>• Foreign trade of the industry<br>• Collaborations in the industry<br>• Standards and certification for the development and use of software<br>• Human resources | Turkish Institute of Strategic Thinking | Turkish |

The report provided in [30] was put together by the Turkish Software Industry Association (acronym in Turkish: YASAD, www.yasad.org.tr) and discusses the software sector as the new strength of the economy. Few excerpts from this report are as follows. The Turkish software sector received a share of 18.9% of national R&D grant funds as of 2007, but there is a reported need for more instruments specialized in supporting it. Also, the software market is claimed to need to grow to compete with hardware and services markets as it owns only 11% of the total information technologies market in Turkey, while this share is usually higher in most other European countries according to the report.

The presentation reported in [31], put together again by YASAD, provided an overview of the software industry in Turkey. Turkish software industry's potential is highlighted as having the best availability scores for qualified engineers and IT skills among Eastern European countries. This potential is reported to be realized with an increase in the volume of software market from 1.6 to 1.8 Billion USD between 2009 and 2011. Software export is featured as a significant factor in software market volume where Germany, USA and UAE are reportedly the leading export countries for Turkey. Author claims that there exists further export opportunities for Turkey via its cultural and geographical proximity to European Union, North Africa, Middle East and Turkic nations in the Central Asia.

According to [20], as of year 2010, the Turkish software industry was worth about $690 million USD. Turkish software firms exported their software products to more than 50 counties in the volume of $250 million USD. These figures are due for changes in age demographics with a 51% of the population being under 25 and more than 1.5 million Turkish Small-Medium Enterprises (SMEs) consuming IT systems in Turkey.

According to a report prepared by Akkaya et al. [32] in the Turkish Institute of Strategic Thinking, a non-governmental organization, there were, as of 2012, around 1,600 software development firms in Turkey with Turkey as their headquarters. There are also many foreign (non- Turkish) software development firms who have R&D offices in Turkey, such as IBM. On a broader scope, there are 10,000+ companies/organizations with IT-based operations

According to [32], software development in Turkey is dominated by small-sized firms (about 51% of the firms) who have less than 10 employees, 35.7% of the firms have between 10-50 employees, 9.8% have 50-250 employees and only 3% have more than 250 employees. The study of Akkaya et al. [32] also shows that 35% of all software development firms are established within technology development zones in Turkey, which have attractive incentives, e.g., lower tax rates. Software is developed targeting various sectors such as government, manufacturing, automotive, telecommunications, electric and electronics, finance and banking, transportation and logistics, textile, education, media, defense, and medical industries.

There are three major national associations related to Information Technology (IT) and SE in Turkey: (1) Turkish Software Industry Association (acronym in Turkish: YASAD), (2) Informatics Association of Turkey (acronym in Turkish: TBD, www.tbd.org.tr), and (3) Turkish Informatics Foundation (acronym in Turkish: TBV, www.tbv.org.tr) which monitor the state of the industry and organize events in this area. There are also more focused SE-related associations such as the Turkish Testing Board (a.k.a., Turkish software testing and quality association, www.turkishtestingboard.org).

There are national conferences and symposiums related to IT and SE, the main ones being: the Turkish National Software Engineering Symposium (acronym in Turkish: UYMS, www.uyms.org.tr), and the National Informatics Symposium (www.citex.org).

## 2.2 Related Work

### 2.2.1 Surveys on Software Engineering Practices in Turkey

Our literature search has identified a number of surveys on SE practices in Turkey [12, 14, 15, 33-36] which have been summarized in Table 3 (sorted by year of study) and are discussed briefly next.

**Table 2-A summary of surveys of the SE practices in Turkey (sorted by year of study)**

| Reference | Year | Topics covered | Authors/Organization | Number of respondents | Language |
|---|---|---|---|---|---|
| [12] | 2001 | • A SPICE-Oriented, SWEBOK-Based Software Process Assessment on a National Scale: Turkish Software Sector Survey<br>• Software requirements<br>• Project management, configuration management, subcontract management and process management<br>• verification and validation, risk management, quality assurance | Specialist Group for Software, Turkish Society for Quality | 68 | English |
| [14] | 2009 | • SE and software management practices in Turkey<br>• Similar to the 2003 survey [12] | Meriç Aykol, Bahçeşehir University, Turkey | 75 | Turkish |
| [15] | 2010 | • Competency level of the software industry in Turkey<br>• Guidelines for enhancement of companies and the sector | Nermin Sökmen, The Scientific and Technological Research Council of Turkey (acronym in Turkish: TÜBİTAK) | 450 sample organizations | Turkish |
| [33-36] | 2011-2013 | • Series of the Turkey software quality reports<br>• From year 2011 until 2014 so far | Turkish Testing Board | Not disclosed | English |

To best of our knowledge, the 2001 survey [12] by Aytaç et al. who were members of the Turkish Society for Quality, was the earliest survey on the topic. Their survey followed the Software Engineering Body of Knowledge (SWEBOK) (version 2004) [37] and the ISO/IEC 15504 standard, also known as the Software Process Improvement and Capability Determination (SPICE) for design of the questions. Practices such as software requirements, project management, configuration management, verification and validation, risk management, quality assurance, subcontract management, and process management were reported.

Later in 2009, Aykol replicated the 2003 survey [12] with some revisions, and published the results as a MSc thesis [14]. The goal was to analyze the changes and trends in the SE practices in the Turkish software industry from 2001 to 2008. State of the practices remained stable between 2001 and 2008. However, this would be misleading as there were no participants from defense industry in 2008 and the author suggests that defense industry has an above average process maturity. Some of the processes whose practices got better are usage of software management tools and techniques, software configuration management and software engineering management, whereas the some that got worse are software testing, software requirements engineering and software implementation.

The study by Sökmen in 2010 [15], sponsored by the Scientific and Technological Research Council of Turkey (acronym in Turkish: TÜBİTAK), was a survey of 450 sample software firms. Most of these companies (62%) have been established after year 2000, indicating the recent growth of the software sector in Turkey. The average employee numbers in a firm is 42 in these installations. When experience levels of software development personnel is investigated; only a quarter of them (25%) have more than 10 years of experience, where rest is distributed between 0-2 years (4%), 3-6 years (45%), and 7-9 years (26%). These quantitative figures for company ages and employee experiences show that Turkish software industry is still young and eager to grow.

Sökmen's study [15] also reported findings for software project characteristics, e.g., (1) average software project duration for the sample companies was around 11 months, and (2) mean and median values of project budget were $235K USD and $25K USD, respectively. The study [15] also assessed the quality awareness in the industry by analyzing quality certification efforts of Turkish software firms. 27% of firms have ISO 9000 family of certificates, 4% have certificates issues by the Turkish Standards Institute, e.g., TS ISO/IEC 15504, and only 1% of firms have CMMI certifications. Moreover, 77% of the companies reported not to have a quality assurance group and 52% do not perform systematic quality assurance at all. The survey [15] also revealed that 36% of the organizations were enthusiastic about being appraised by the CMMI or SPICE. As of this writing (July 2014), there are 28 Turkish organizations which have been appraised for the CMMI as listed by CMMI Institute [38]. There are also four organizations who have achieved the ISO 15504 (SPICE) rating according to the Turkish Standards Institution [39].

Last but not the least, there has been a series of annual surveys, called the Turkish software quality reports, have been conducted by the Turkish Testing Board (TTB) since year 2011 [33-36]. TTB is the regional body representing and supporting software testing professionals in Turkey. These annual surveys focus only on software testing, have been conducted by soliciting input from professional testers in Turkey, and present various statistics about the following topics: test techniques, test competency, test objective, exit criteria, cloud testing, mobile testing, maturity and standardization, exploratory testing, and skill set. However, since the basic statistical parameters such as number and characteristics of respondents have not been published in the annual surveys of TTB, the significance of the results are not clear.

### 2.2.2 Surveys on Software Engineering Practices World-wide

A number of surveys have been conducted on the subject of SE practices in various scales and countries. Our literature search has identified a number of studies [2-15, 17, 18, 40], which have been summarized in Table 3 and discussed briefly next. Note that this list includes only papers conducting surveys on the entire SE spectrum, and not surveys focusing on only one sub-area of SE, e.g., testing [23]. Surveys studying sub-areas of SE are discussed in Section 2.2.3. Let us also note that this list is not exhaustive, but is only a representative list, since an in-depth review of all surveys on SE practices will be outside the scope of this paper.

It is interesting to see that more and more surveys of this type have been conducted since 1995, denoting the emergence of the need for surveys of this type as the software industry is becoming more mature. Among the surveys in this list, the American software community, with six articles, is the most active in conducting and reporting surveys on SE practices.

**Table 3-A summary of surveys on the subject of SE practices (sorted by year of study)**

| Paper Reference | Scale/region | Year | Number of respondents | Goal/Focus area |
|---|---|---|---|---|
| [2] | Dallas-Fort Worth area USA | 1983 | 63 | SE practices |
| [3] | USA, Japan | 1984 | 30 companies | SE practices in USA and Japan |
| [4] | USA | 1988 | 17 large projects (97 interviews) | A field study of the software design process for large systems |
| [5] | Japan, USA | 1990 | 26 projects from US and 16 projects from Japan | Differences between USA and Japan in software development practices and performances |
| [6] | Western Europe, Japan, USA | 1996 | 147 projects | Differences between three regions in tools, technology and practices and how development speed is affected |
| [7] | Canada | 1997 | 394 (6+13+8+367) | Examination of SE work practices |
| [8] | United Kingdom | 1997 | Over 50 | Adoption rate of SE techniques in UK |
| [9] | Europe | 1999 | 397 | Adoption levels of SE practices in European countries |
| [10] | New Zealand | 2000 | 24 companies | Software development practices |
| [11] | India, Japan, USA, Europe | 2003 | 104 projects | Regional differences in software development practices and performances |
| [12] | Turkey | 2003 | 68 | A SPICE-Oriented, SWEBOK-Based Software Process Assessment on a National Scale: Turkish Software Sector Survey |
| [13] | Global | 2007 | 57 | State of practice in software development for medical devices |
| [14] | Turkey | 2009 | 75 | SE practices in Turkey |
| [15] | Turkey | 2010 | Around 450 company executives | Turkish software development industry, companies and practices |
| [40] | Italy | 2010 | 62 projects from 28 Italian companies | Evaluating the perceived effect of software engineering practices in the Italian industry |
| [17] | New Zealand | 2012 | 195 | Software development practices in New Zealand |
| [18] | Netherlands | 2012 | 99 | Software engineering practices in Netherlands |
| The current survey | Turkey | 2013 | 202 | SE practices in Turkey |

One of the earliest studies surveying SE practices was performed by Beck and Perkins in the USA in 1983 [2]. They investigated company profiles, usage of SE methods and tools and problem encountered in software industry. In terms of effort spent in the software development life cycle (SDLC) phases, they reported that companies allocate most of the effort for design, implementation and testing. Another interesting finding is that larger companies have a higher tendency to use systematic SE techniques.

Another study was reported in 1984 by Zelkowitz et al. [3] which surveyed the SE practices in Japan and USA. As expected in those early years of SE, they reported a very low tools usage rate except for code compilers, text editors and unit test tools. Additionally, [3] also reported poor review processes and data collection and analysis, and low usage rates of incremental development and prototyping.

In a 1988 study, Curtis et al. [4] performed interviews with participants from 17 projects with diverse characteristics. They analyzed and clustered problems stated by interviewees and identified three most distinct and common problems in developing large systems as; insufficient domain knowledge, volatile and conflicting requirements and lack of communication and coordination.

In a 1990 study, Cusumano et al. [5] aimed to compare Japanese and USA projects in terms of process and performance by utilizing standard forms for collecting data. Results show some similarities between two countries in terms of the types and size of products developed, languages, tools and hardware used, and experience of engineers. An interesting difference is that Japanese were reportedly allocating more time on design and testing activities and less on coding when compared to US. Some shared findings for both countries suggest that code reuse rate and higher productivity are associated and also using testing tools are correlated with lower error rates.

In a 1996 study, Blackburn et al. [6] provided comparative research findings on how tools, technologies and practices correlate with software development productivity in the Western Europe, Japan and the USA. Percentage of effort allocated to development phases, a noteworthy data, is also provided. With respect to a total of 147 project data, all three regions allocate almost same amount of effort to design activities (about 20%). However, there exist significant differences between regions in percentage of effort spent for implementation and testing/integration stages ranging from 27% to 37% and 17% to 26% respectively. Though, results should be interpreted with caution, since the participating projects being selected only from larger companies indicate a possible bias in survey results.

In a Canadian study [7] work practice data of software engineers including data from an individual engineer, engineering groups and company-wide tool usage were collected and analyzed. Data from six software engineers revealed that 50% of software engineers write code, 17% perform design, 17% perform testing and 17% review others' work during their routine work practices.

Holt [8] surveyed and reported the SE practices in the United Kingdom (UK) in 1997. Although the study dates back to the late 90s, about only half of the participants of the survey (56%) used any CASE (Computer-Aided Software Engineering) tools. Moreover, 31% of them used no structured design approach and none of them used any kind of a formal design method. Considering the participants' high degree of awareness in SE issues, as they were invited from a special interest group and a newsletter in SE, these findings indicate surprisingly poor practices. Another important finding is that Waterfall (30%), V-Model (24%) and Spiral (20%) were the popular life-cycle models used in UK in that time.

In 1999, Dutta el al. [9] made use of a survey with 42 yes-or-no questions to learn about the adoption levels of SE practices in the European countries. They reached a conclusion via survey results that France and UK were best performers in SE practices whereas Sweden, Spain and Belgium were the weak ones. In terms of target sectors, companies in aircraft and spacecraft, finance and insurance and telecom had high adoption of best practices. Some interesting findings exist for average adoption rates of practices such as testing every function (55%), regression testing (32%), procedures for controlling changes (65%), formal procedures for estimation (50%), and periodic status reviews by management (80%). These adoption rates seem high and misleading since choice range for questions was limited to two options (yes-or-no), which might have lead participants to report some practices that are not performed frequently. However, the results are valuable for readers who would interpret answers as No=Never and Yes=Seldom-to-Always.

In [10], software development practices used in New Zealand were reported using data collected from organizations through a questionnaire and interviews. Results show that large organizations tend to have more established processes, allocate more time on gathering requirements and perform more formal testing activities. Authors also reported that participating organizations spend about 25% of their effort on requirements specification (including high level design in some cases) and 27% on integration and testing.

In 2003, Cusumano et al. [11] focused on comparing use of software development practices and performances in different regions. The study was an updated and extension of a prior study by Cusumano and Kemerer in 1990 [5]. This comparison surveyed a total of 104 projects selected from 27 major companies located in USA, Japan, India and Europe. Performance comparison relied on effort spent and defects reported per lines of code (LOC). Authors also investigated usage of eleven development practices for sake of practice comparison. Findings suggest that in USA; architecture, requirements and design documentation are less popular. Another other major finding was that 36% of projects followed a waterfall-like approach in software development. Pair programming (35%), daily builds (36%) and paired testers (41%) were less popular choices whereas beta releases (73%) and regression testing (84%) were more commonly practiced.

The work of Denger et al. [13] utilized an online questionnaire for investigating the software development practices performed by companies producing medical devices. Results indicated a low level of process improvement model usage in this sector since only 10% of the companies followed CMMI and 4% followed SPICE. Also it is notable that all CMMI following companies have at least 50 employees. On contrary to the low process improvement model usage rate, 50% of the companies follow a defined process in software development always or frequently. An interesting finding is that respondents believe 63% of the software-related problems arise from requirements engineering activities. Another noteworthy result shows that natural language is the most popular requirements specification notation whereas formal notations are the least.

In an Italian study [40] effects of software engineering practices on the outcome of software projects were analyzed. Authors suggest that precise requirements, experienced project managers, risk management, collaborating teams and not compromising quality are significant factors for successful projects.

In a 2012 study in New Zealand, Kirk and Tempero [17] conducted a survey for software development practices. Survey included questions about company and participant profile and usage frequencies of practices for software development. They provide data and findings about practices in several phases such as planning, requirements, design, implementation, testing, support and delivery. Participants' opinions on beneficial and unhelpful practices were also investigated via open-ended questions within the scope of the survey. Results showed that organizations in New Zealand tend to adopt standard process models (e.g., Waterfall, Scrum) and process descriptions to suit organizational contexts rather than following them consistently. Another important finding is that most of the organizations claim to be agile and use iterations indeed, but fail to establish strong customer involvement in software development processes. Authors suggest that two of the areas that most issues are centered around were requirements and verification. Issues about requirements were about clarity and accessibility of requirements whereas verification-related issues included ineffective or missing reviews, unit tests, testing releases and patches and establishment of a designated testing environment.

In order to determine whether there is a gap between the current state-of-the-practice and state-of-the-art in SE, Vonken et al. conducted and reported in 2012 a SE survey in Netherlands [18]. From the analysis of the data that the authors obtained from 99 respondents, they extracted 22 interesting observations, some representing unexpected insights from an academic point of view, e.g.:

- Developers using waterfall are less positive about the quality of the development process and the quality of the software product than developers using agile methods.
- Agile developers share more team responsibilities with respect to product and process than waterfall developers.
- Traceability between requirements and software design seems important, but not to everyone.

As for the surveys in the Turkish context, we already discussed the following works [12, 14, 15] in Section 2.2.

### 2.2.3 Surveys on sub-areas of Software Engineering World-wide

There also exist other survey studies that have focused on specific sub-areas of SE such as requirement engineering, testing or quality or on a specific industry such as medical devices. Since this work is slightly related to those works as well, we searched for related papers in such related areas and industries as well. For brevity, we do not provide a comprehensive literature review of those surveys here, but instead we discuss several randomly-selected studies [23-25, 41-44] that include interesting findings related to our research as shown in Table 4 (sorted by year of study)

**Table 4-A summary of surveys on the subject of practices in sub-areas of SE (sorted by year of study)**

| Paper Reference | Scale/region | Year | Number of respondents | Goal/Focus area |
|---|---|---|---|---|
| [41] | USA | 1992 | 23 projects from 10 companies | Requirements modeling |
| [42] | Finland | 2000 | 12 companies | Software requirements engineering practices in Finland |
| [43] | Mostly USA | 2003 | 194 | Software requirements engineering practices |
| [23] | Province of Alberta, Canada | 2010 | 60 | Software testing practices in the province of Alberta, Canada |
| [44] | Finland | 2012 | 408 | Usage of Agile and lean methods in Finland |
| [24] | Canada | 2013 | 246 | Software testing practices in Canada |
| [25] | Turkey | 2013 | 136 | Software testing practices in Turkey |

The authors of [41] reported a survey-based review of the state of the practice in requirements modeling with data collected from 23 projects from 10 companies in 1992. Some notable findings from that study are as follows: (1) organizations use general-purpose CASE (Computer-Aided Software Engineering) tools with as much satisfaction as using specific-purpose tools for requirements engineering, (2) about 33%of the projects utilized prototyping usually for user interfaces, (3) most of the projects perform requirements validation and, (4) requirements volatility is considered a significant factor in requirements engineering as stated by all participants of the study.

State of the practice of requirements engineering in Finland was surveyed by Nikula et al. in 2000 [42]. An interesting finding is that: none of the 12 surveyed companies used formal specifications whereas all of them used natural language for specifying requirements most of the time.

In 2003, Neill and Laplante [43] surveyed requirements engineering practices in the USA. Survey included 194 participants who were graduate students. Considering the participants' above average academic profile, results might have included some bias in favor of more sophisticated practices, although there is not a clear-cut way to describe which practices are categorized as sophisticated. Prototyping is reported to be very popular with a usage rate of 60%. Among requirements specification notations, usage rate of informal representations such as natural language (51%) is followed by user scenarios or use cases (50%) and object-oriented techniques (30) whereas usage of formal languages is low (7%).

Adoption and usage of Agile and lean methods in Finnish software organizations is investigated by Rodriguez et al. [44]. Among 408 participants, 75% of them work in companies with less than 100 employees. So, we can say that smaller companies dominate this survey, which might be important since company size is usually believed to co-relate with selection of agile versus traditional methods. Findings of the study suggest that 55% of the participants use agile methods. A few interesting average usage frequency values, with a Likert scale measure of 1 (never) to 5 (systematically), were reposted in this research for unit testing (3.7), test-driven development (2.7), pair programming (2.4), and refactoring (3.4).

Garousi and his research reported [24] a Canada-wide survey in 2013 for investigating software testing practices in Canada. According to this study, which is a follow-up research among a sequence of surveys on testing practices in Canada, including two earlier regional provincial studies [23, 45], test-last development (59%) was found to be around three times more popular than test-driven development (20%) approaches. Other interesting findings from various practices as reported in [24] include: (1) a high percentage of participants (39%) do not use any test-case generation technique, (2) manual testing is more common than automated testing, (3) 43% of the respondents do not use any coverage metrics, and (4) testers are usually outnumbered by developers. Another important finding was that Agile methods are mostly implemented in smaller companies whereas traditional methods are more popular in larger companies.

Last but not least, the authors of the current article presented in a national Turkish conference in 2013 [25] the results of a survey on the software testing practices in Turkey. The current article is the extended version of the survey results with data from all 202 participants and presents the data and findings for all the SE practices.

When unfavorable practices are expressed, we hope that those items could be seen as "open problem" for the research community to work on.

## 3 SURVEY GOAL, DESIGN AND EXECUTION

In this section, we discuss the following aspects of our survey:

- Survey goal and research questions (Section 3.1)
- Survey design and questions (Section 3.2)
- Survey execution (Section 3.3)

### 3.1 SURVEY GOAL AND RESEARCH QUESTIONS

The approach we used in our survey study is the Goal, Question, Metric (GQM) methodology [46]. Stated using the GQM's goal template [46], the goal of our survey is:

- to characterize the SE practices in Turkey for the purpose of identifying the trends,
- and also to provide a view on the latest SE techniques, tools and metrics used by practitioners (professionals) and the challenges faced by them,
- to benefit both SE professionals and also researchers both in Turkey and world-wide,
- for observing the latest trends in the SE industry and identifying the areas of strength and weakness and encouraging more academia-industry collaborations.

Based on the above goal, we raise the following research questions (RQs):

RQ 1. What are the profiles and demographics of the practitioners, companies and projects taking part in the survey, e.g., academic degrees of professionals, industry sectors of the companies, and project sizes?

RQ 2. What are the overall characteristics of the Software Development Life Cycle (SDLC) and development processes in the companies?

RQ 3. What types of software requirement practices are used?

RQ 4. What types of software design practices are used?

RQ 5. What types of software development practices are used?

RQ 6. What types of software testing practices are used?

RQ 7. What types of software maintenance practices are used?

RQ 8. What types of software configuration management, release planning and support practices are used?

RQ 9. What types of software project management practices are used?

RQ 10. What are the trends in using SE tools?

RQ 11. What types of software quality assurance practices are used?

RQ 12. How much in-house research and also interaction with academia are conducted in firms?

Each of the above RQs is used to derive several "survey questions" and metrics in the following section.

### 3.2 Survey Design and Questions

We systematically designed an online survey based on our past survey experience in the Canadian context [23, 45]. We also benefited from the SWEBOK (version 2004) [37] in categorizing our survey and its questions. The 2004 version of the SWEBOK divides the SE discipline into 10 knowledge areas which correspond, in a one to one relationship, to our RQs 2-11 above.

The criteria we used in designing our survey's question set was to make sure that the questions are relevant to the industry and also to capture the most useful information. In order to develop a survey that would adequately cover the latest topics in SE while at the same time permitting us to economize on the number of questions, we reviewed the similar past surveys [2, 3, 5, 6, 8-15, 17] (as discussed in Section 2.2) and designed a draft set of questions. The authors then consulted with several industrial practitioners to do a careful peer review on the draft set of questions. Getting feedback from industrial partners in design of surveys is an approach followed in previous surveys as well (e.g., [23, 45]). The goal behind this phase in our survey design was to ensure that the terminology used in our survey was familiar to the participants. This step was necessary since, unfortunately, the SE terminology used in academia versus industry can sometimes be slightly different or even confusing. The feedbacks from the industrial practitioners were used to finalize the set of survey questions.

After the iterative design and improvement of the questions, we finally had 46 questions. The complete list of the questions used in the survey is shown in Table 5. Note that, for each of the RQs raised in Section 3.1, several "survey questions" and their corresponding metrics have been derived. Table 5 also shows the traceability among the RQs and the survey questions.

Most of our questions had quantitative pre-designed multiple-choice answers (e.g., the 1st question regarding the participant's current position), while a few had qualitative (free text) answers, e.g., "Q 46-Please provide a list of top three challenges that you have been seeing in your projects".

For brevity, we are not showing the pre-designed multiple-choice answers of the questions in Table 5, but they can be found in an online resource [47]. As we can see in Table 5, several questions (e.g., #14) had answers of type 5-point Likert scale. For those questions, the five choices were: never (1), seldom (2), sometimes (3), frequently (4), and always (5).

**Table 5-List of the questions developed and used in the survey**

| RQ | Aspect | Survey Questions (and Metrics) | Type of Answers | | | | | |
|---|---|---|---|---|---|---|---|---|
| | | | Single answer | Multiple answers could be chosen | Likert scale | Integer | Free text field | Binary answer (yes or no) |
| 1 | Profiles and demographics of practitioners, | Practitioners: | | | | | | |
| | | 1. What is (are) your current position(s)? | | x | | | x | |
| | companies and projects | 2. How many years of work experience do you have in IT and software development industries? | | | | x | | |
| | | 3. What is your highest academic degree? | x | | | | | |
| | | 4. What is (are) your university degree(s) in? | | x | | | x | |
| | | 5. Please choose the city that you work in? | | x | | | x | |
| | | Companies: | | | | | | |
| | | 6. What is the target sector of the products developed by your company? | | x | | | x | |
| | | 7. What is the size of your company (number of employees)? | | x | | | | |
| | | Projects: | | | | | | |
| | | 8. What kind of software do you develop? | | x | | | | |
| | | 9. How do you classify the software packages that you develop in your company? | | x | | | x | |
| | | 10. In case you have measured the size measure in your latest development project, please indicate the methodology used (COSMIC Function Points, IFPUG Function Points, Use-Case Points, Story Points, Lines of Code, etc.) and the approximate size of your recent project? | | x | | | x | |
| | | 11. What is the number of software projects undertaken in your company the past two years? | | x | | | | |
| 2 | Overall characteristics of the Software Development Life Cycle (SDLC) and development processes | 12. How much effort in % is spent by your team in the past projects on different SDLC phases? | | | | x | | |
| | | 13. How much are you involved in different SDLC phases? | | | x | | | |
| | | 14. How much challenge do you experience in each of the SDLC phases/tasks? | | | x | | | |
| | | 15. What kind(s) of software development methodologies do you use in your team? | | x | | | x | |
| | | 16. Do you formally follow any of the well-known process improvement models? | | x | | | x | |
| | | 17. How often are various SE process practices carried out by your organization? | | | x | | | |
| 3 | Software requirement | 18. How often are various software requirement practices carried out by your organization? | | | x | | | |
| | | 19. What type of notation does your team use to document software requirements? | | x | | | x | |
| 4 | Software design | 20. How often are various software design practices carried out by your organization? | | | x | | | |
| | | 21. What types of software design activities do you perform? | | | x | | | |
| | | 22. How important is each of the software quality factors when you design a software system? | | | x | | | |
| 5 | Software development | 23. Which programming languages do you use in your company? | | x | | | x | |
| | | 24. How often are various software development practices carried out by your organization? | | | x | | | |
| 6 | Software testing | 25. How often are various software testing practices carried out by your organization? | | | x | | | |
| | | 26. In terms of the type and phasing of testing during the SDLC, what is the type of your test activities? | | | x | | x | |
| | | 27. How often do you conduct the different types of test activities? | | | x | | x | |
| | | 28. In your current or most recent software project, what test technique did the team use to generate test cases? | | x | | | x | |
| | | 29. Overall in all of your past projects, how much automated versus manual testing have you done? | | | | x | | |
| | | 30. How often do you use various code (test) coverage metrics in your test activities? | | | x | | x | |
| | | 31. How often do you use various other test and quality metrics do you explicitly measure in your projects? | | | x | | x | |
| | | 32. Test management: Please identify the ratio of testers to developers (testers: developers) in your current or most recent project. For example, if you had on average one tester for every 2 developers, the answer would be 1:2. | | x | | | | |
| | | 33. What criteria are used in your projects to decide that the testing activities are terminated/completed? | | x | | | x | |
| 7 | Software maintenance | 34. How often are various software maintenance practices carried out by your organization? | | | x | | | |
| | | 35. What is the level of challenge you experience in various software maintenance practices? | | | x | | | |
| | | 36. If you have reported challenges in the above question, please summarize the challenges. | | | | | x | |
| | | 37. What types of challenges have you noticed in software maintenance tasks? | | x | | | x | |
| 8 | Software configuration management, release planning and support practices | 38. How often are various software release and delivery practices carried out by your organization? | | | x | | | |
| | | 39. What types of practices are carried out by your organization while supporting your software projects? | | | | | | x |
| 9 | Software project management | 40. How often are various software project management practices carried out by your organization? | | | x | | | |
| | | 41. What types of software project management practices are carried out by your organization? | | | x | | | |
| 10 | Usage of SE tools | 42. Which tools are used for different software engineering tasks? | | x | | | x | |
| 11 | Software quality assurance | 43. What types of software quality assurance practices are carried out by your organization? | | | x | | | |
| 12 | Research Activities and Interaction with Academia | 44. Does your company have a dedicated R&D department/unit, specializing on SE research and aiming at developing new better ways/techniques to develop software? | | x | | | | |
| | | 45. How often do you interact (talk) to SE university researchers about your problems/challenges. | | x | | | | |
| | | 46. How often do you read technical papers (articles) published in SE journals, conferences or workshops? | | x | | | | |

### 3.3 Survey Execution

The survey was hosted on an online survey hosting service called SurveyMonkey (www.surveymonkey.com). Research ethics approval for the survey was obtained from the Middle East Technical University's Applied Ethics Research Center in May 2013. The survey was available to participants for three months during the summer of 2013. Participants were asked to complete the survey online and participation was voluntary and anonymous. Respondents could withdraw their results from the survey at any time and, as per the ethics guidelines, researchers agreed to publish only summary and aggregate information from the survey.

To ensure that we would receive as many responses as possible, we followed the following survey invitation strategy. We sent email invitations to our network of partners/contracts in Turkish software companies. The sample included companies with development offices in Turkey, whether those offices were the company's head office or not. Since development offices (conducting software development and testing tasks) were of interest to us in studying their SE practices, whether or not the company had a head office in Turkey was not an important factor in our study. The same criterion was also applied in the two earlier surveys [23, 45] as well. We also made public invitations to the Turkish software engineering community by placing messages in social media, e.g., LinkedIn. 202 software practitioners across Turkey took part in our online survey and our survey data includes all those responses.

## 4 Survey Results and Findings

We report in this section the survey results and analyze the findings, in the following order:

- Profiles and demographics of practitioners, companies and projects (Section 4.1)
- Overall characteristics of the SDLC and development processes (Section 4.2)
- Software requirement (Section 4.3)
- Software design (Section 4.4)
- Software development (Section 4.5)
- Software testing (Section 4.6)
- Software maintenance (Section 4.7)
- Software configuration management, release planning and support practices (Section 4.8)
- Software project management (Section 4.9)
- Usage of SE tools (Section 4.10)
- Software quality assurance (Section 4.11)
- Research activities and interaction with academia (Section 4.12)

### 4.1 Profiles and Demographics

RQ 1 wanted to determine the profiles and demographics of the practitioners, companies and projects taking part in the survey, e.g., academic degrees of professionals, industry sectors of the companies, and project sizes. We discuss the results next.

#### 4.1.1 Participants Profile

202 software engineers from Turkish software industry participated in our survey. Here we report the following results:

- Participants positions (Q 1)
- Work experience in software development (Q 2)
- Highest academic degree (Q 3)
- Academic major (Q 4)
- Geographical location (Q 5)

##### 4.1.1.1 Participants Positions (Q 1)

The positions of respondents in this survey are shown in Figure 1. Note that since this was a multiple-choice question, multiple roles could be recorded, e.g., a person can be a developer and a tester at the same time. As the figure shows, most of the participants were "software developers" and "software engineers". Note that the axes showing the numbers in all figures in this section are number of participants (unless otherwise specified).

We should note that, as it has been established in studies on information quality (for example by Garvin [48]), people in different positions see and rate importance of different issues differently and in general have varying viewpoints on SE and related processes.

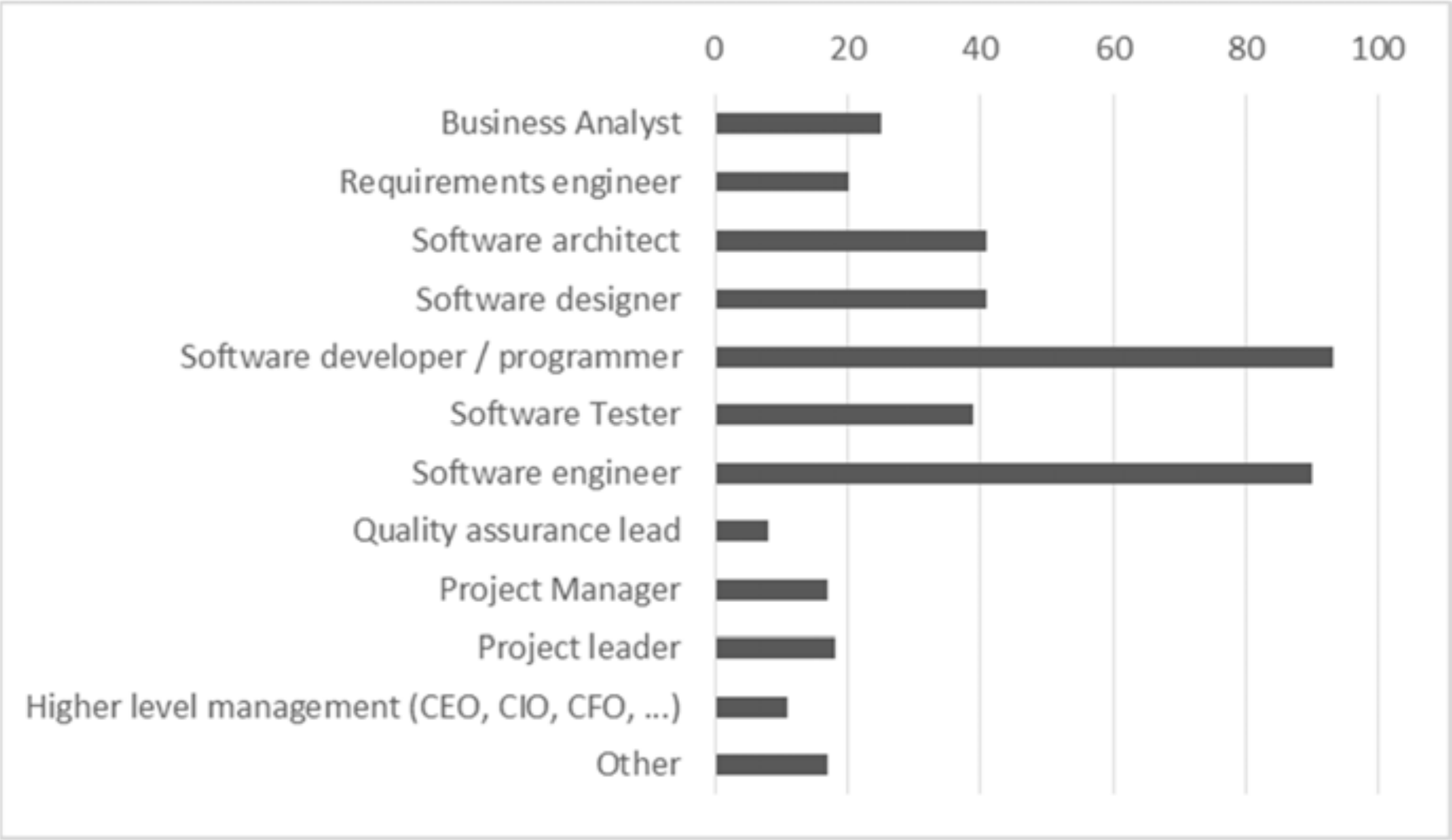

**Figure 1-Respondent positions**

Since this question was a multiple-choice question, we measured the frequency of the cases in which a respondent had reported more than one position. Results are shown in Figure 2. More than half of the respondents reported to be in only one position. Surprisingly, fourteen respondents reported to have been assigned to six or more roles concurrently.

To compare our results with the Canadian survey conducted in 2010 [24], the data from that study have also been shown. We can notice that there the two trends are quite similar in that most participants in both studies have one single position.

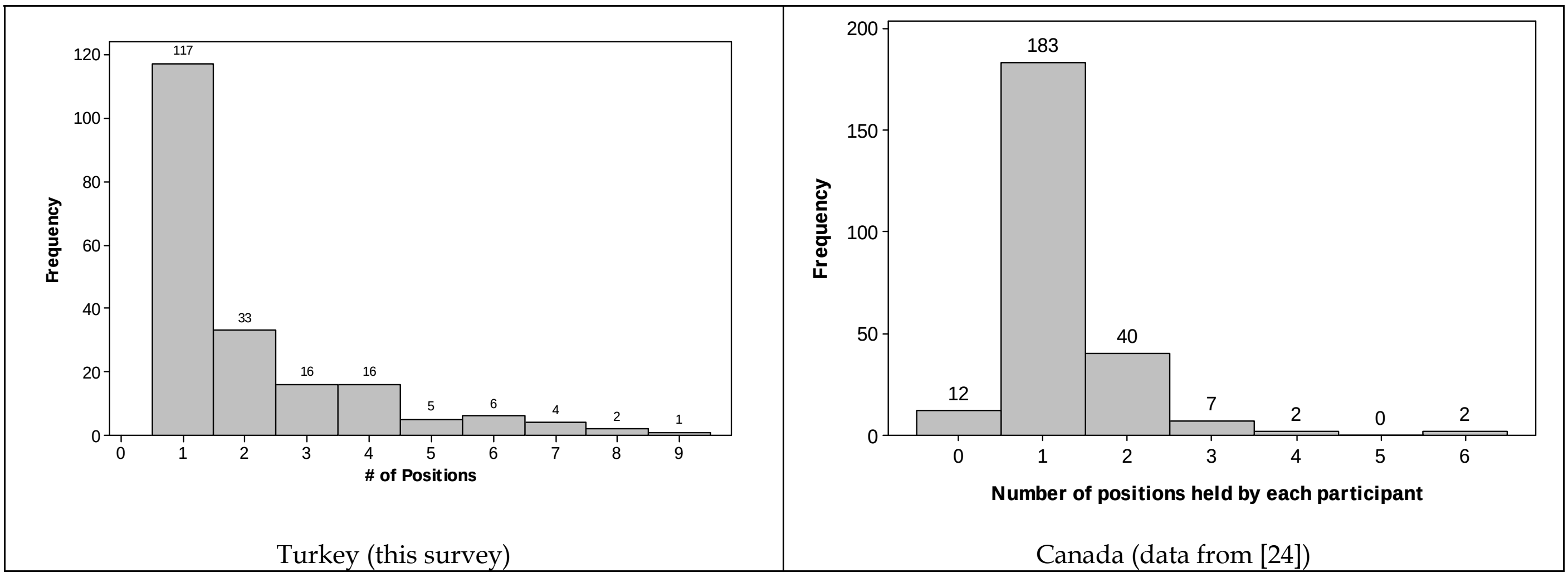

**Figure 2-Histogram of number of positions held by each participant**

### 4.1.1.2 Work Experience in Software Development (Q 2)

The left-side of Figure 3 shows, as an individual-value plot, the distribution of participants' years of work experience in the software industry. The average and median values are 7.8 and 6 years, respectively.

As we can visually observe in Figure 3, most of the graph is skewed downwards, denoting the "young" trend of the participants' pool. Around 44% and 79% of respondents had fewer than 5 and 10 years of working experience in the software industry, respectively. Only 21% of the respondents had more than 10 years of experience. We acknowledge the positive impact of diversity in participants' population as responses from participants with various lengths of work experiences can help the survey to gather valuable inputs from a wider audience base.

To compare our results with the Canadian survey [24], the data from that study have also been shown. The average and median work experience values of the Canadian respondents were 9.5 and 8 years, respectively. Thus, we can see that the Turkish SE practitioner community is, in general, slightly younger than the Canadian counterpart.

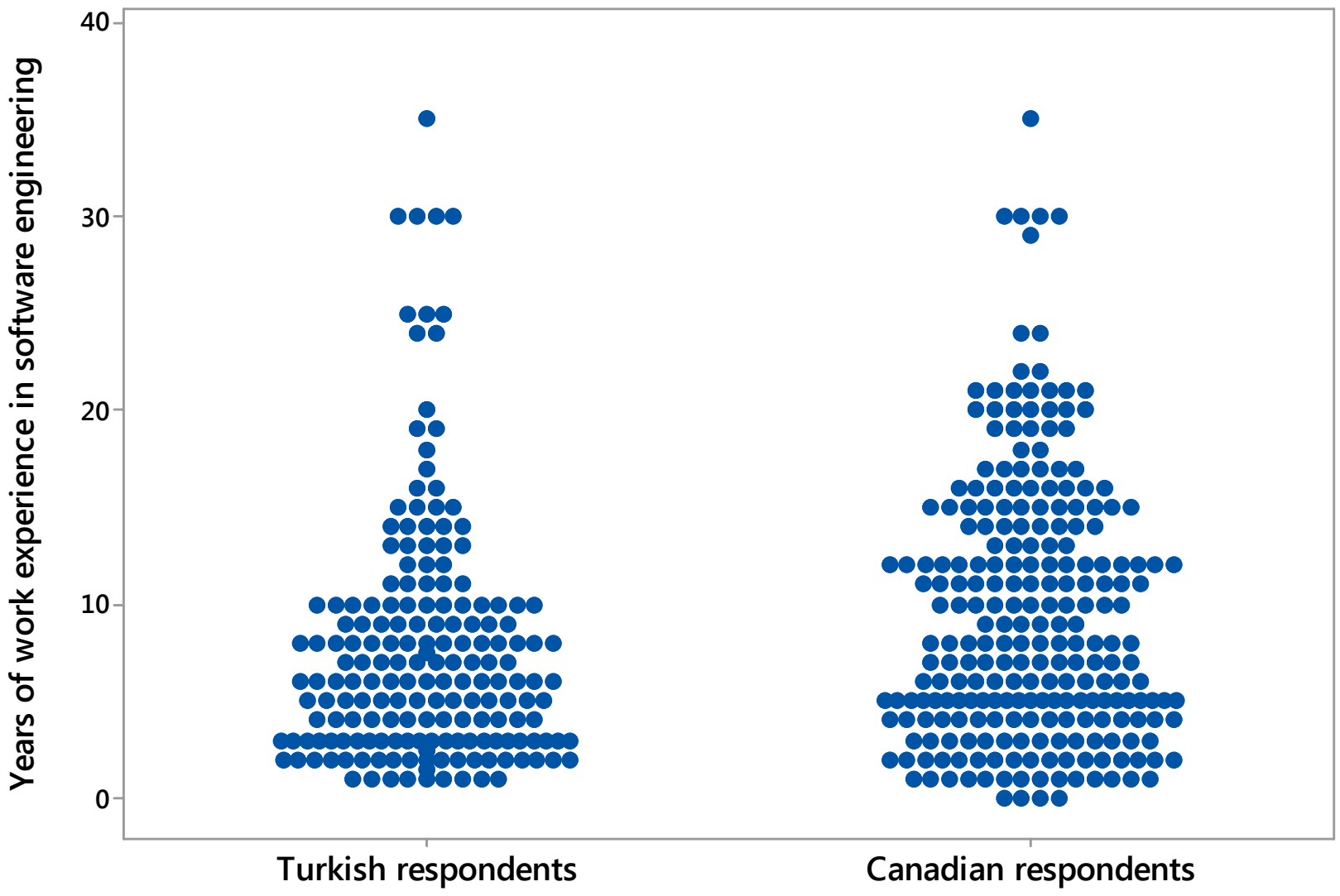

**Figure 3-Respondents' work experience (number of years) in SE as an individual-value plot**

#### 4.1.1.3 Highest Academic Degree (Q 3)

In order to understand the respondents' educational background, participants were asked to provide their highest academic degree. Shown in Figure 4, the result reveals that 54% and 40% of respondents have a Bachelor's and a Master's degree, respectively. 5, 4, and 4 respondents had High school or lower, college and PhD degrees respectively.

Data from the Canadian survey [24] have also been shown. We can notice that the two trends are quite similar, with the exception that there are a few PhDs in the Turkish respondents, but none in Canadian respondents' pool.

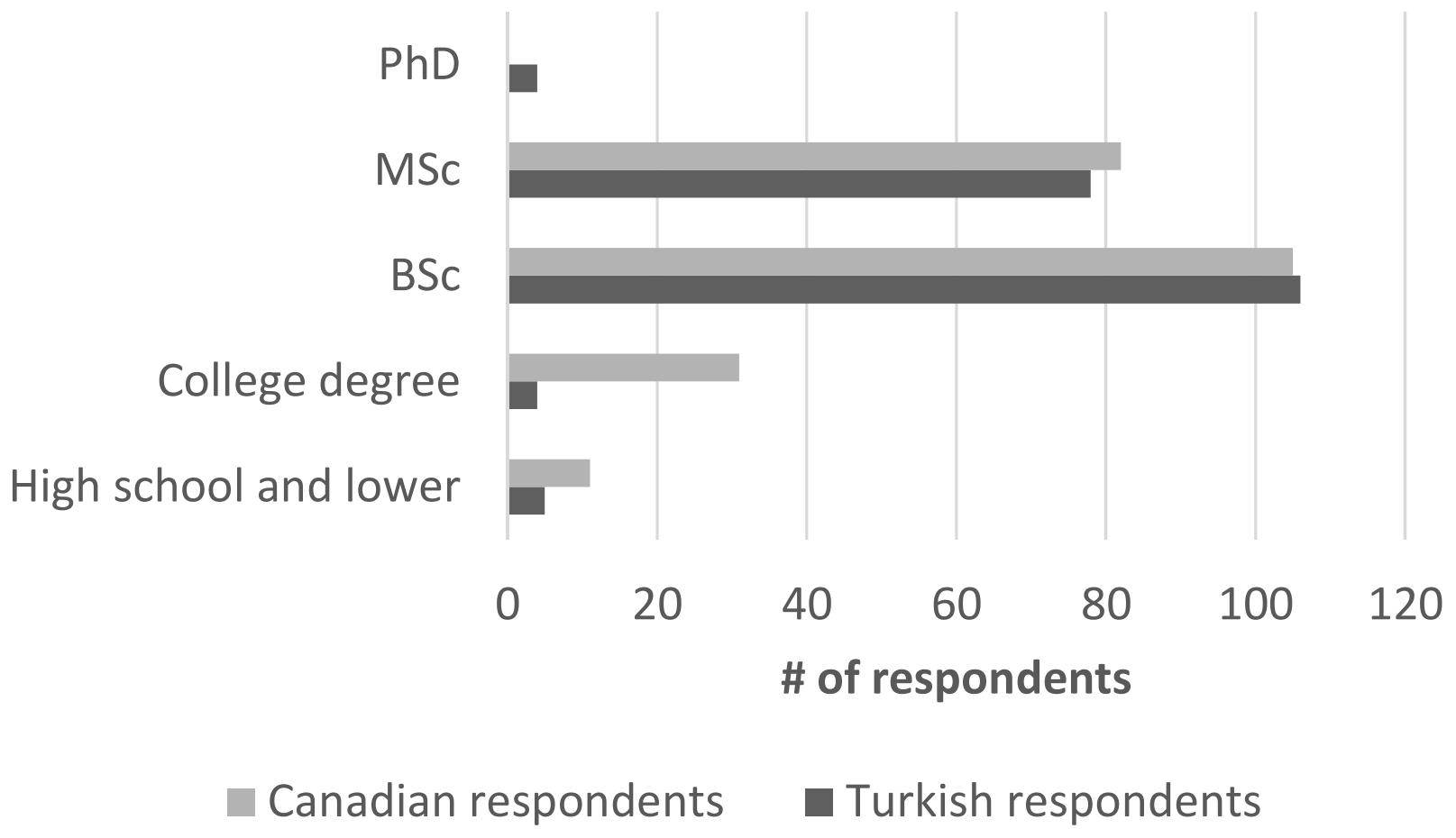

**Figure 4-Respondents Highest Academic Degrees**

#### 4.1.1.4 Academic Major (Q 4)

As a follow-up question to the previous question, in order to understand the respondents' educational skill-set, participants were asked to provide their last academic major (e.g., software engineering, and computer engineering). The results are shown in Figure 5.

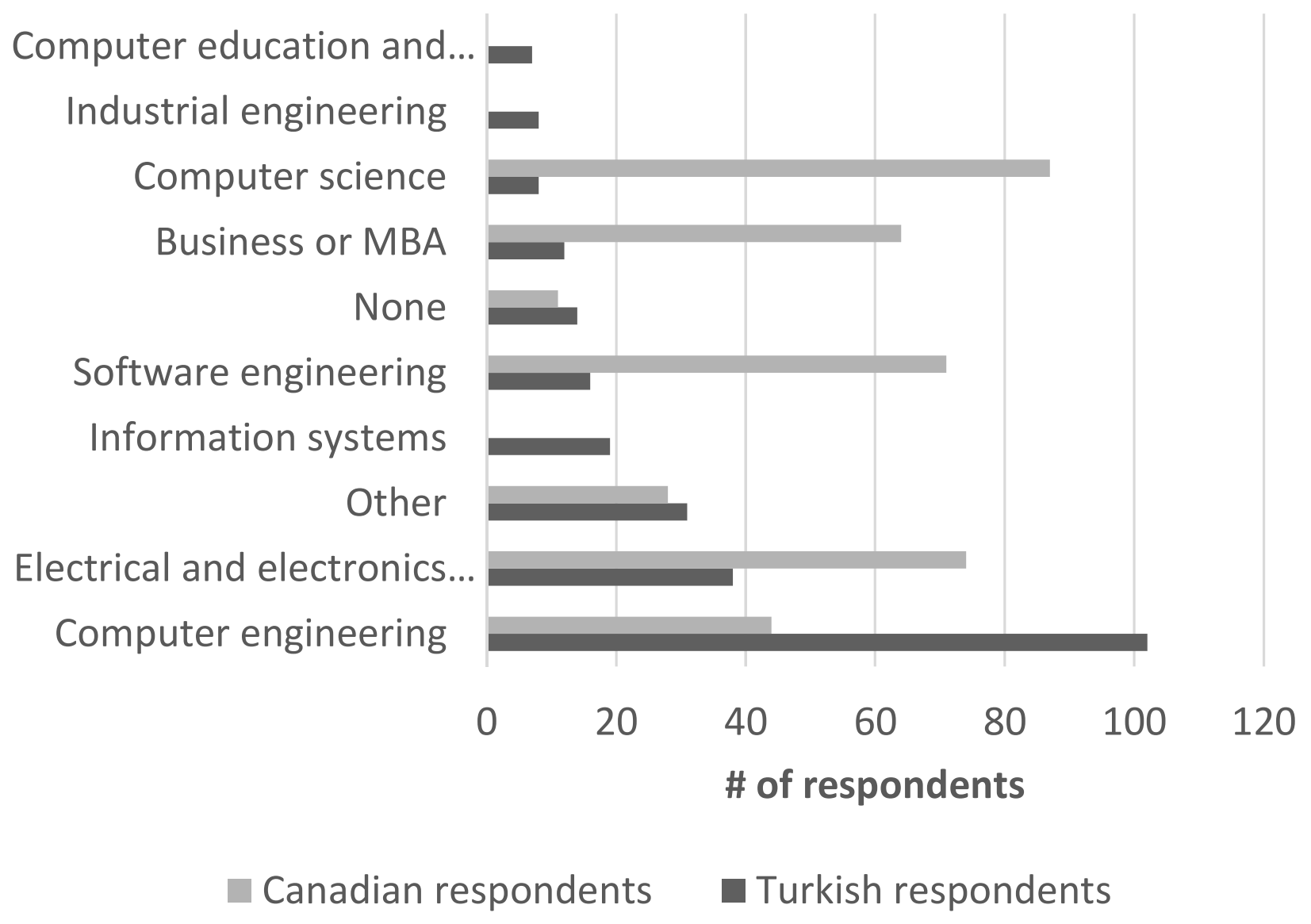

**Figure 5-Respondents Academic Major**

Computer Engineering (with 40.0% of the participants), Electrical and Electronics Engineering (14.9%) and other (12.2%) are the top three academic majors in Turkey. Other degrees included majors in Mathematics, Physics, Statistics, Economics and other engineering disciplines (e.g., Mechanical, Communication, and Geological). From our personal discussion with several practitioners from these backgrounds, we found that such practitioners are invaluable assets to their teams since they carry a wealth of knowledge in their "business domains". By comparing the Turkish and the Canadian trends [24], we can notice that there are differences between the two trends. Turkish software practitioners mostly hold Computer Engineering university degrees, while the Canadian counterparts hold Computer Science degrees. This can be explained as most schools in Turkey have Computer Engineering departments which correspond to the Computer Science departments in the North America. Also, we can observe that there are more participants with the software engineering degree in Canada compared to Turkey. The root cause for this is perhaps since there are only a handful number of Turkish universities who have started to offer software engineering degrees in recent years.

#### 4.1.1.5 Geographical Location (Q 5)

Similar to our previous Canada-wide survey [24], our survey had a question to record the city of residence of each respondent. Our goal was to reach out to as many cities across Turkey as possible and to ensure that all regions where there is a presence of software industry are reasonably well represented in the data set.

Regional breakdown of the respondents are shown in Figure 6. Not surprisingly, Ankara and then Istanbul are the cities where most of the respondents are located. Due to the location of the authors (Ankara) and the fact that most of our contacts were local, about 68% of the respondents were from Ankara. The "other" cities in Figure 6 included: Afyon, Balıkesir, Eskişehir, Gebze, Kocaeli and Samsun. According to the survey report of Akkaya et al. [32], as of 2012, 47% and 33% of the Turkish software firms were located in Istanbul and Ankara, respectively. The remaining 20% were located in other cities across Turkey. The participation ratio in our survey from various cities (percentage values visible in Figure 6) are slightly different than the data by Akkaya et al. [32], however there is slight similarity denoting reasonable and realistic mix of respondents from across Turkey in our survey.

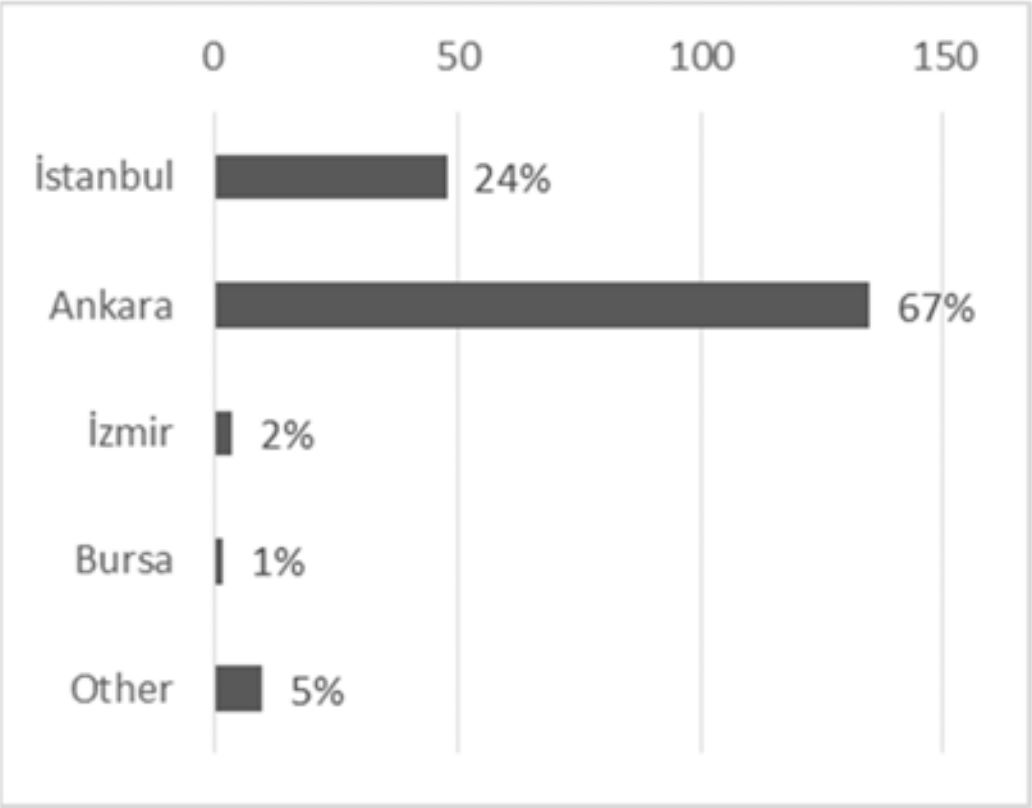

**Figure 6- Regional Breakdown of the Respondents**

Figure 7 shows the map of Turkey highlighting the survey respondents' locations. As we can see, we can see the distribution of the respondents are from north-west quadrant of Turkey. This somewhat matches the reality as per other reports, e.g., work of Akkaya et al. [32], and also the authors' personal observations. We can say that our respondents were mostly located in the country's north-west quadrant.

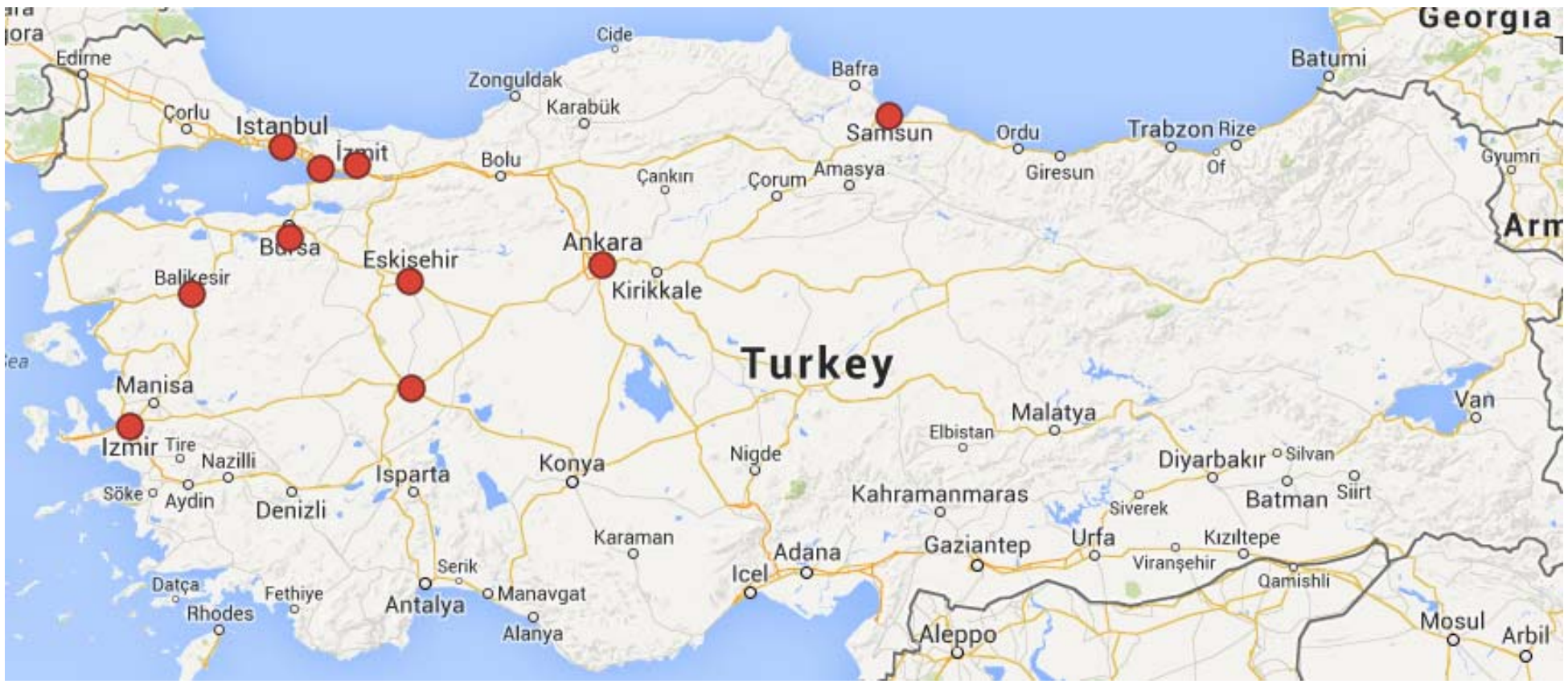

**Figure 7- Map of Turkey showing the survey respondents**

### 4.1.2 Profile of Companies

Next we report the following results:

- Target sector of products (Q 6)
- Company size (Q 7)

#### 4.1.2.1 Target Sector of Products (Q 6)

The next question was about the type of products developed by the software organizations (Figure 8). Ten possible choices were pre-provided in the questionnaire (as shown in Figure 8), which were designed in discussions with our industry partners. The most popular target sector was: military and defense. The 2nd and 3rd ranks were IT and Telecommunication, and the Government sector (not including military and defense).

The software industries of the military and defense domains are quite strong in Turkey, especially in Ankara the capital city, and many SE practitioners work for those firms. The "other" category in Figure 8 as reported by the respondents included: food, e-commerce, automotive, textile, gaming and game technologies, education, transportation and logistics. As we can see in Figure 8, there is a good mix of respondents from various application domains.

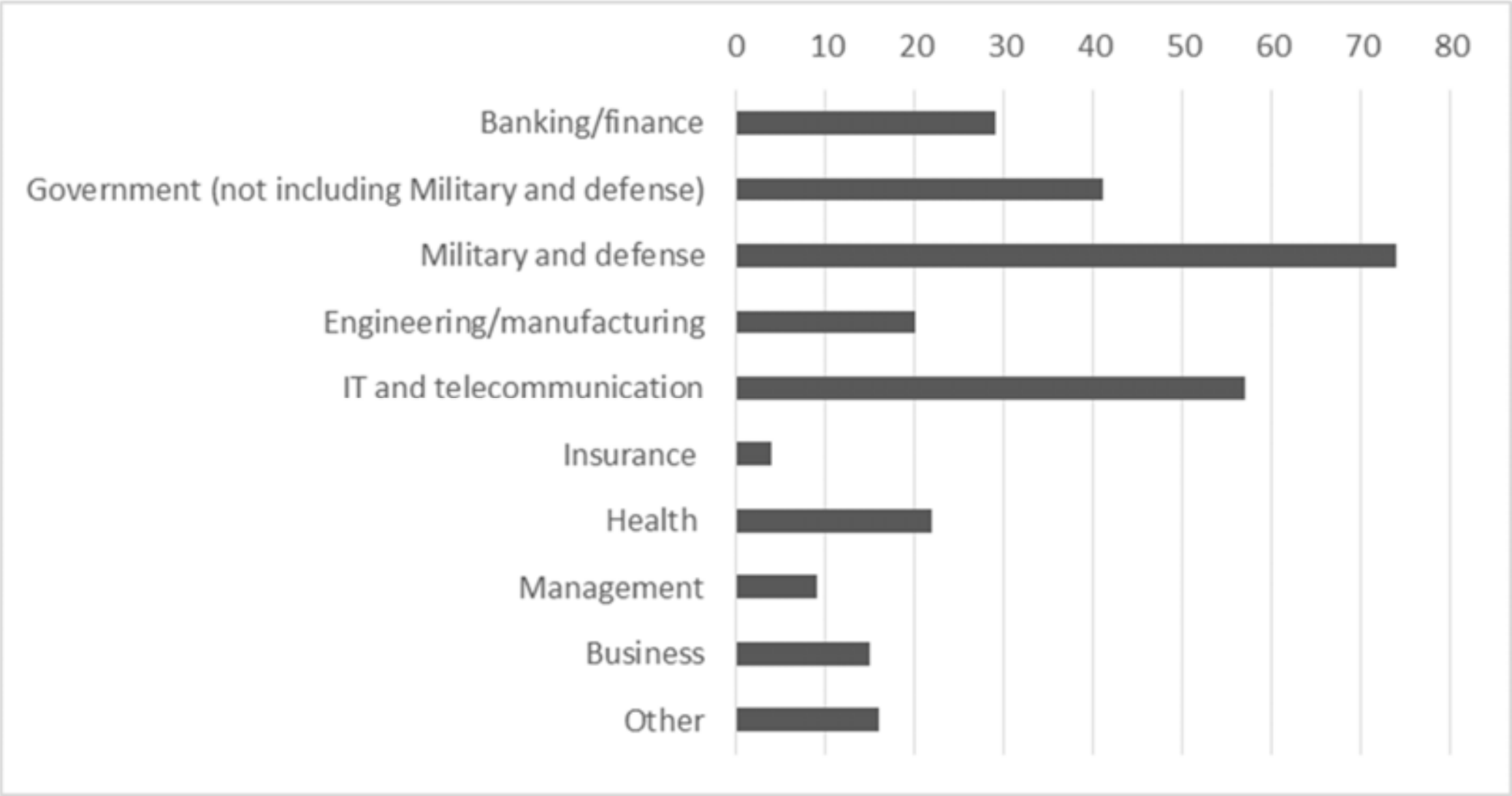

**Figure 8-Type of products developed by software organizations surveyed**

### 4.1.2.2 Company Size (Q 7)

It was important to also know the size of the companies to ensure that we could correlate the survey results with the number of employees in our survey. The histogram of company sizes in terms of number of employees is shown in Figure 9. It is interesting that more than 80 participants from large corporations with 500+ employees completed our survey. A good mix of respondents from other ranges was also present in our survey pool. This in turn would enable our analysis to cover a wider spectrum of inputs in terms of company sizes.

In comparison of the Turkish and the Canadian trends [24] in Figure 9, we can notice that there are differences between the two trends. In terms of percentage, the current Turkish survey seems to include more large companies (with 500+ employees) than small companies, compared to Canada.

| | |
|---|---|
| 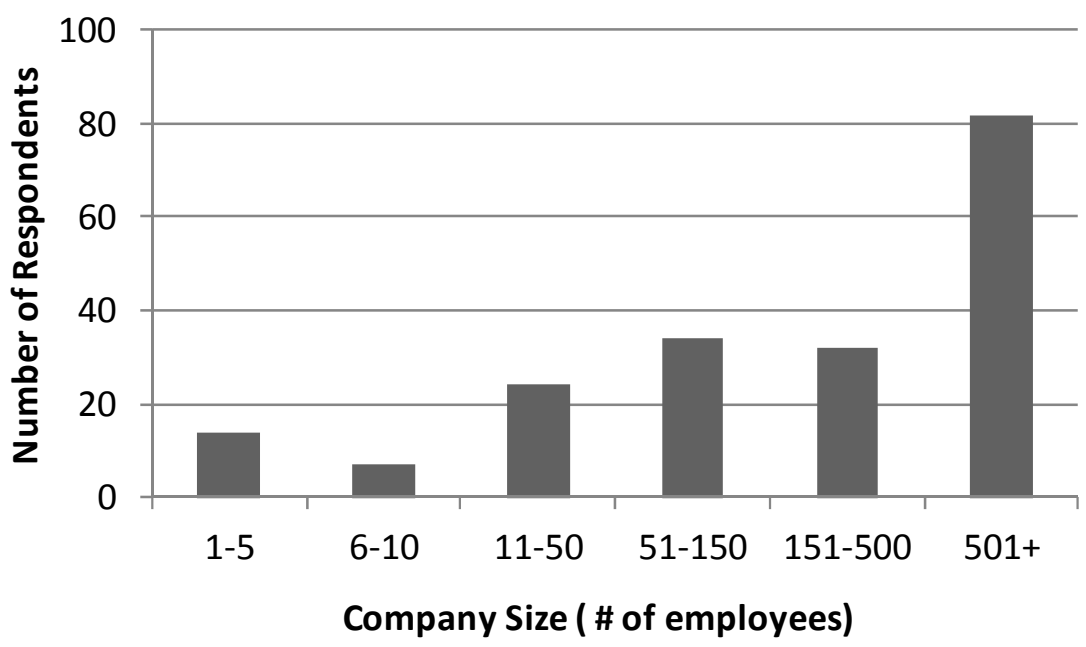 | 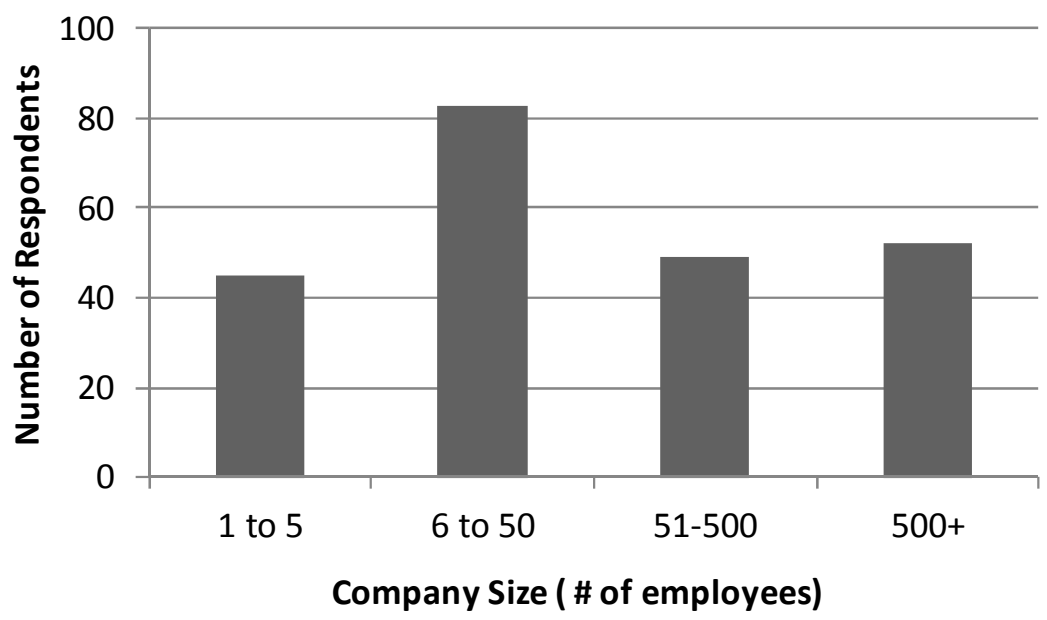 |
| Turkey (this survey) | Canada (data from [24]) |

**Figure 9- Company Size**

### 4.1.3 Software Projects Profile

We report the following results in the next sections:

- Intended industry types (Q 8)
- Software types (Q 9)
- Size measurement (Q 10)
- Number of projects (Q 11)

### 4.1.3.1 Intended Industry Types (Q 8)

For understanding the profile of software projects that survey participants contribute to, first question was about the intended industry types of the products produced. As Figure 10 shows, participants mostly produce proprietary or in-house software. Development of general-use software products is the lowest with 16%. Participants were allowed to provide multiple answers, but only 22% mentioned they develop products for more than one intended industry type.

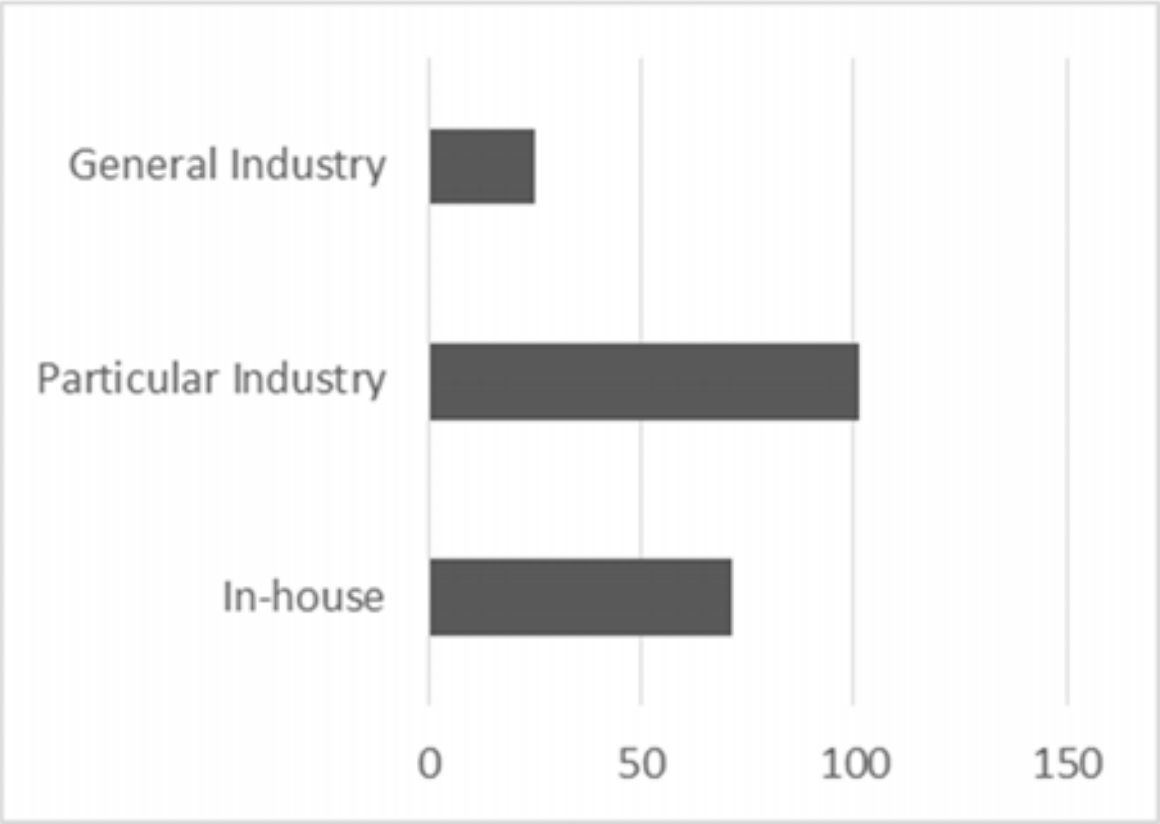

**Figure 10- Intended industry types of software products**

**4.1.3.2 Software Types (Q 9)**

Following the question above, participants were asked to categorize the software packages developed in their companies in a given set of categories. Results are shown in Figure 11. Top categories were: (1) business applications software, (2) safety-critical and mission-critical software, and (3) web applications. Similar to the previous question, respondents were able to provide multiple answers. 42% of the participants reported their companies develop more than one type of software. The "other" category in Figure 11 include: gaming, animation, software for hardware testing, telecommunication software, data management, data analysis, and data visualization software. This question is slightly related to question 6 (target sector of products).

It is logical to say that the type of software products developed in the Turkish firms should match the type of demands. Thus, we can say that business applications and mission-critical software have high market demands among our respondents.

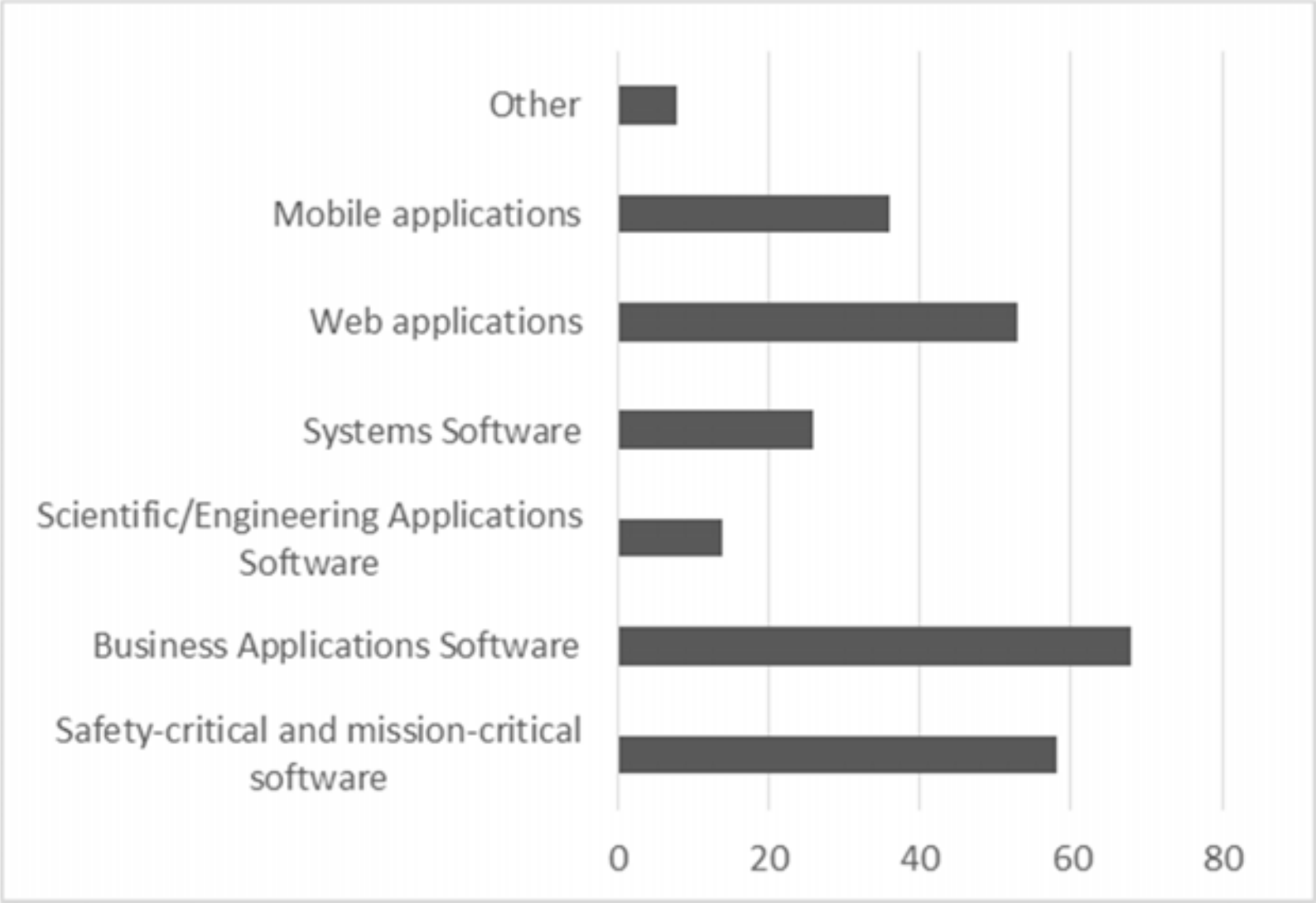

**Figure 11- Software types**

**4.1.3.3 Size Measurement (Q 10)**

We were interested to know the types of size measurement methods in use in Turkey and also the approximate size of software projects. 135 out of the 202 respondents answered this question and 73 (54%) of those respondents reported not using any size measurement methods, whereas 33% mentioned that they have measured lines of code (LOC) and 15% used use-case points. Utilization of story points (4.4%), COSMIC (3%) and IFPUG (1.5%) are low among the participants. Figure 12 visualizes the trend.

Among the respondents who reported using software size measurement methods, 15 respondents provided the approximate size of their recent software products. 14 respondents provided the approximate LOC of their projects, while one respondent reported size in use-case points (500 use-case points). A box-plot of KLOC sizes of projects is shown in Figure 12. The KLOC sizes ranged from 500 LOC to 3.5M LOC, with an average value of 844 KLOC.

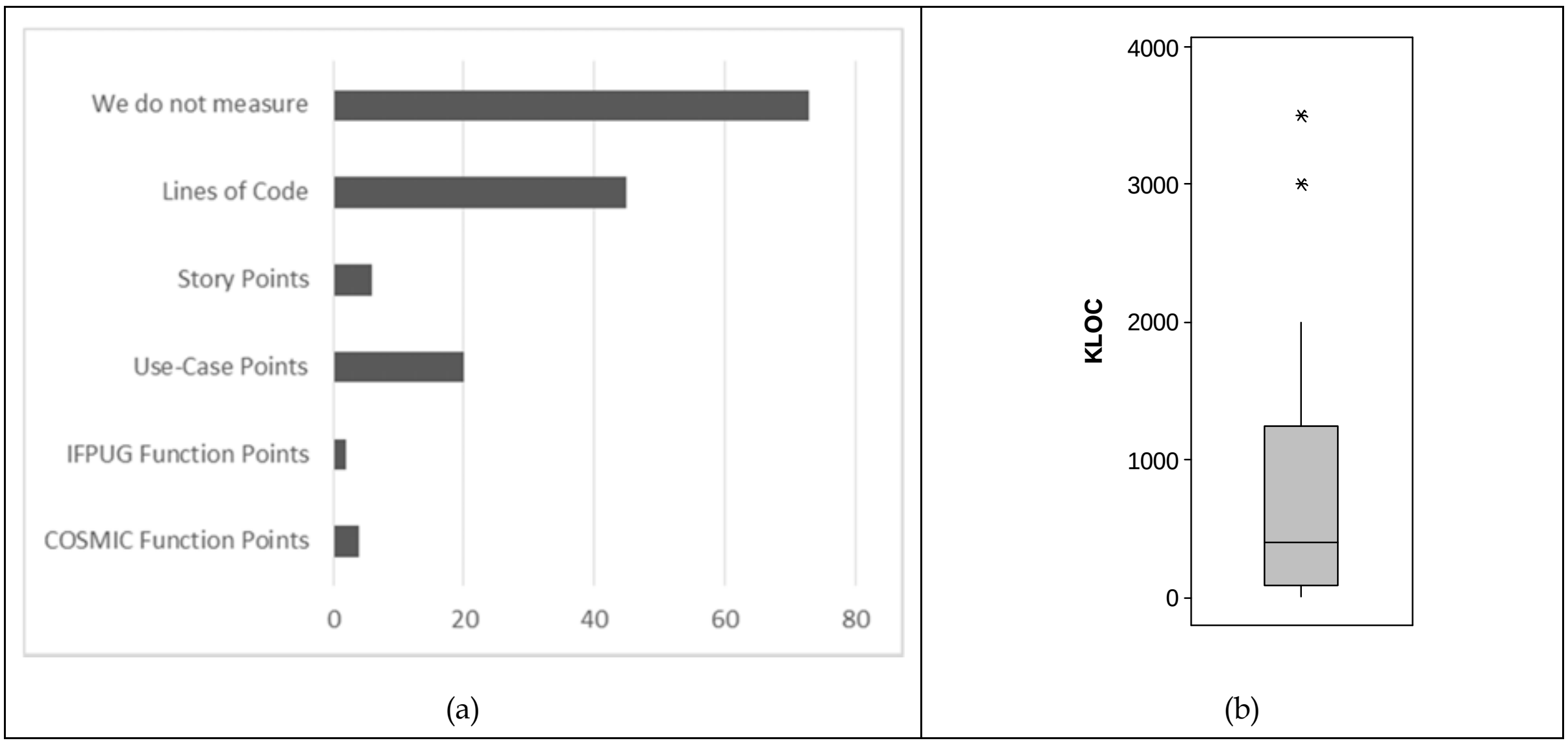

**Figure 12- (a): Usage of size measurement methods. (b): box-plot of KLOC sizes of projects (# of data points=14)**

#### 4.1.3.4 Number of Projects (Q 11)

This question asked the number of projects undertaken in last two years (2011-2013) by respondents' companies. Most of the companies carried out less than 10 projects (63%) in the last two years. As depicted in Figure 13, as the number of projects decrease (down the vertical axis), number of participants increase (on the horizontal axis). In other words; participants, who work in companies that undertake high number of projects, are represented less in our survey data.

Since several companies might be represented multiple times in our data set for this question (i.e., having several respondents from one single company), Figure 13 does not provide an overall picture for the types of companies, but it provides a profile of respondents in terms of the types of companies they work in.

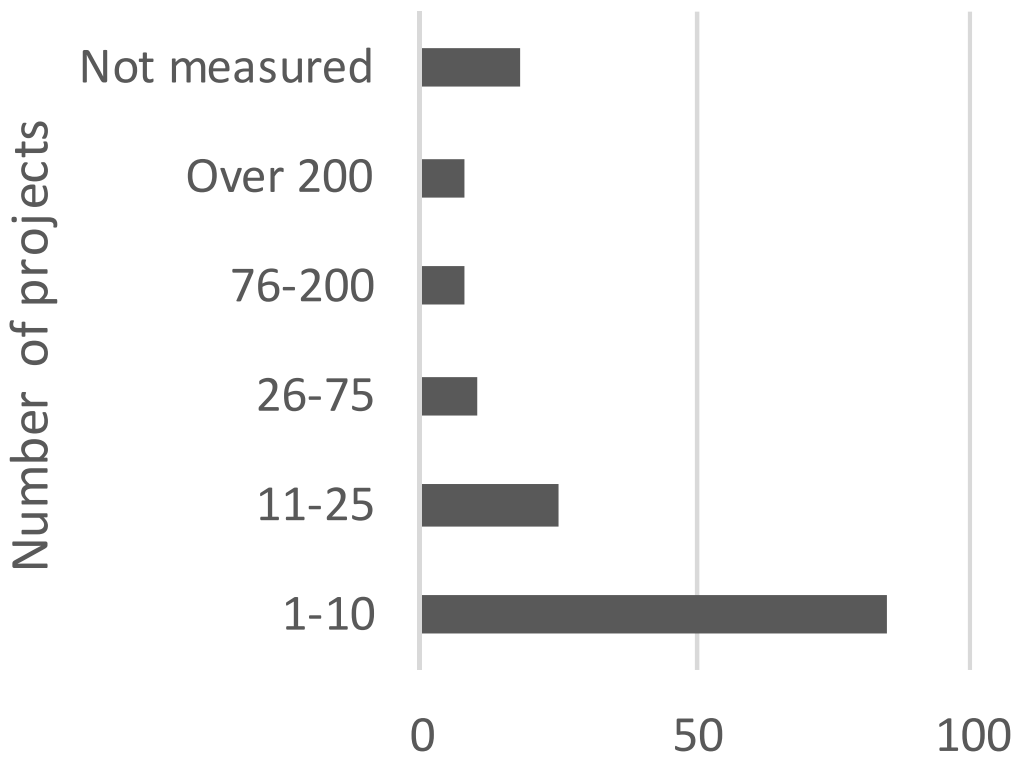

**Figure 13- Number of projects completed during 2011-2013 by companies**

### 4.2 Overall Software Development Life-Cycle (SDLC) and Processes

We report the following results:

- Efforts spent on SDLC phases (Q 12)
- Degree of participant's involvement in phases (Q 13)
- Level of challenge experienced in phases/tasks (Q 14)
- Software development methodologies (Q 15)

- Following process improvement models (Q 16)
- Process-related practices (Q 17)

### 4.2.1 Efforts Spent on SDLC Phases (Q 12)

The first question of this section was about the percentages of average effort spent by teams on each SDLC phase. As shown in Figure 14 as a box-plot, the development phase is the phase where teams spend the most effort on (with an average of 31%). We can state the following noteworthy observations from these trends:

- Software testing, requirements, design and maintenance have similar average values (14%, 12%, 12% and 11% respectively).
- Respondents report having spent significantly lower effort amounts on project management, technical documentation, configuration management, process management and quality assurance phases.
- The effort ranges for some of the phases are so "wide", while the ranges for some are quite narrow. This difference in range sizes could denote that different software engineers and companies spend different amounts of efforts on different SDLC phases, as one would expect. To analyze correlative factors for this trend, cross-factors analysis of SDLC-phases and other metrics (e.g., industry types) is needed in future works.

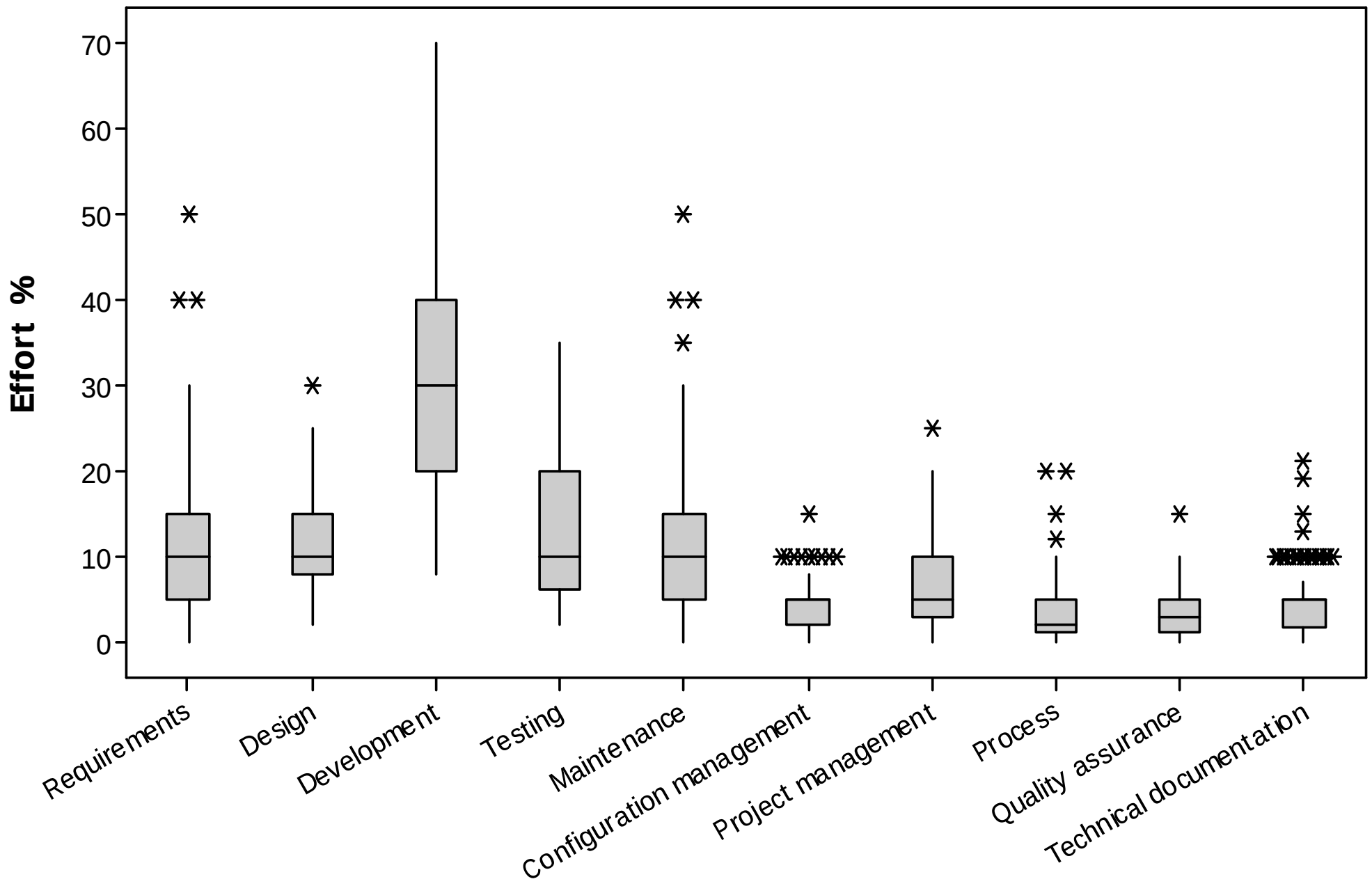

**Average values:**

| | | | | | | | | | |
|---|---|---|---|---|---|---|---|---|---|
| 12.51% | 12.00% | 31.28% | 13.85% | 11.05% | 4.28% | 6.95% | 3.95% | 3.60% | 4.98% |

**Median values:**

| | | | | | | | | | |
|---|---|---|---|---|---|---|---|---|---|
| 10 | 10 | 30 | 10 | 10 | 5 | 5 | 2 | 3 | 5 |

**Figure 14- Box-plot showing the effort spent on SDLC phases**

### 4.2.2 Degree of Participant's Involvement in Phases (Q 13)

Respondents were asked about their involvement in SDLC phases. Recall from Section 3.2 that, for this question, a 5-point Likert scale was utilized for providing the answers. These five choices were: never (1), seldom (2), sometimes (3), frequently (4), and always (5).

Results are shown in Figure 15. Respondents have reported that they participate the most in implementation (development), requirements, design and testing phases. Interestingly, these results are in line with those of the previous question, i.e., these four phases are the same four phases that teams spend most effort for.

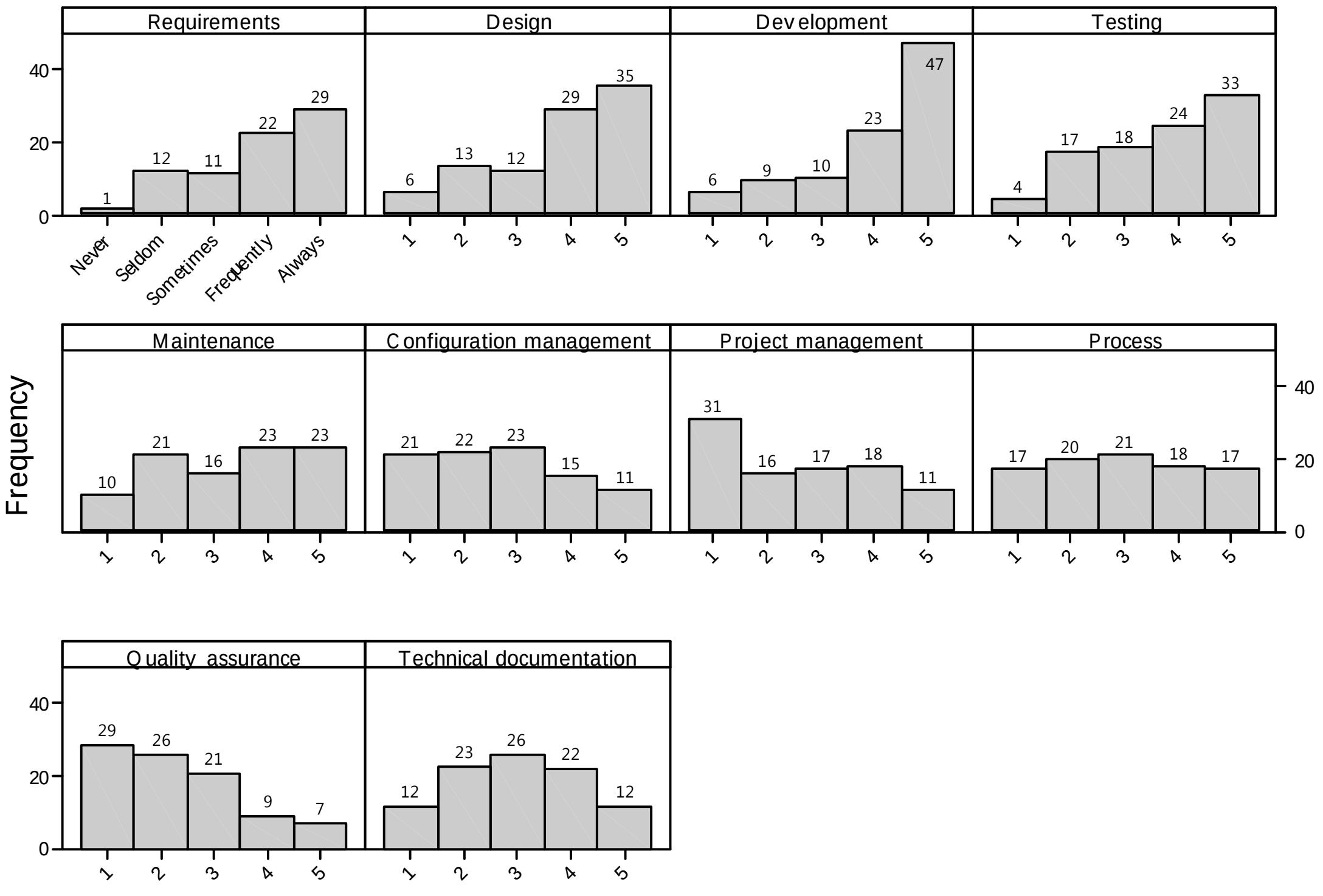

**Average values:**

| | | | |
|---|---|---|---|
| 3.88 | 3.78 | 4.01 | 3.68 |
| 3.30 | 2.71 | 2.59 | 2.98 |
| 2.34 | 2.99 | | |

**Median values:**

| | | | |
|---|---|---|---|
| 4 | 4 | 4 | 4 |
| 3 | 3 | 2 | 3 |
| 2 | 3 | | |

**Figure 15- Participants' involvement in SDLC phases**

### 4.2.3 Level of Challenge Experienced in Phases/Tasks (Q 14)

This question was about the level of challenge experienced in a given set of phases and tasks. Figure 16 shows the results. A 5-point Likert scale, same as for the previous question, was utilized for providing the answers.

As shown by the average values in Figure 16, respondents experience the most challenge in requirements. Design, communication with management, testing and communication with end users are the next runner ups in challenges experienced by participants. The implication of these results is that, companies should spend more efforts and training on these challenging tasks.

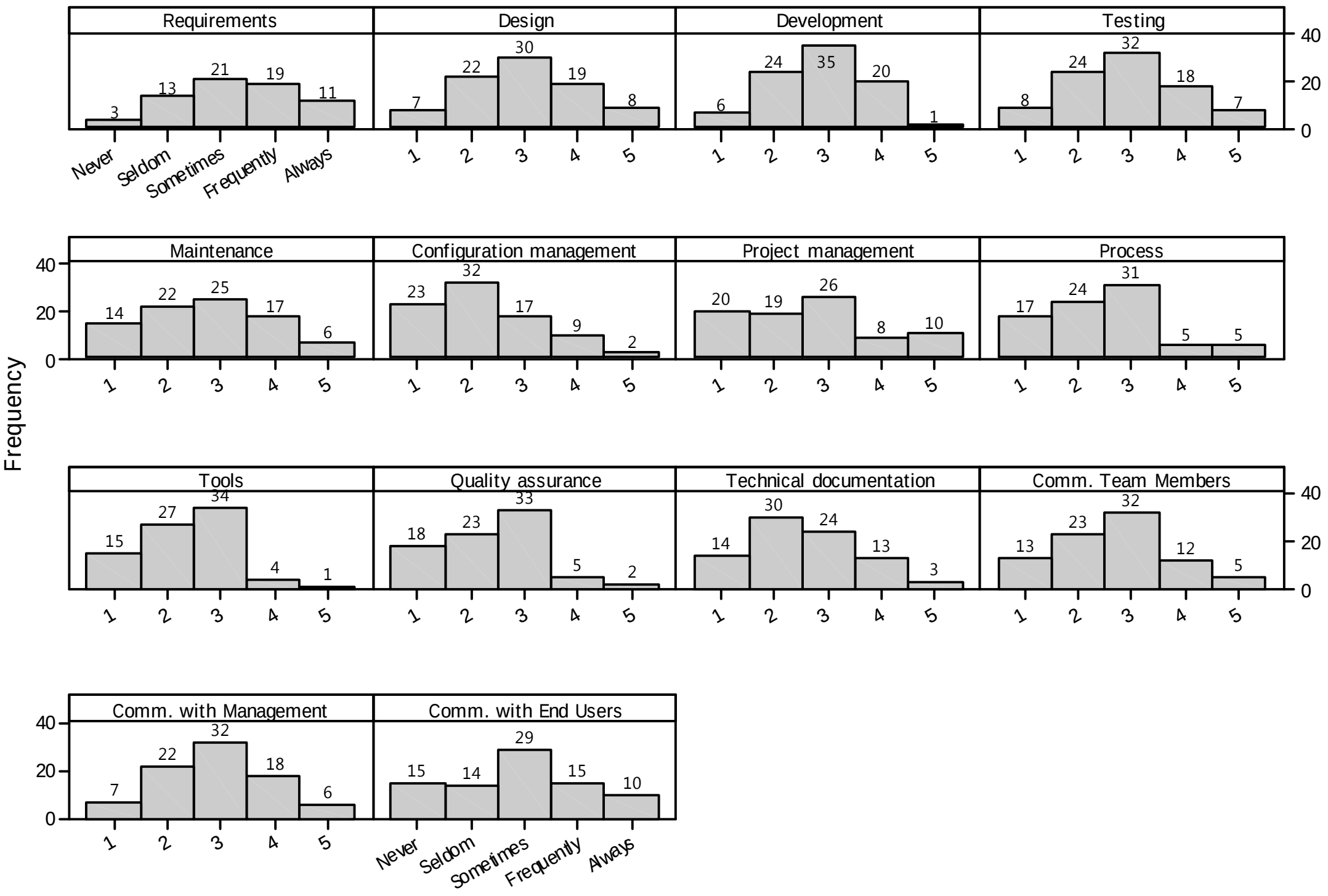

**Average values:**

| | | | |
|---|---|---|---|
| 3.33 | 2.99 | 2.84 | 2.91 |
| 2.75 | 2.22 | 2.63 | 2.48 |
| 2.37 | 2.38 | 2.54 | 2.68 |
| 2.93 | 2.89 | | |

**Median values:**

| | | | |
|---|---|---|---|
| 3 | 3 | 3 | 3 |
| 3 | 2 | 3 | 2 |
| 2 | 2 | 2 | 3 |
| 3 | 3 | | |

**Figure 16- Levels of challenge experienced in phases/tasks**

### 4.2.4 Software Development Methodologies (Q 15)

Q 15 was about the software development methodologies used by participants. Respondents were allowed to select more than one methodology. Results are shown in Figure 17.

Waterfall, as a rather old but still widely-used lifecycle model, is the methodology that more than half of the respondents (53%) use. The next widely-used lifecycle models are Incremental and Agile/lean development models with usage rates of 38% and 34%, respectively. The other methodologies has lower usage rates, namely: prototyping (28%), Scrum (20%), spiral (16%), extreme programming (16%), and product-line development (12%). The reason that we have included both Agile and also Agile-related practices, such as XP and Scrum, in the pre-defined list of practices to be chosen by the respondents is that, in our discussions with practitioners, we have seen that some software engineers just choose to follow Agile without choosing any specific method such as XP and Scrum.

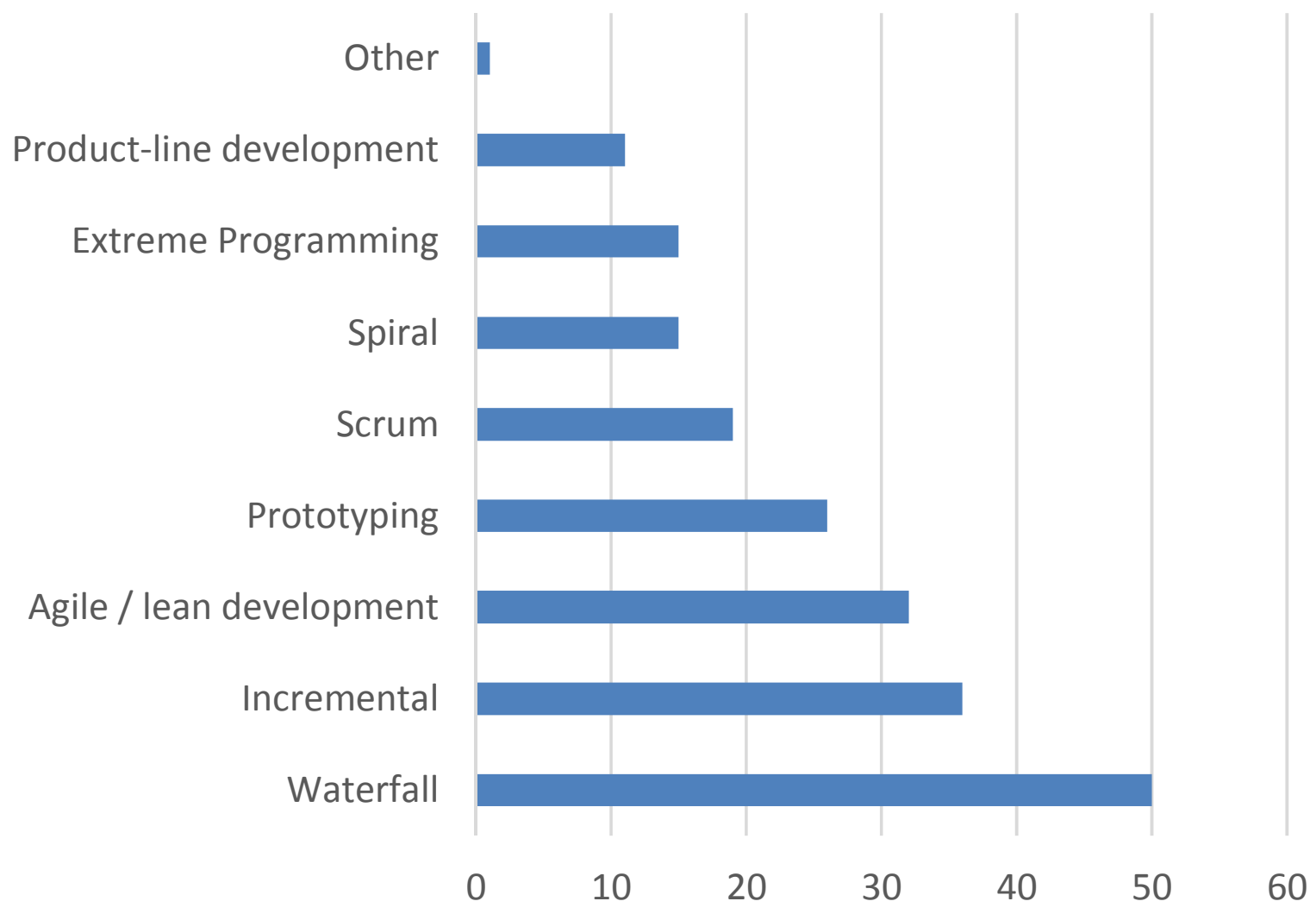

**Figure 17- Usage of software development methodologies**

### 4.2.5 Following Process Improvement Models (Q 16)

The goal of Q 16 was to investigate the usage of process improvement models. Respondents were again free to report more than one model. Results are shown in Figure 18.

Among teams who reported following process improvement models, the ISO 9000 family of standards and Capability Maturity Model Integration (CMMI) are widely followed with adoption rates of 62% and 61%, respectively. Usage of ISO/IEC 15504, also known as Software Process Improvement and Capability Determination (SPICE), is rather low. Others that are reported to be followed are: DO-178B (an avionics software standard), AS 9100 (a quality management system standard for the aerospace industry) and Six Sigma.

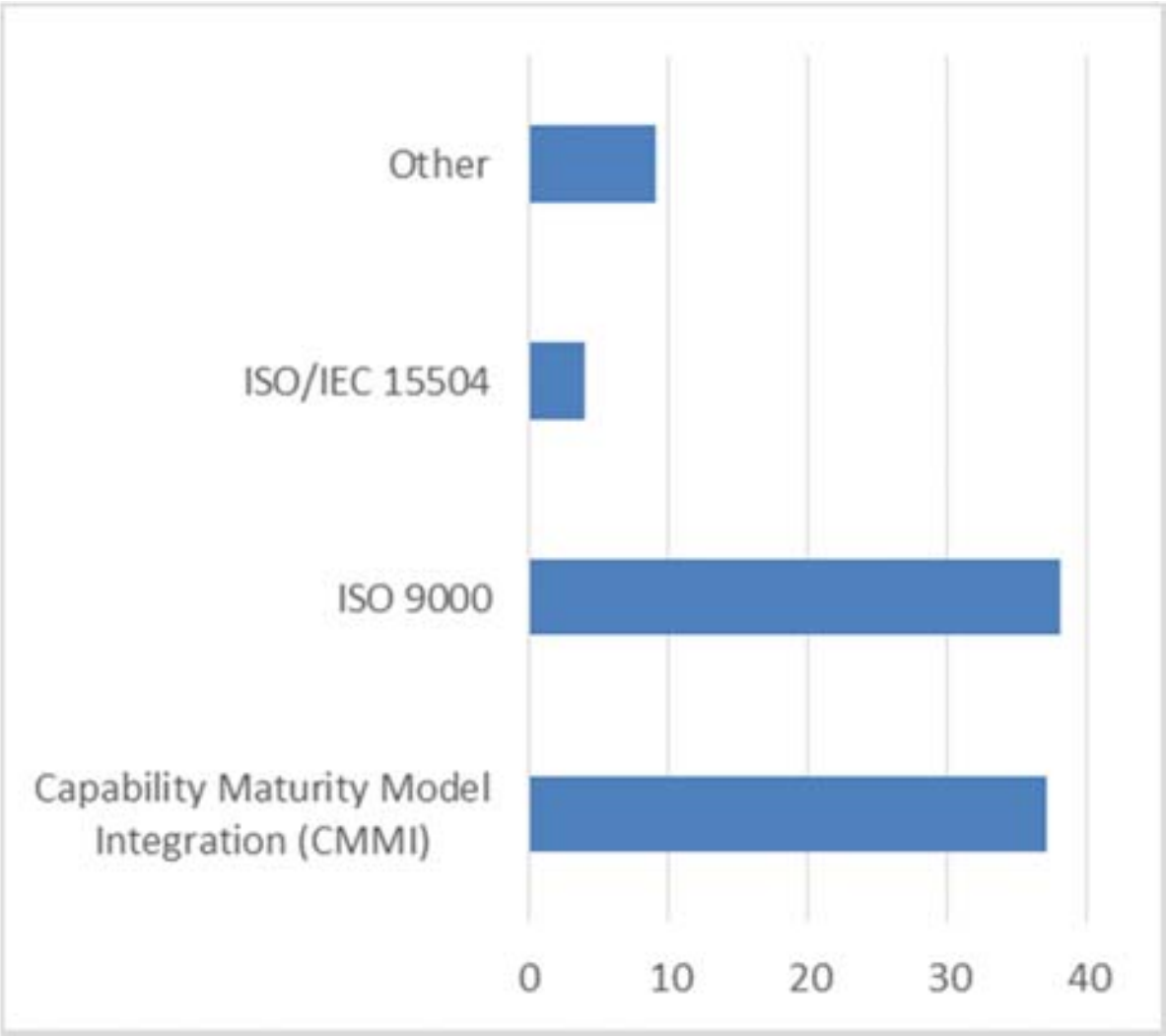

**Figure 18- Usage of process improvement models**

### 4.2.6 Process-related Practices (Q 17)

This question asked respondents to report about the practices carried out by their organizations in terms of SE processes. The following items were provided to be rated using a 5-point Likert scale.

- We perform tasks based on a defined standard process in the quality manual
- We perform tasks as defined by the project management team which are not defined in a quality manual
- We use systematic monitoring and assessment methods during the processes, e.g., using metrics

- We use systematic approaches to continuously learn from previous projects to optimize our performance in the upcoming projects

Results are shown in Figure 19. Using systematic approaches to continuously learn from previous projects is a slightly widely used practice, which seems similar to the CMMI practices. The other process-related practices are also being used to some degree in the Turkish software industry.

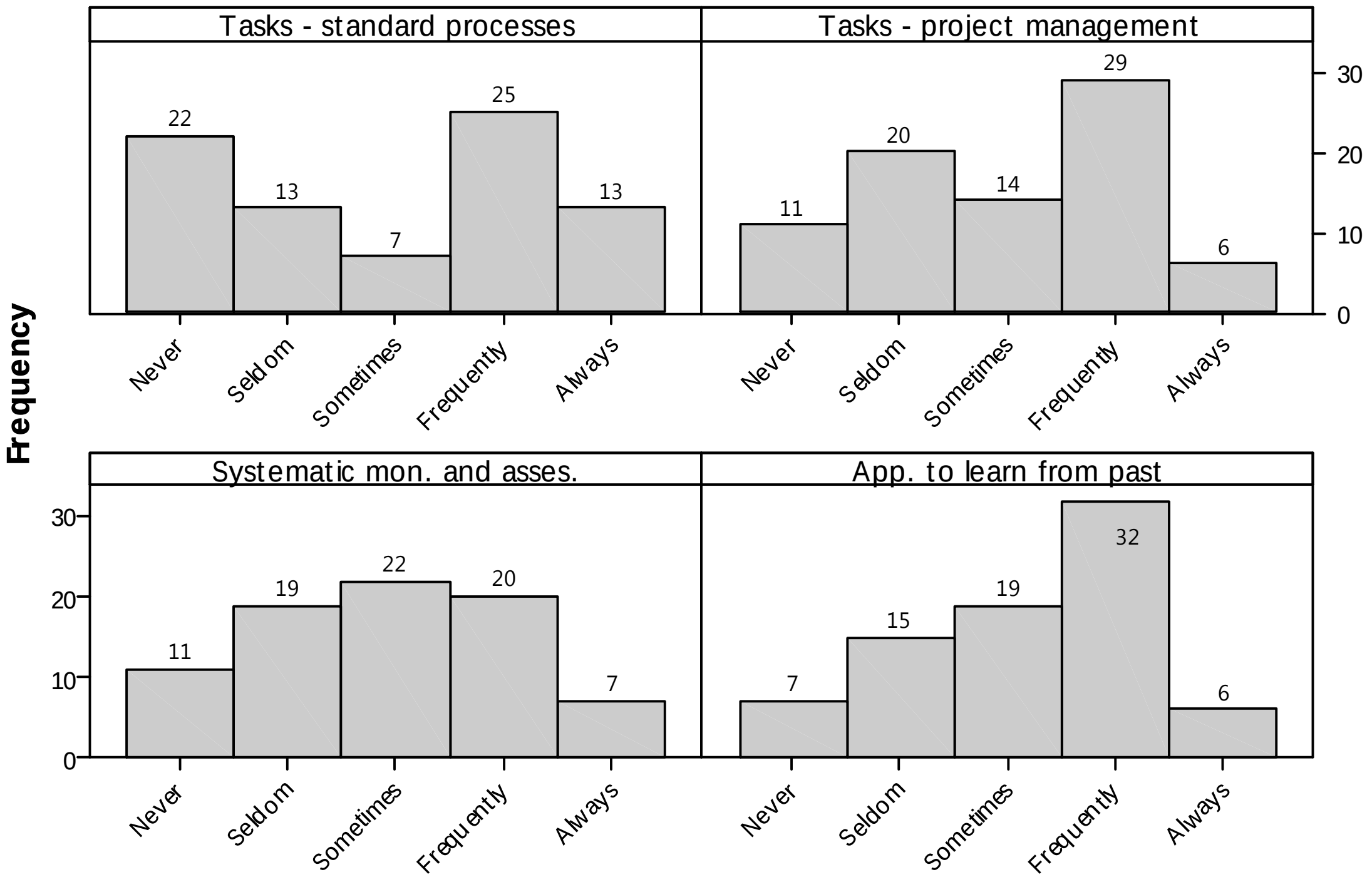

**Average values:**

| | |
|---|---|
| 2.93 | 2.99 |
| 2.91 | <u>3.19</u> |

**Median values:**

| | |
|---|---|
| 3 | 3 |
| 3 | 3 |

**Figure 19- Process related practices**

## 4.3 SOFTWARE REQUIREMENTS

We report the following results next:

- Requirements engineering practices (Q 18)
- Notations used for documenting requirements (Q 19)

### 4.3.1 Requirements Engineering Practices (Q 18)

In order to understand the requirements-related practices, we asked participants how frequently they carry out the following ten practices using 5-point Likert-scale:

1. We receive product requirements from outside the group.
2. We rely on the innovation of a few people.
3. Requirements are clearly documented.
4. Details of requirements are mostly in peoples' heads.
5. We plan releases well in advance.
6. Everyone in the team is kept informed about product strategy and direction.
7. We implement requirements traceability (tracking requirements).
8. Features are formally documented and reviewed.
9. We formally analyze requirements (e.g., using classifications, negotiations, and conceptual modeling)
10. We validate software requirements (e.g., by doing inspections, peer reviews, or joint walkthroughs)

Results are shown in Figure 20. The top practice in this category is implementing requirements traceability to other requirements or work products. Less popular practices are receiving requirements from outside the group, relying on innovation of a few people, and keeping everyone informed about product strategy and direction.

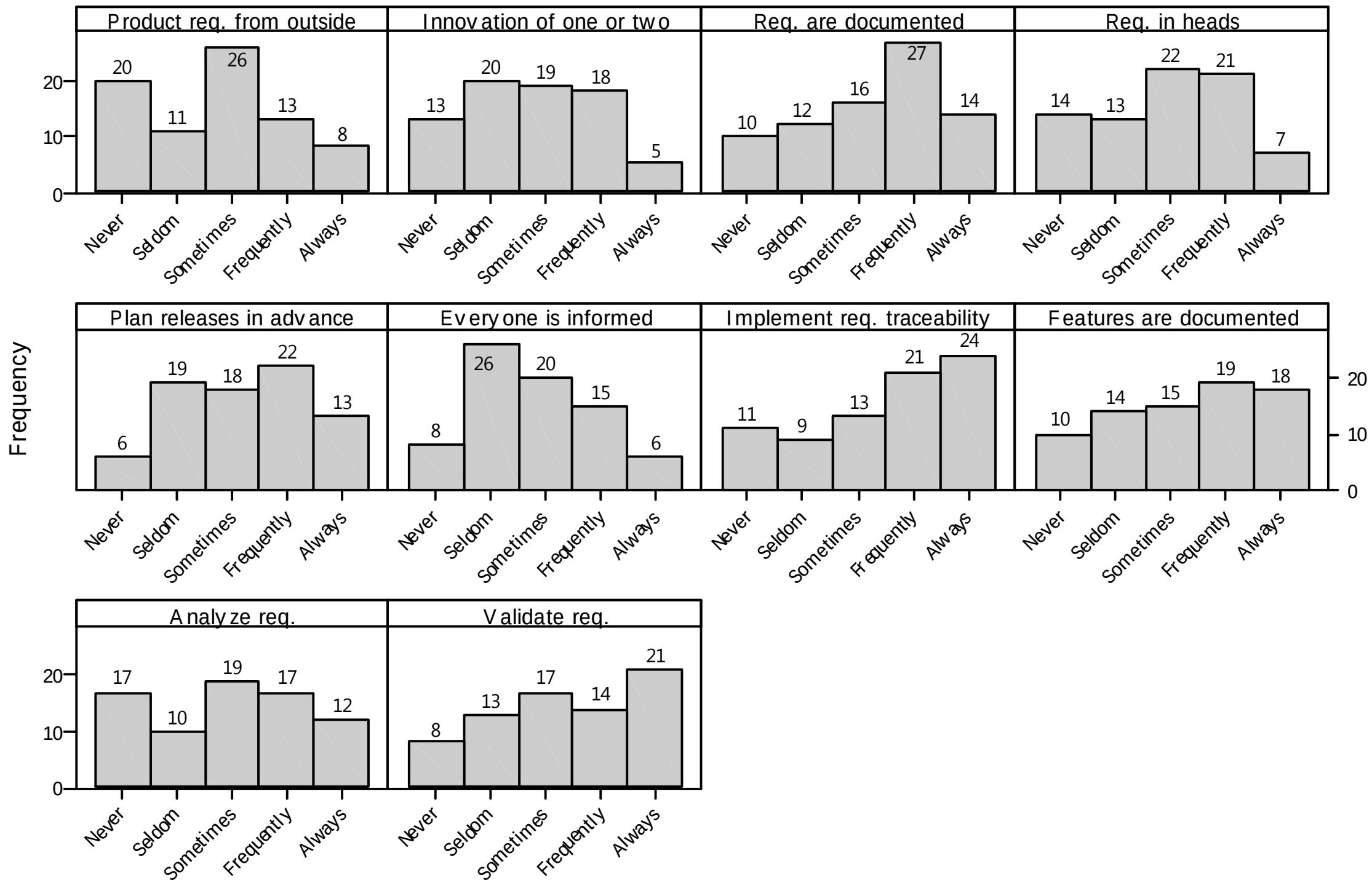

**Average values:**

| | | | |
|---|---|---|---|
| 2.71 | 2.76 | 3.29 | 2.92 |
| 3.21 | 2.80 | 3.48 | 3.27 |
| 2.96 | 3.36 | | |

**Median values:**

| | | | |
|---|---|---|---|
| 3 | 3 | 4 | 3 |
| 3 | 3 | 4 | 3 |
| 3 | 3 | | |

**Figure 20- Requirements engineering practices**

By reviewing the data, we found that it would be interesting to conduct cross-aspect analyses of requirements engineering practices. As a preliminary step, we looked at two such selected analyses:

- "Requirements are clearly documented" (item #2 in the above list of practices) versus "requirements are mostly in peoples' heads" (item #3): Our hypothesis to analyze this pair of aspects is that, the more requirements are clearly documented, the less one would expect that the team members would need to memorize the requirements (in their heads).
- "We formally analyze requirements" (item #9 above) versus "we validate software requirements" (item #10): Our rationale to analyze this pair of aspects is that they are somewhat interrelated and with increase in one aspect, one would expect an increase in the other aspect.

The bubble charts in Figure 21 visualize the above two cross-aspect analyses.

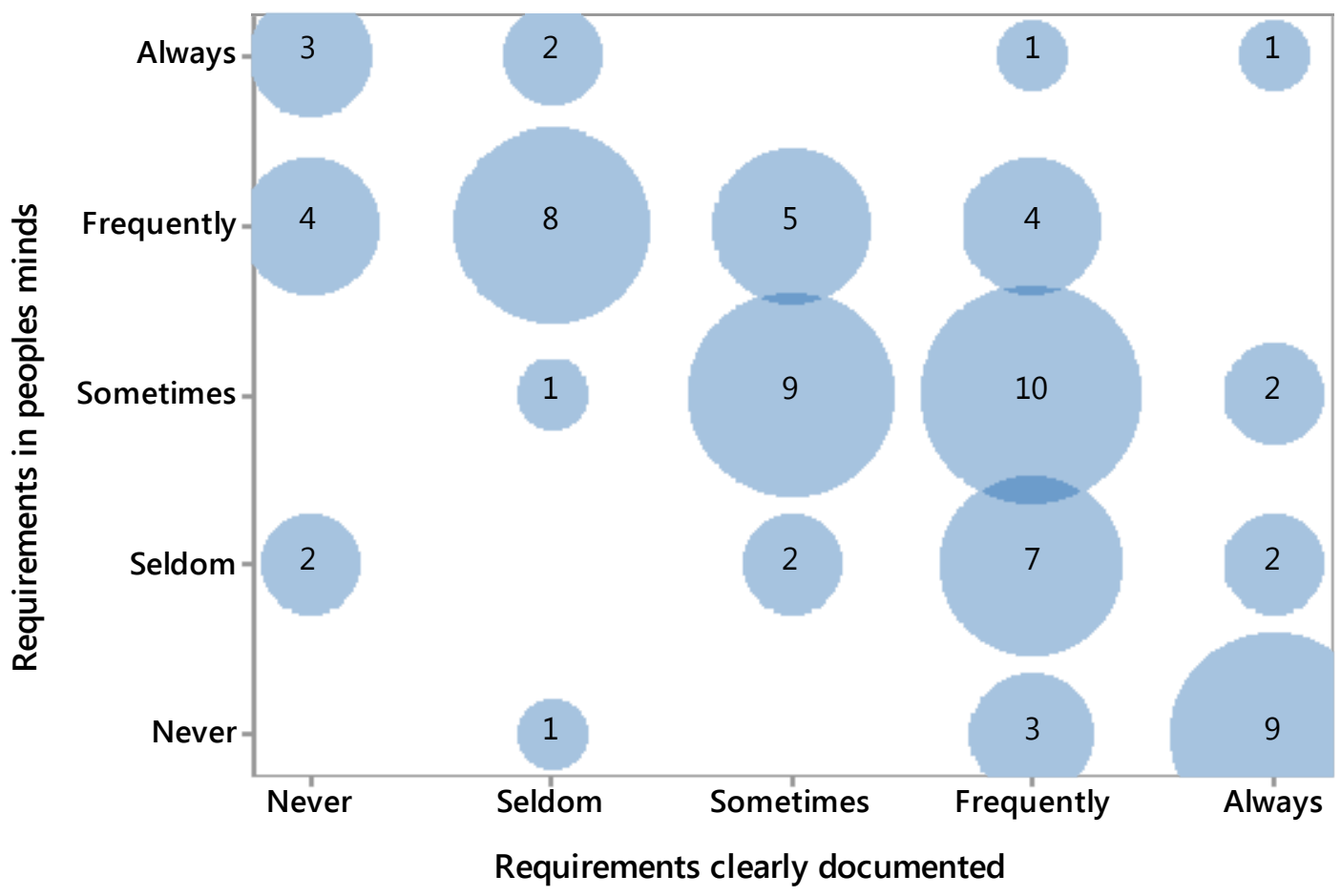

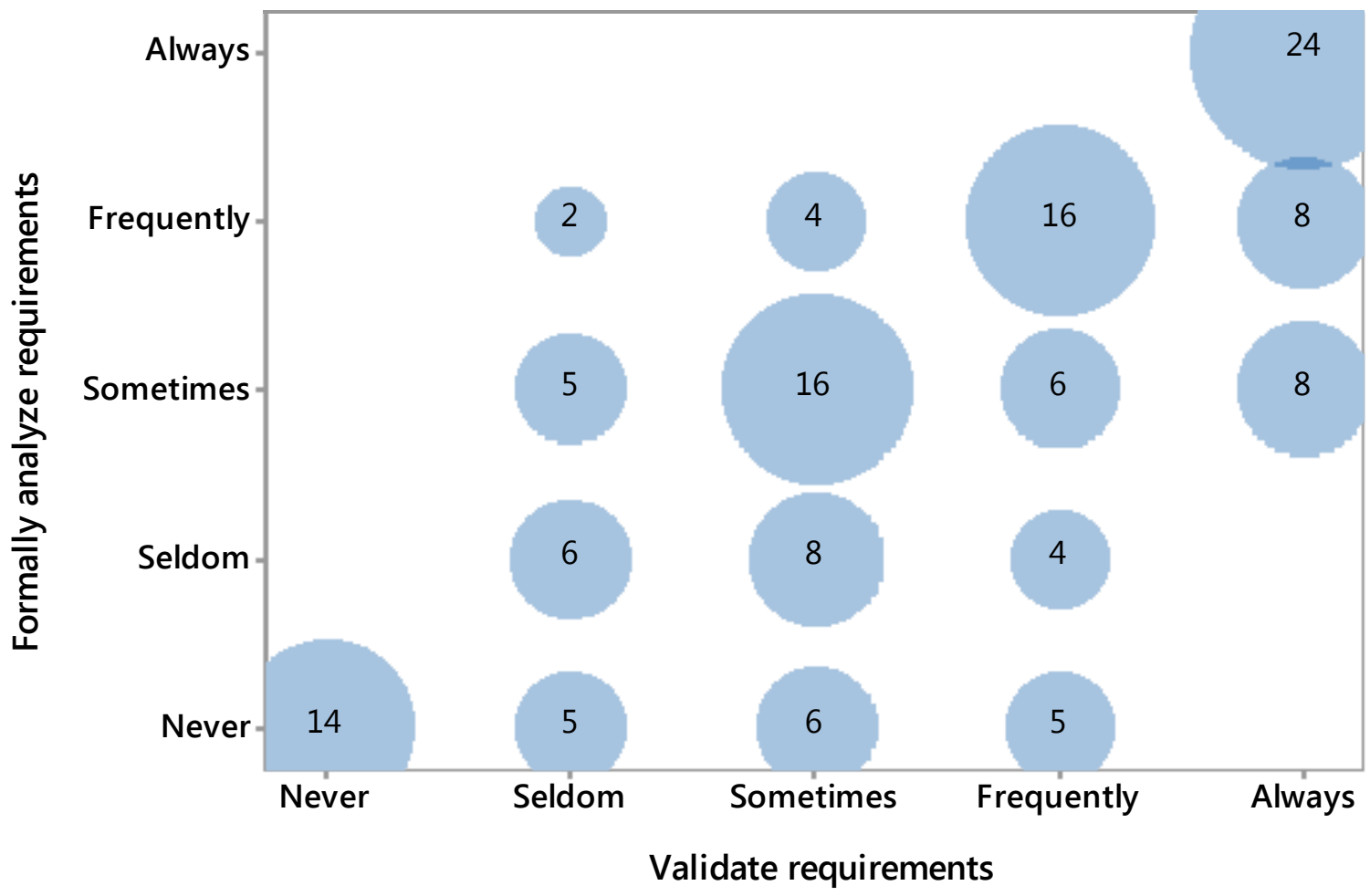

**Figure 21- Selected cross-aspect analyses of requirements engineering practices (size of bubbles=number of participants with each opinion)**

We discuss each of the above two cross-aspect analyses based on the bubble charts shown in Figure 21. The size of bubbles (the labels) denote the number of participants which have casted their opinions on the above pair of aspects, e.g., three participants mentioned that, in their projects, requirements are *never* clearly documented and team members always *memorize* the requirements.

To statistically investigate the above two issues, we calculated the Pearson's correlation coefficients for the two data sets shown in Figure 21. The coefficients are -0.51 and 0.78, respectively, denoting (1) a negative correlation between the case when requirements are documented, and whether team members memorize the requirements, and (2) positive correlation between formal analysis and validation of software requirements in companies. The Pearson's correlation coefficients reveal that, as one would expect, both the hypotheses raised above are supported.

### 4.3.2 Notations used for Documenting Requirements (Q 19)

Participants were asked about the requirements specification notations they use. They could choose more than one notation. As presented in Figure 22, using natural (textual) language (61%) and use cases (54%) are the most widely used notations. Formal requirement specification notations are the least used ones (8%). Only 84 out of 202 respondents replied to this question and 13% of those participants, did not report documenting the software requirements at all. This is an important finding that requires further analysis for causes (is it because of low quality awareness or emergence of Agile methods or something else?). Also, it would be important to analyze the potential effects of not documenting requirements (e.g., how is the end product affected by not documenting requirements?), which is outside the scope of our study at this time. These figures are somewhat similar to the ones provided by Aykol [12], where use cases (69%) is the most and formal specification languages (8%) is the least popular notations for requirements specification in Turkey as of 2008.

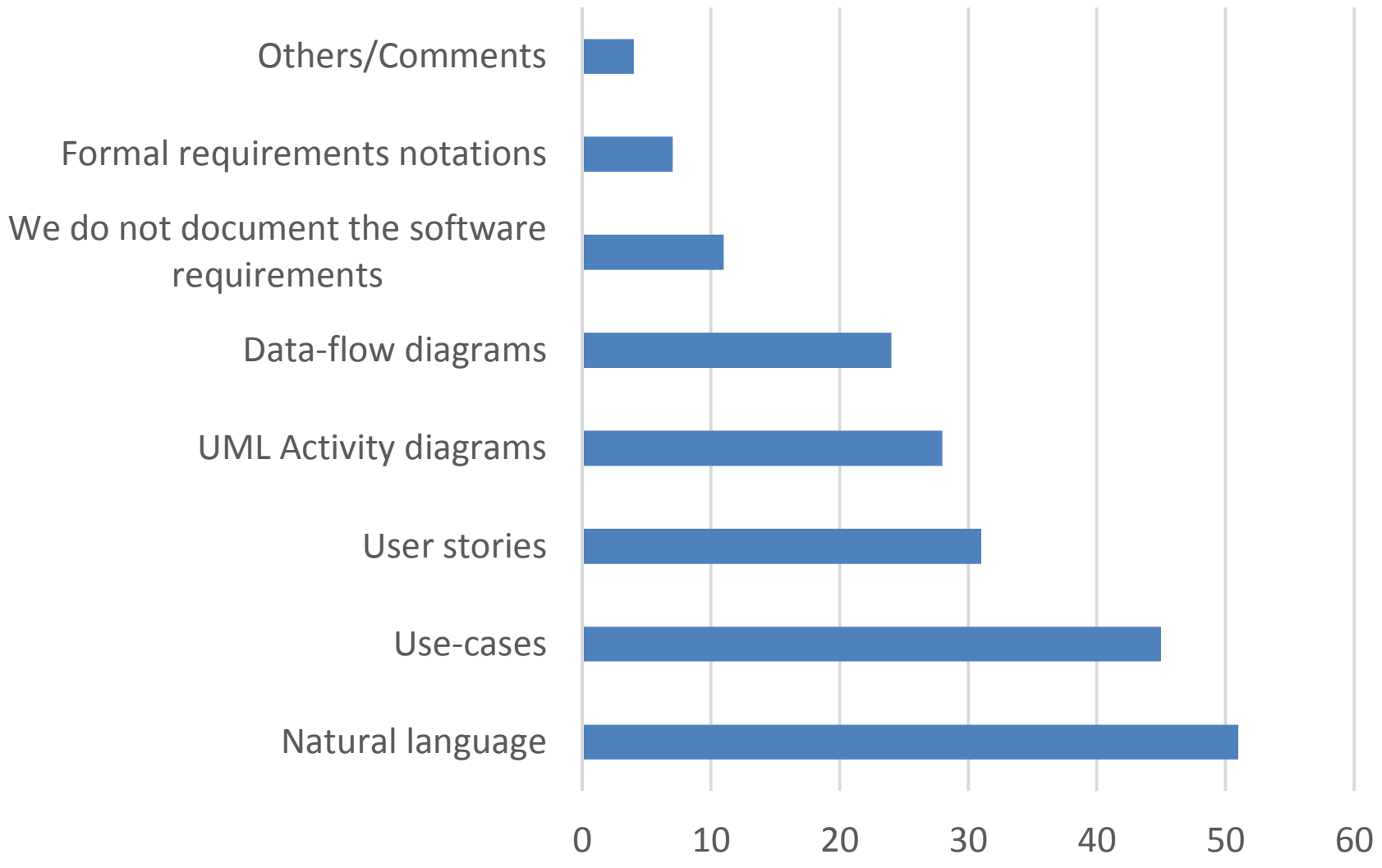

**Figure 22- Requirements specification notations**

## 4.4 Software Design

We report the following results:

- Design-related practices (Q 20)
- Design types (Q 21)
- Importance of design-related quality attributes (Q 22)

### 4.4.1 Design-related Practices (Q 20)

We asked how often the respondents employ the following practices in the design phase:

1. We have dedicated people for architecture and/or high-level design.
2. We follow a strict design review process.
3. We review designs informally within the team.
4. We often need to be creative because features and/or enhancements are incompletely documented in the earlier requirements phase.

The answers were in a 5-point Likert scale where 0 represents "never" and 5 represents "always". Results are shown in Figure 23. The distributions show that most respondents use the practices #3 and #4 "sometimes".

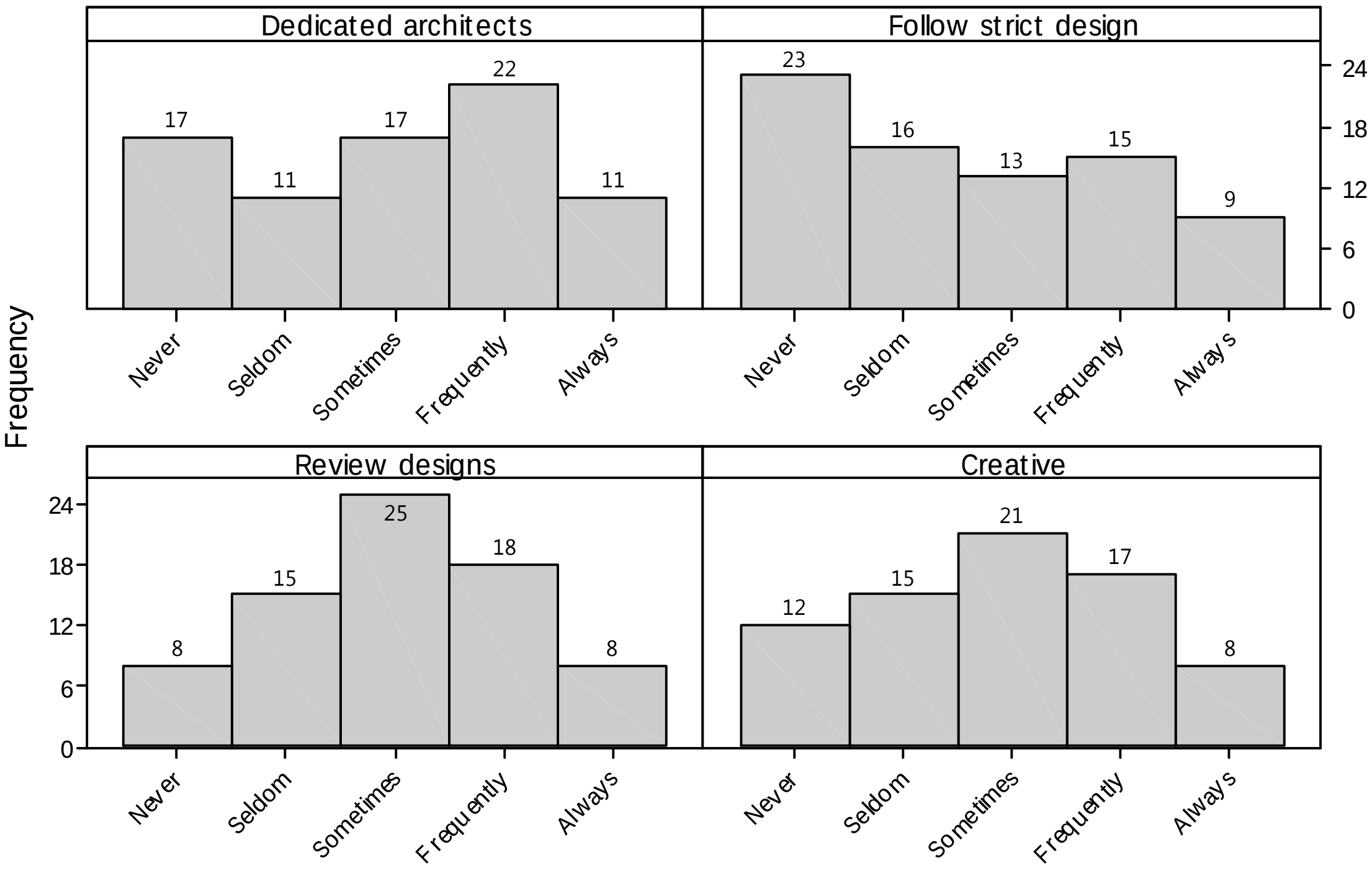

**Average values:**

| | |
|---|---|
| 2.99 | 2.62 |
| 3.04 | 2.92 |

**Median values:**

| | |
|---|---|
| 3 | 2 |
| 3 | 3 |

**Figure 23- Design practices**

### 4.4.2 Types of Design Activities (Q 21)

Participants were asked about the frequency of design activities. A list of 13 pre-determined design activities was provided. Results are shown in Figure 24. Class design and architecture design are the two most frequently performed design activity types; whereas network design and design-by-contract are the least popular. Class and architecture design are two primary tasks in the object-oriented (OO) SE and they have been adopted quite well among the respondents.

Design-by-contract, also known as contract programming, is an approach for designing software, which prescribes that software designers should define formal, precise and verifiable interface specifications for software components. According to many online forum discussions among practitioners, e.g., [49, 50], design-by-contract is not that popular in the software. This is due to various reasons, e.g., there is limited support for it in most modern programming languages [49]. Perhaps, these are among the reasons why design-by-contract is not that popular in Turkey.

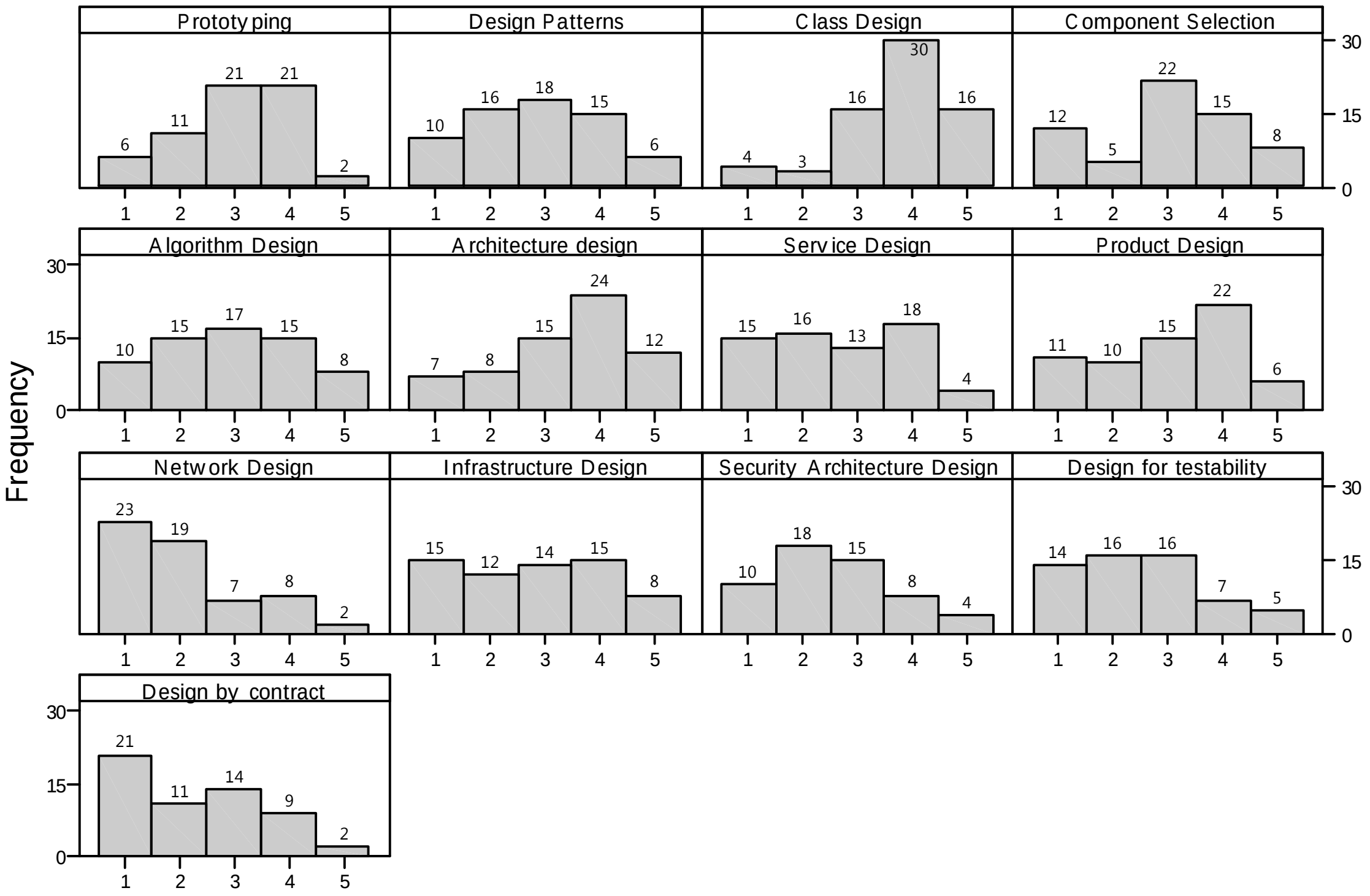

**Average values:**

| | | | |
|---|---|---|---|
| 3.03 | 2.86 | 3.74 | 3.03 |
| 2.94 | 3.39 | 2.70 | 3.03 |
| 2.10 | 2.83 | 2.60 | 2.53 |
| 2.30 | | | |

**Median values:**

| | | | |
|---|---|---|---|
| 3 | 3 | 4 | 3 |
| 3 | 4 | 3 | 3 |
| 2 | 3 | 2 | 2 |
| 2 | | | |

**Figure 24- Design activity types**

### 4.4.3 Importance of Design-related Quality Attributes (Q 22)

This question was about importance of nine given aspects in software design. The participants provided their answers in a 5-point Likert scale where 1 represents not important and 5 represents very important. Results are shown in Figure 25. We can see that all graphs are skewed towards right, which indicates that most participants considered these aspects important in designing software systems. Modularity and reliability were voted as the most important design-related quality attributes.

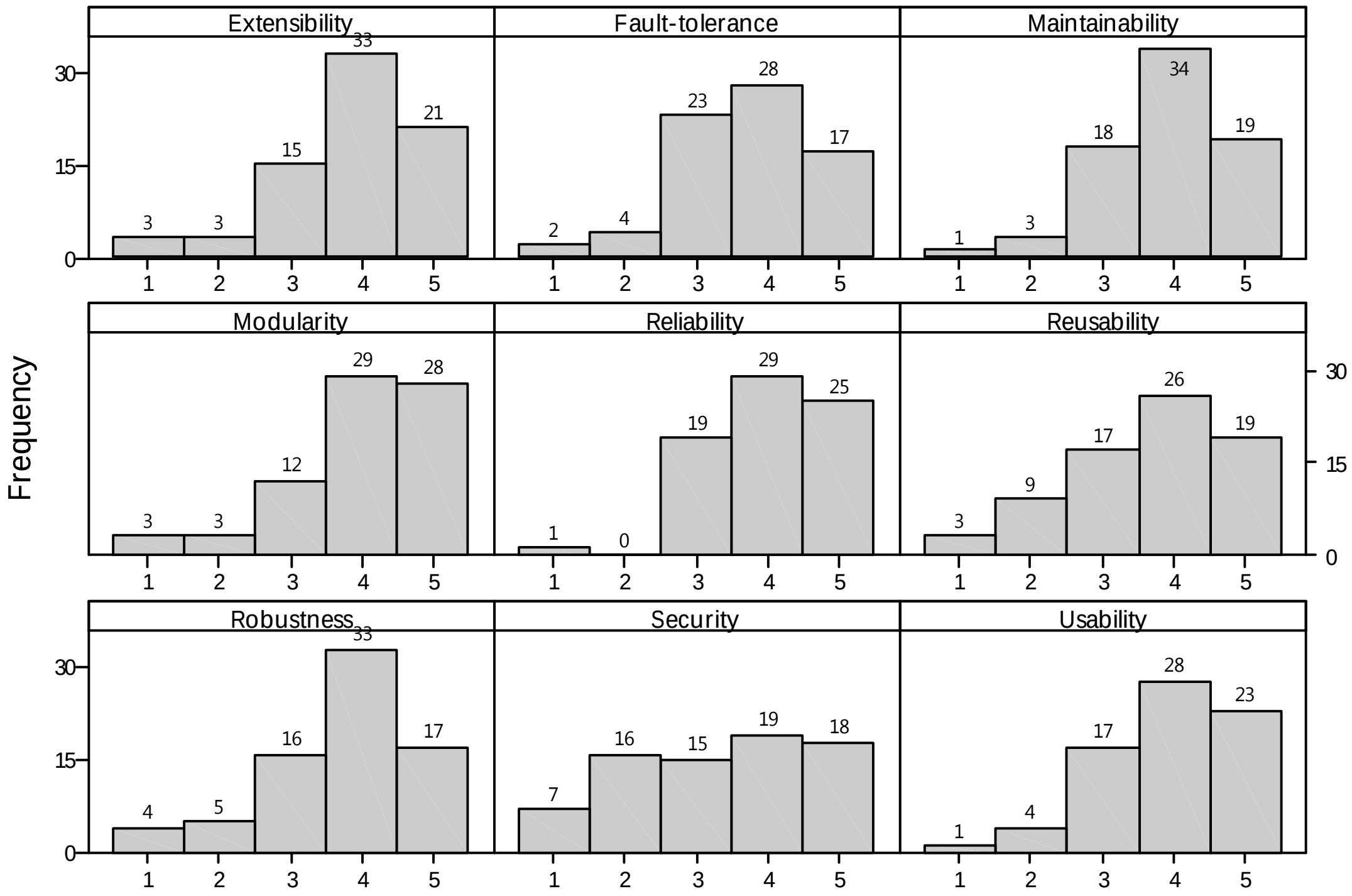

**Average values:**

| | | |
|---|---|---|
| 3.88 | 3.73 | 3.89 |
| 4.01 | 4.04 | 3.66 |
| 3.72 | 3.33 | 3.93 |

**Median values:**

| | | |
|---|---|---|
| 4 | 4 | 4 |
| 4 | 4 | 4 |
| 4 | 3 | 4 |

**Figure 25- Importance of design-related quality attributes**

## 4.5 Software Development/Implementation Phase

We report the following results next:

- Programming languages (Q 23)
- Development-related practices (Q 24)

### 4.5.1 Programming Languages (Q 23)

This question gathered the choice of programming languages used for development. Participants were allowed to report as many languages as they use. As shown in Figure 26, Java is the most commonly used language, which is closely followed by .Net Family languages and C/C++. Other programming languages, as reported by participants, are Python, PL-SQL, COBOL, PL/I, JavaScript, Delphi, Objective-C, and Ruby.

More than half of the participants (59%) use more than one programming language. Also we see that none of the programming languages have a strong dominance over all others in the Turkish software industry. From these findings, we conclude that the Turkish software engineers are not inclined towards a single specific language.

For comparison purposes, we also show in Figure 26 the data from two earlier surveys (Canada-wide [24] and the province of Alberta, Canada [23]) and also the TIOBE index [51] for the popularity of programming languages. The TIOBE index is calculated from number of search engine results for queries containing the name of the languages. As we can notice, Java seems to be slightly more popular in Turkey compared to Canada and the rest of the world (according to the TIOBE index).

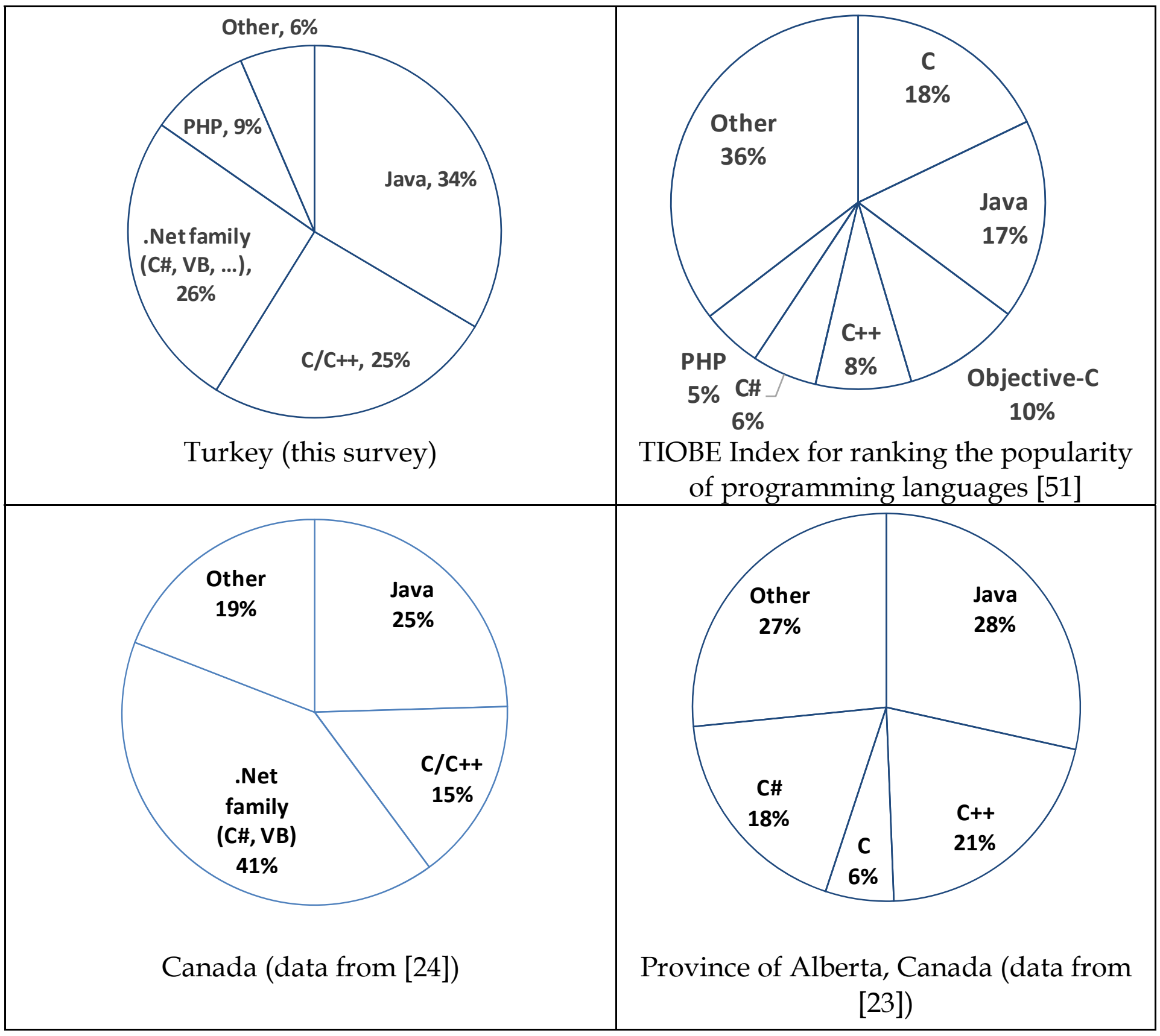

**Figure 26- Programming languages used by the respondents**

### 4.5.2 Development-related Practices (Q 24)

In Q 24, we asked how often, in a 5-point Likert scale, the following development-related practices are carried out in the organizations. Figure 27 shows the corresponding histograms.

- We practice pair programming
- We practice refactoring
- We practice code inspections / peer reviews
- We perform static code analysis using automated tools
- We practice systematic built-in code documentation (using comments and frameworks such as JavaDocs)
- We measure, manage and minimize code complexity
- Developers are responsible for high-level design and implementation
- Design and coding are carried out together
- We build the product at least once per day
- We build when a feature or features are completed
- Our build process is documented and followed
- We have a dedicated build person
- Code is reviewed before submitting to the build

Three of the graphs with the highest average values (underlined in Figure 27) are skewed (partially) towards right and indicate that they belong to popular practices. Those practices are: (1) building (the code-base) when a feature is completed, (2) developers are responsible for high-level design and implementation, and (3) performing design and coding together.

The least popular practice by far is pair programming. This is somewhat a surprising finding since pair programming is an Agile practice and as indicated in findings for Question 15, Agile development is widely used in the Turkish industry.

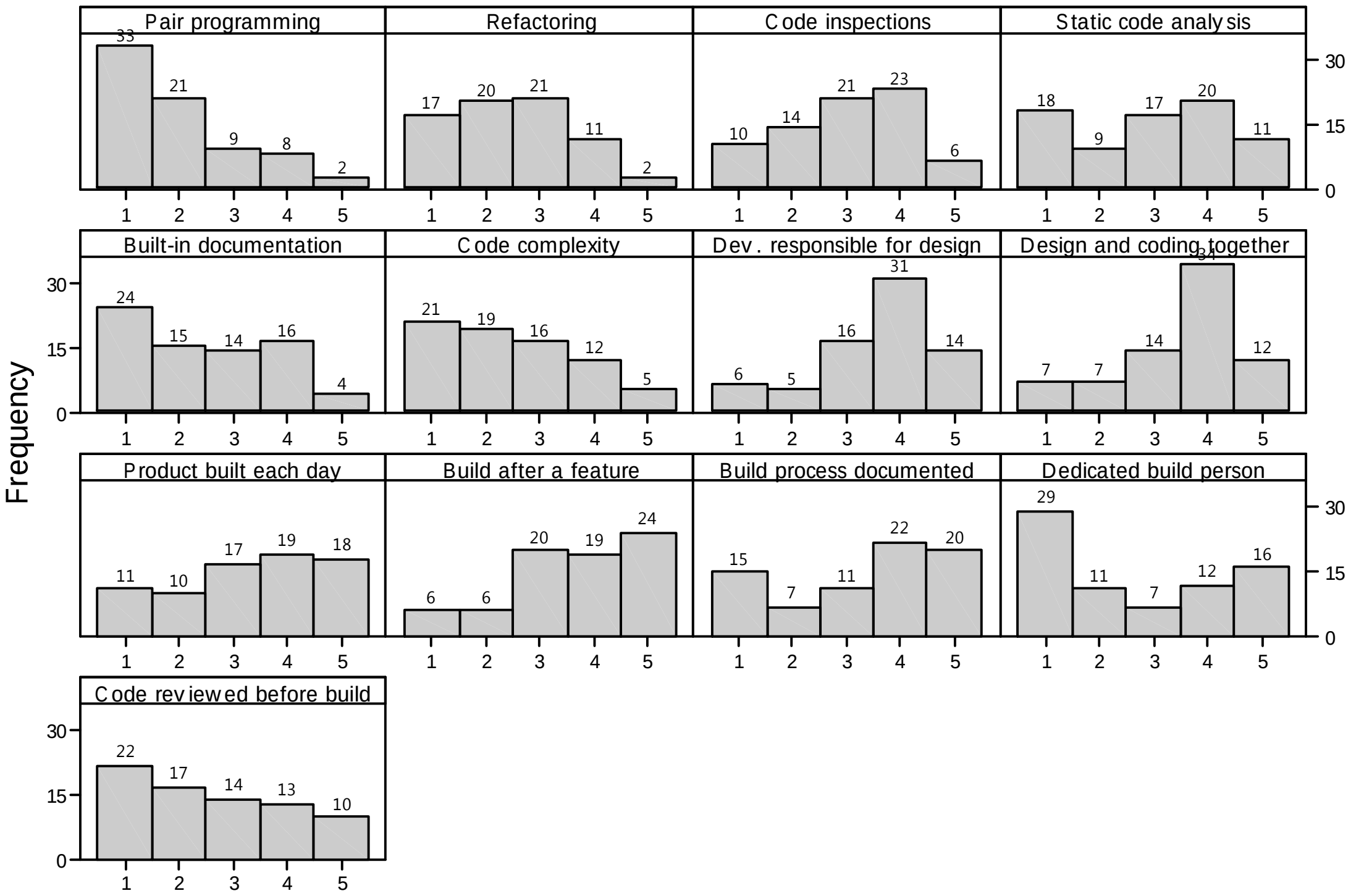

**Average values:**

| | | | |
|---|---|---|---|
| 1.97 | 2.45 | 3.01 | 2.96 |
| 2.47 | 2.47 | 3.58 | 3.50 |
| 3.31 | 3.65 | 3.33 | 2.67 |
| 2.63 | | | |

**Median values:**

| | | | |
|---|---|---|---|
| 2 | 2 | 3 | 3 |
| 2 | 2 | 4 | 4 |
| 3 | 4 | 4 | 2 |
| 2 | | | |

**Figure 27- Development-related practices**

Similar to what we did in the analysis of requirements engineering practices (Section 4.3.1), after reviewing the data for Q 24, we found that it would be interesting to conduct cross-aspect analyses of development-related practices. As a preliminary step, we looked at the cross-aspect analysis of conducting pair programming versus code inspection/peer reviewing (the first and the third items in the above list of practices). We wanted to know how often these two activities are conducted and with which frequency (recall the 5-point Likert scale we have used: never, seldom, sometimes, frequently, and always). The bubble chart in Figure 28 visualizes the cross-aspect analysis. We can state the following noteworthy observations from these trends:

- To statistically investigate the above cross-aspect issue, we calculated the Pearson's correlation coefficient on the data set shown in Figure 28. The coefficient value is 0.10, meaning that there is almost no correlation between the two practices, i.e., given that a practitioner (or a team) is conducting pair programming does not necessarily imply that s/he is also conducting code inspections/peer reviews.
- Pair programming has low adaptation ratio in the Turkish software industry, while code inspection/peer reviewing has reasonable usage.
- There is no vote by any respondent for the "always-always" combination ("always" conducting both practices). However four respondents stated that they "frequently" conduct both practices.

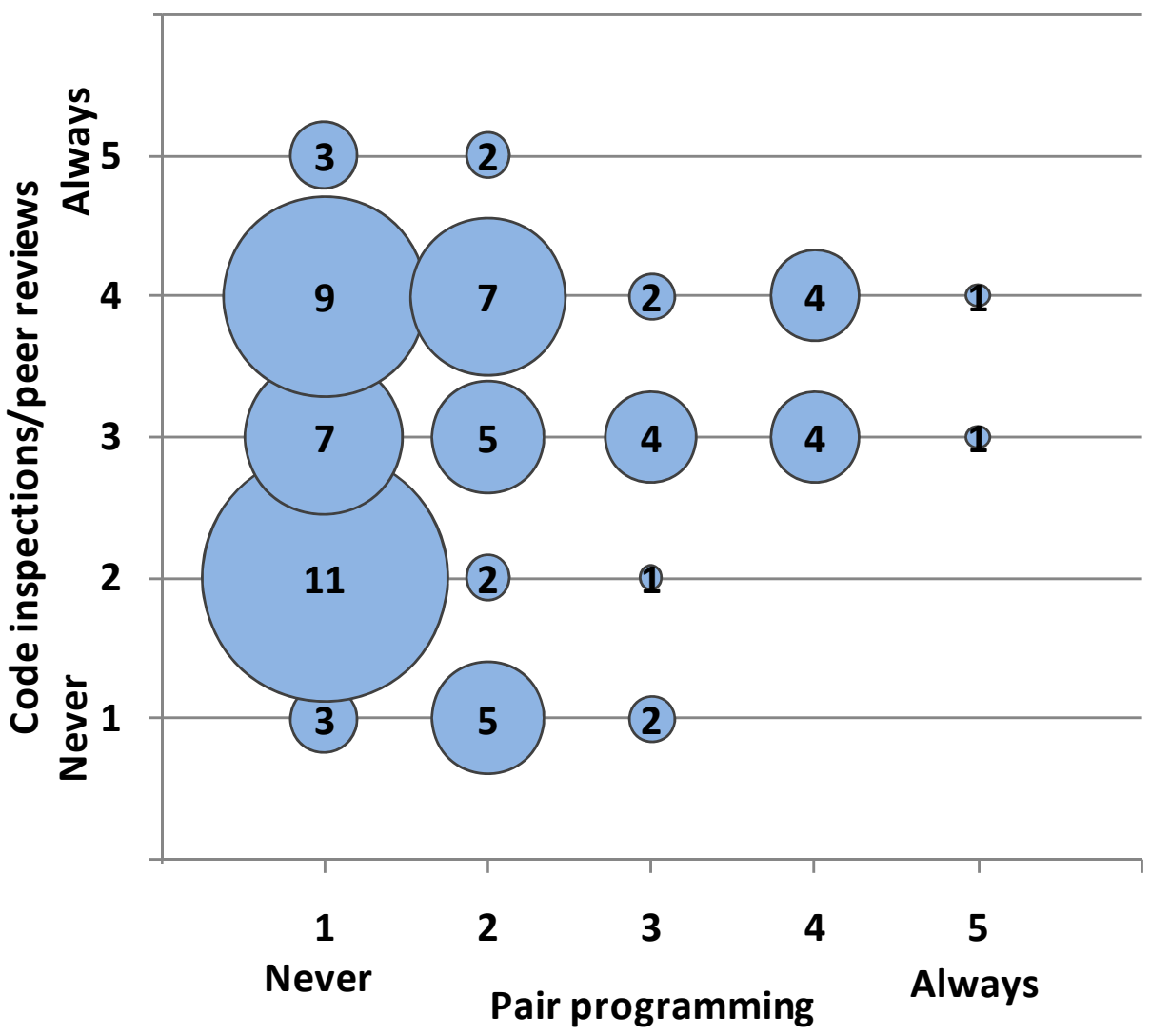

**Figure 28- Selected cross-aspect analyses of development-related practices: conducting pair programming versus code inspection/peer reviewing (size of bubbles=number of participants with each opinion).**

## 4.6 Software Testing

Here we report the following results next:

- Testing-related practices (Q 25)
- Phasing of testing during the SDLC (Q 26)
- Test types/levels (Q 27)
- Test-case design techniques (Q 28)
- Test automation (Q 29)
- Test-related metrics (Q 30)
- Other quality-related metrics (Q 31)
- Ratio of testers to developers (Q 32)
- Criteria for terminating testing (Q 33)

### 4.6.1 Testing-related Practices (Q 25)

Q 25 asked participants the frequency of practices performed during testing. Figure 29 shows the responses where "1" means never and "5" means always. Average values of the votes have also been shown. We had provided the following seven practices for this question.

- We develop unit tests for code.
- Unit tests are formally reviewed.
- We have a separate team for testing.
- Developers and testers work closely together.
- Developers test the product before release.
- A manager, client advocate or support person helps test the product.
- All new features are independently tested by a test team.

The most and least popular activities in this list are: "Developers test the product before release", and "Unit tests are formally reviewed", respectively. 35 respondents mentioned that they have separate testing teams.

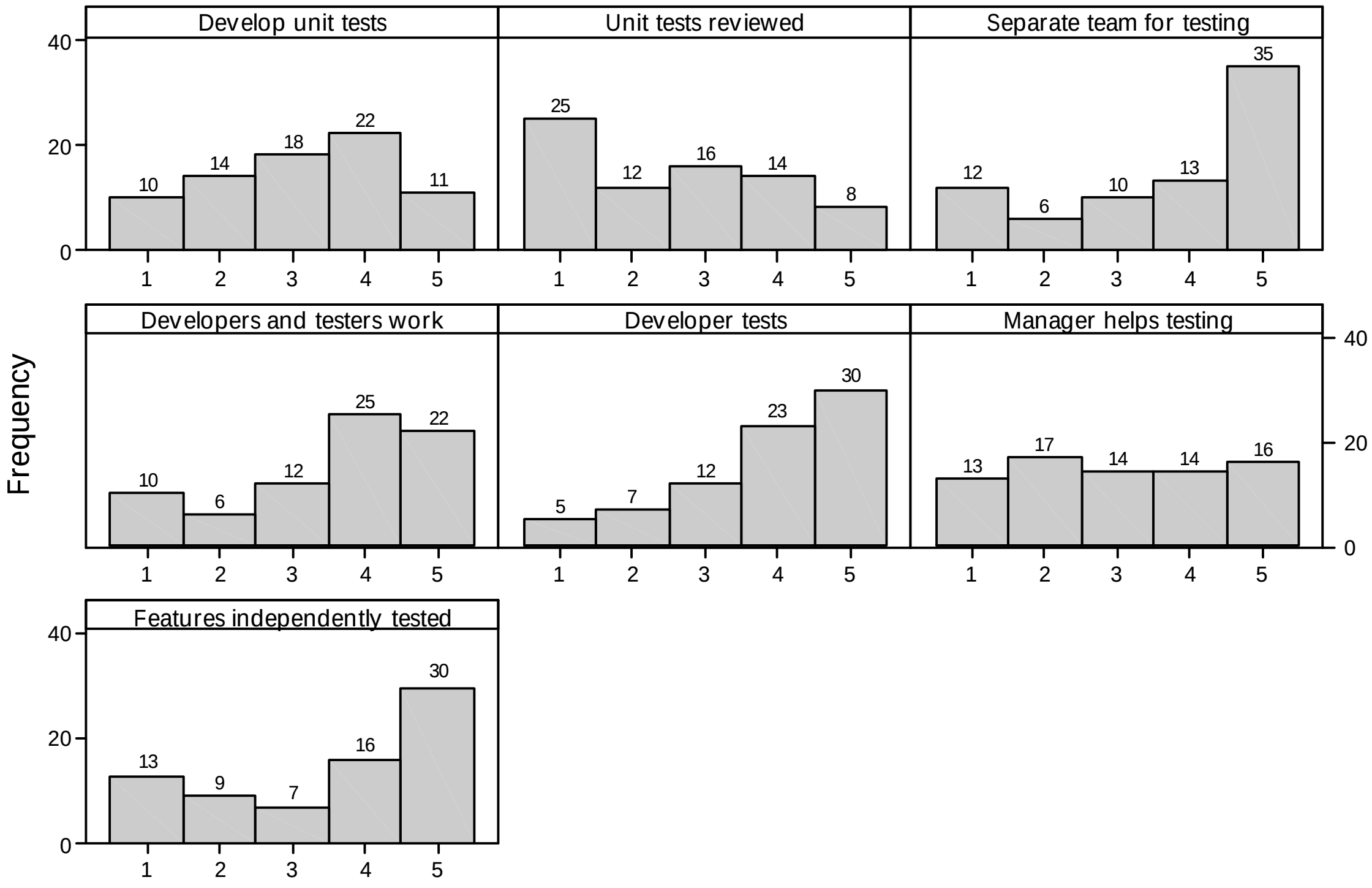

**Average values:**

| | | |
|---|---|---|
| 3.13 | 2.57 | 3.70 |
| 3.57 | 3.86 | 3.04 |
| 3.55 | | |

**Median values:**

| | | |
|---|---|---|
| 3 | 3 | 4 |
| 4 | 4 | 3 |
| 4 | | |

**Figure 29- Testing-related practices**

### 4.6.2 Phasing of Testing During the SDLC (Q 26)

Participants were then asked about the phasing of their testing effort during the SDLC processes. Results are shown in Figure 30, together with results of the Canadian survey [24]. We can see that Test-last Development (TLD) approaches (i.e., testing after development) are still much more popular than Test-driven (-first) Development (TDD). The result indicates that traditional test-last development approach is still dominant among the respondents. In comparison of the Turkish and the Canadian trends [24], we can see that in both countries, TLD is more popular than TDD.

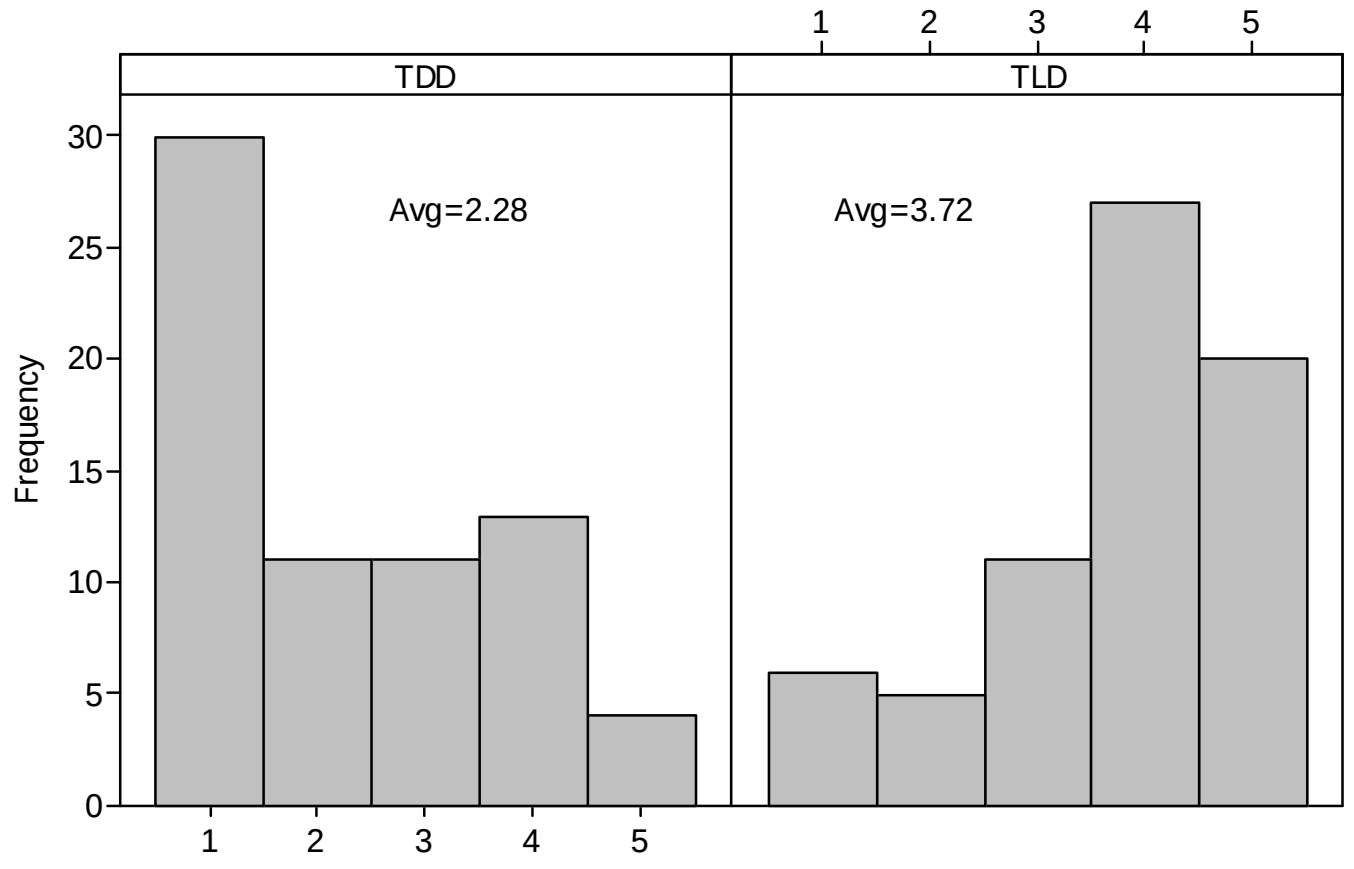

Turkey (this survey)

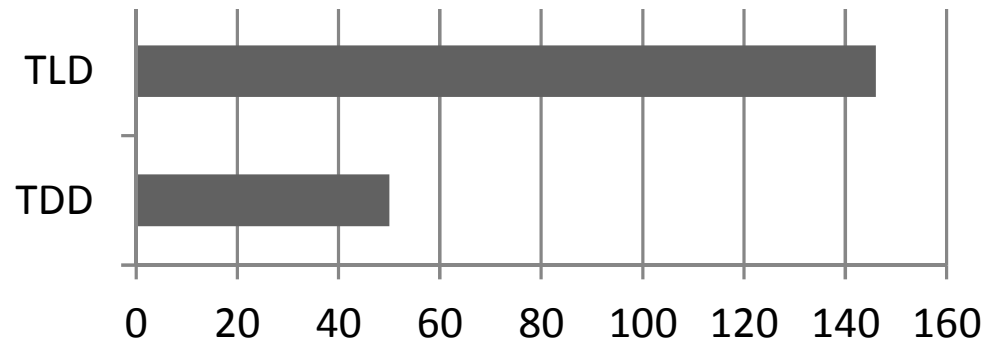

Canada (data from [24])

**Figure 30- Test-driven development (TDD) vs. Test-last development (TLD)**

Similar to what we did in the analysis of requirements engineering practices (Section 4.3.1) and development-related practices (Section 4.5.2), we found that it would be interesting to conduct cross-aspect analyses of TDD's adoption versus TLD's. We wanted to know how often each pair of these two test-phasing styles is conducted. The bubble chart in Figure 31 visualizes this cross-aspect analysis. We can state the following noteworthy observations from these trends:

- To statistically investigate the above cross-aspect issue, we calculated the Pearson's correlation coefficient on the data set shown in Figure 31. The coefficient value is -0.17 (a very weak negative correlation), meaning that there is almost no correlation between the two practices, i.e., high usage of TLD by a tester (or a team) does not necessarily imply low usage of TDD.
- Many respondents (17 of them) reported no usage of TDD and usage of TLD practice in all their projects. Since the adoption of TDD in the Turkish industry is very low, it seems that most Turkish SE practitioners are unaware of TDD's advantages. Thus, we recommend to the practitioners reviewing the success stories of adopting TDD by major international software firms such as IBM [52] and Microsoft [53].
- There is only one vote by one respondent for the "always-always" combination ("always" conducting both practices).
- Interestingly, four respondents stated that they "never" conduct TDD nor TLD.

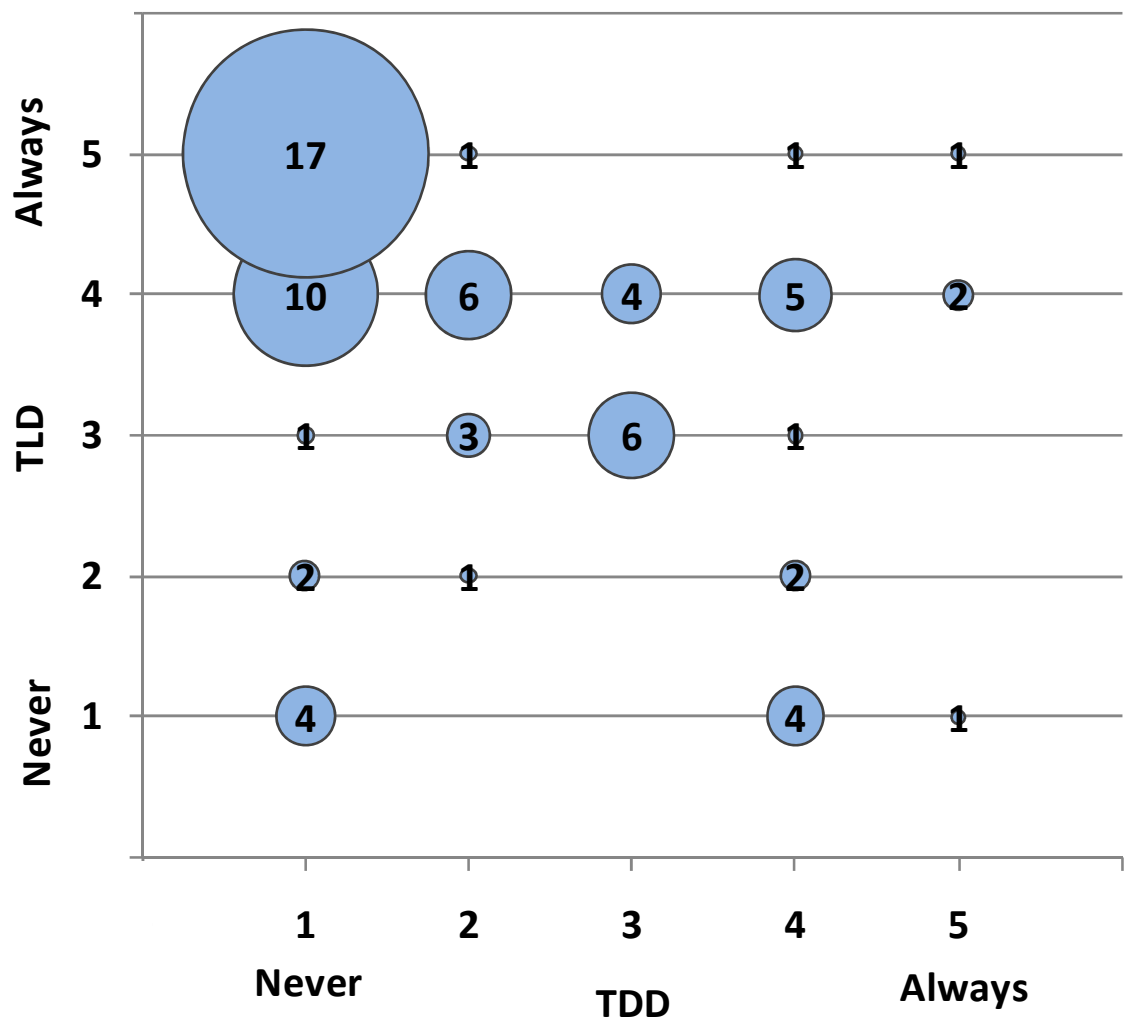

**Figure 31- Comparative analysis of TDD's adoption versus TLD's**

### 4.6.3 Test Types/Levels (Q 27)

Q 27 asked about the types of test activities. The responses were gathered using the Likert scale again. Results are shown in Figure 32. Average values of the votes have also been shown. Functional/system testing, user acceptance testing, and integration testing are the three most widely used test approaches, with average vote values of 4, 4, and 3.7 out of 5, respectively. Load/stress testing and security testing are the least used techniques.

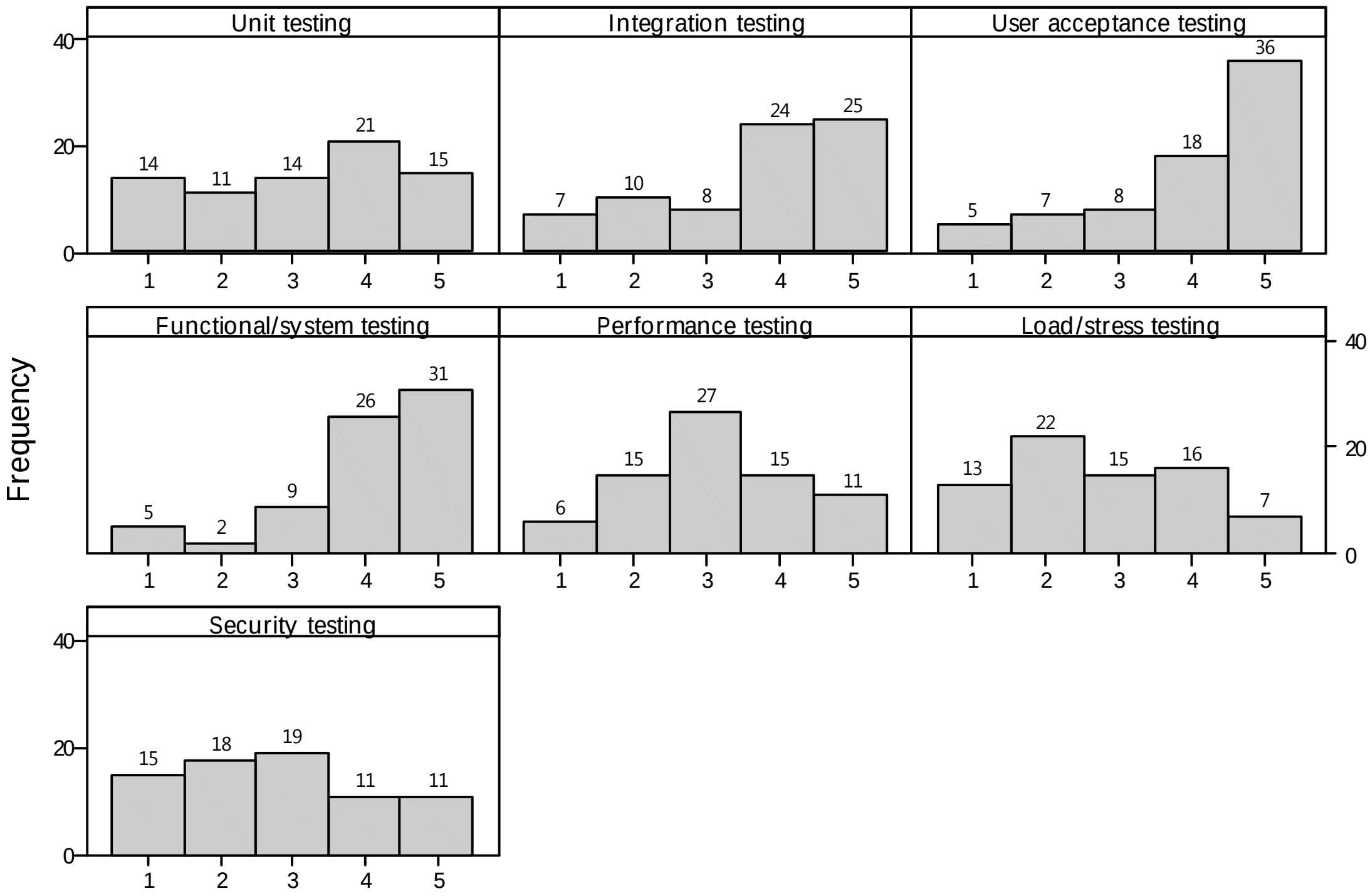

**Average values:**

| | | |
|---|---|---|
| 3,1 | 3,68 | 3,9 |
| 4,0 | 3,1 | 2,7 |
| 2,79 | | |

**Median values:**

| | | |
|---|---|---|
| 3 | 4 | 4 |
| 4 | 3 | 3 |
| 3 | | |

**Figure 32- Test types/levels**

### 4.6.4 Test-case Design Techniques (Q 28)

Test-case design is an important activity in software testing. Similar to our recent Canada-wide survey [24], we asked in the survey about techniques used to design test cases. Results of the two countries are shown and compared in Figure 33, together with data from the survey of the Turkish Testing Board (TTB) in year 2013 [33].

The x-axis values are percentages out of the respondents' pool. According to our survey, usage rate of each of the four formal approaches vary between 22% and 33%, which is slightly better than the Canadian numbers (between 15% and 27%). On the other hand, while about 17% of respondents from Canada mentioned that they have not used any explicit test-case design technique in their projects, this number is significantly higher (49%) in Turkey.

The TTB survey also provided measures on three of the test-case design approaches which have also been incorporated in Figure 33. Interestingly, the TTB survey ratio for usage of boundary-value analysis is higher than our survey's, which could be perhaps due to the fact that the TTB survey audience was a somewhat specialized pool of test experts who are members of the TTB organization, and who have high chances of using formal test-case design approaches.

Aykol's 2009 study [14] had also asked about test-case design techniques and the results were as follows: usage of boundary value technique: 25%, and usage of equivalence partitioning: 0%.

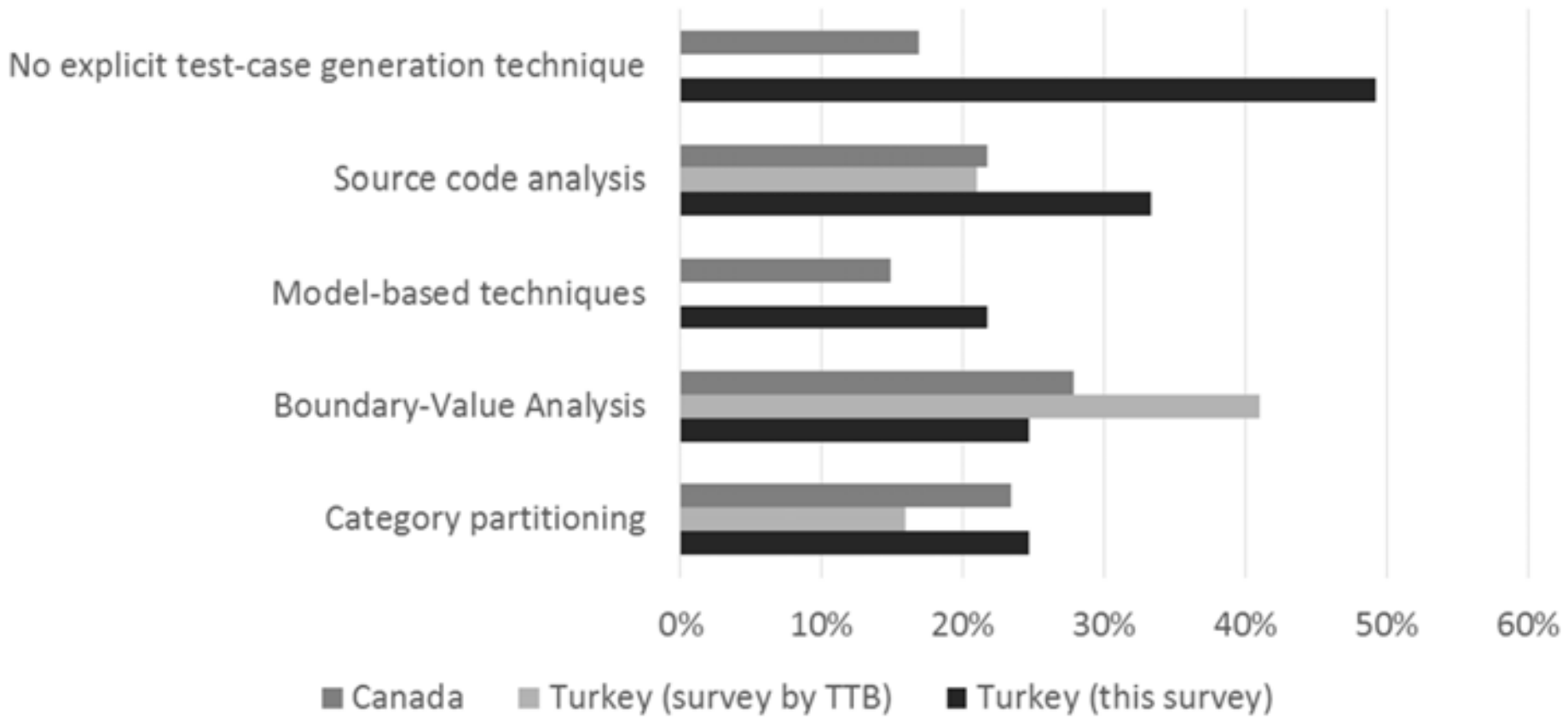

**Figure 33- Test-case design techniques**

### 4.6.5 Test Automation (Q 29)

The next question intended to measure the frequency of conducting manual versus automated testing. Since the data are also available from the Canadian survey, we compare the results for both countries in Figure 34. We can see that in both countries, there is generally wide spectrum for both test approaches (anywhere between 'never' to 'always'). This denotes that different respondents have very different practices in this context, i.e., some heavily practice automated testing, while others favor manual testing.

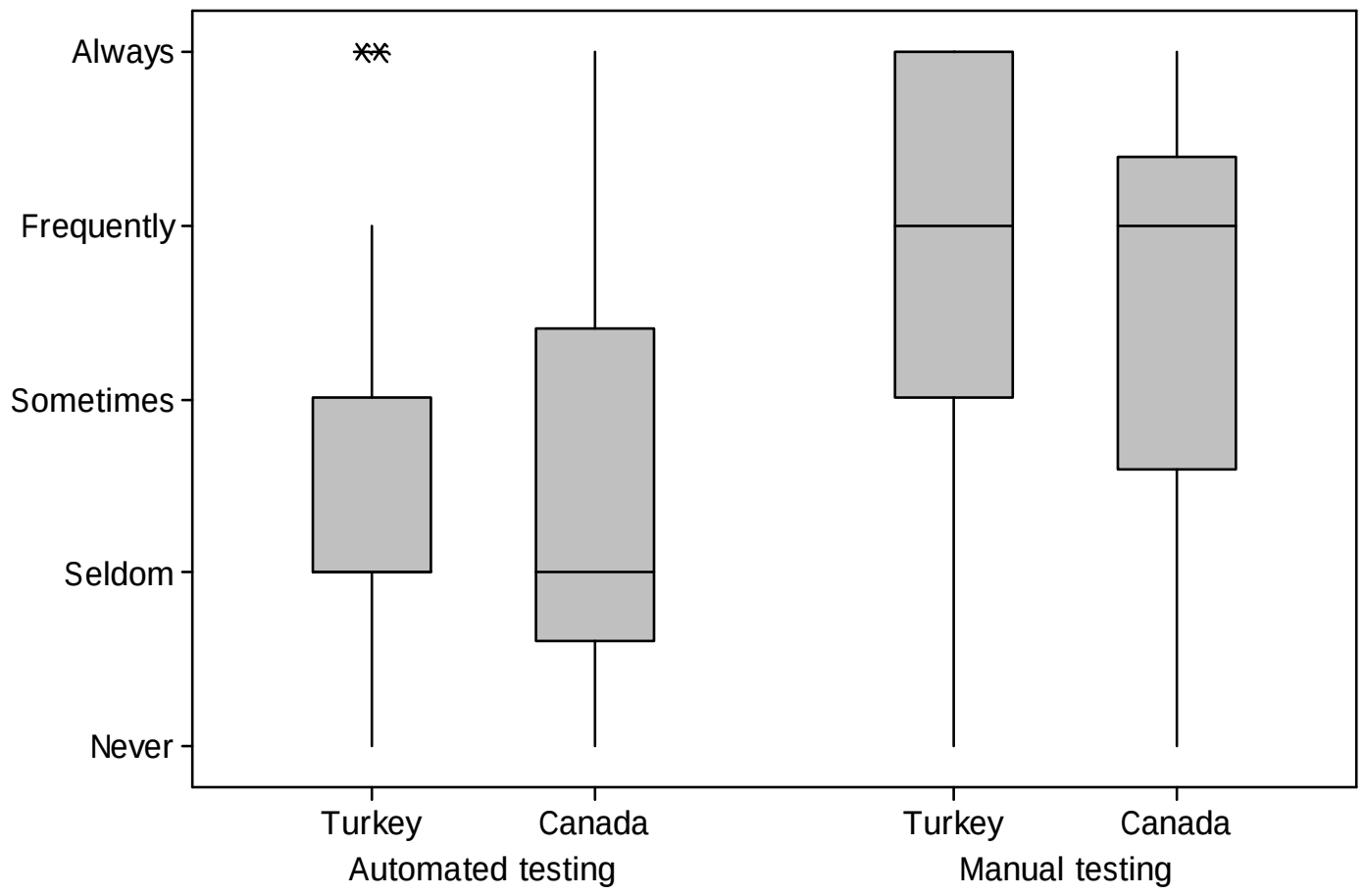

**Figure 34- Manual versus automated testing**

Similar to what we did in the comparative analyses of TDD's adoption versus TLD's (Section 4.6.2), we found that it would be interesting to conduct cross-aspect analyses of manual versus automated testing. We wanted to know how often each pair of these two test approaches is conducted. The bubble-chart in Figure 35 visualizes this cross-aspect analysis. We can state the following noteworthy observations from the bubble-chart:

- To statistically investigate the above cross-aspect issue, we calculated the Pearson's correlation coefficient on the data set shown in Figure 35. The coefficient value is -0.48, meaning that there is a slight negative correlation between the two practices, i.e., high usage of manual testing by a tester (or a team) somewhat implies the low usage of automated testing by that tester (or team).
- Many respondents (13 of them) reported no usage of automated testing and regular ("always") usage of manual testing practice in all their projects. Since the adoption of automated testing in the Turkish industry is very low, it seems that most Turkish SE practitioners are unaware of the advantages of automated testing. Thus, we recommend to the practitioners reviewing the success stories of adopting automated testing, e.g., [54-57]
- There is only one vote by a respondent for the "always-always" combination ("always" conducting both manual and automated testing).

- There is only one vote by a respondent for the "never- never" combination, which perhaps corresponds to a respondent who has never conducted neither manual nor automated testing.

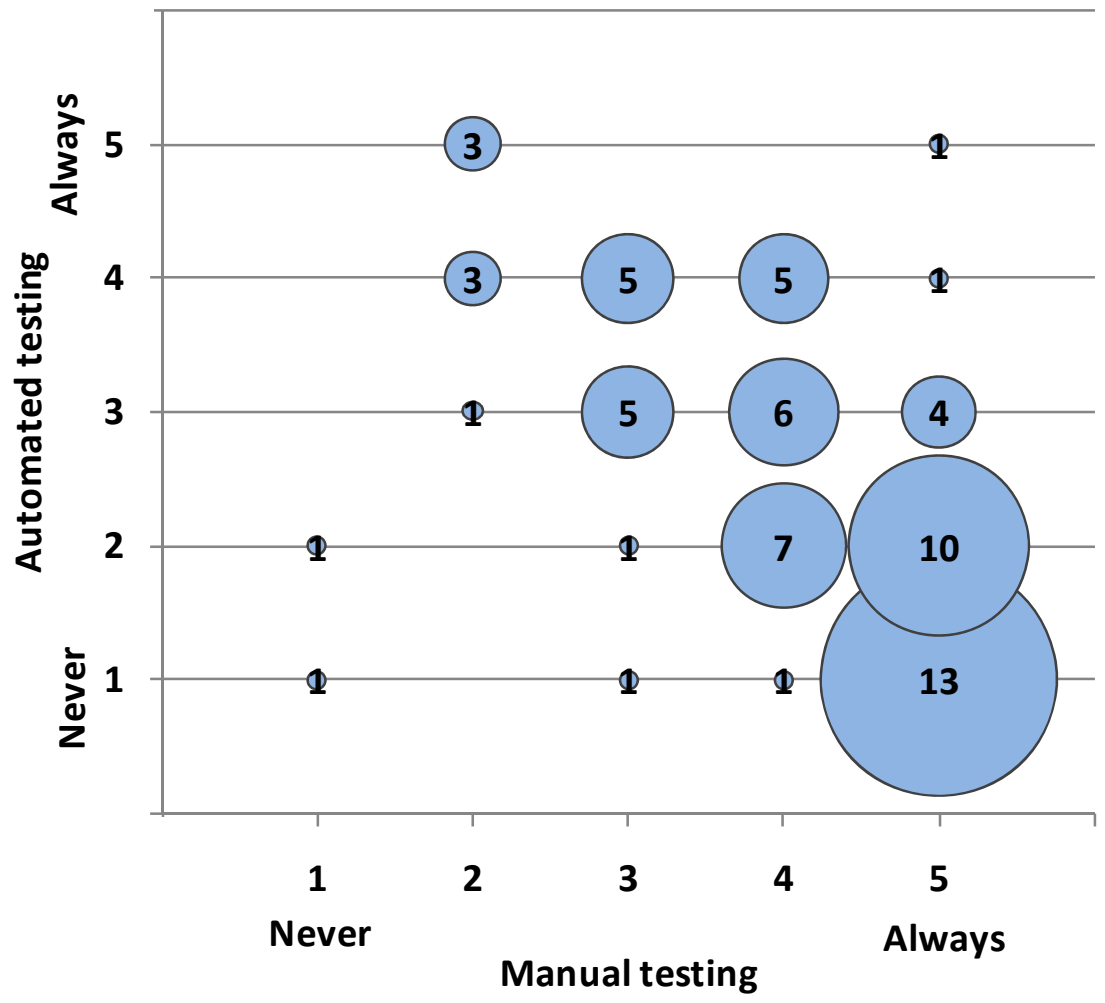

**Figure 35- Comparative analysis of manual versus automated testing**

### 4.6.6 Test-related Metrics (Q 30)

Code (or test) coverage metrics are important means during both test-case design and test effectiveness measurement. The four histograms in Figure 36 depict the results of the survey on the usage of four of the widely used metrics. All four histograms are generally skewed towards left, implying the generally low popularity of coverage metrics in the Turkish software industry. We find out that most Turkish SE practitioners are unaware of the advantages of code coverage metrics. Thus, we recommend to the practitioners reviewing the success stories of using code coverage metrics.

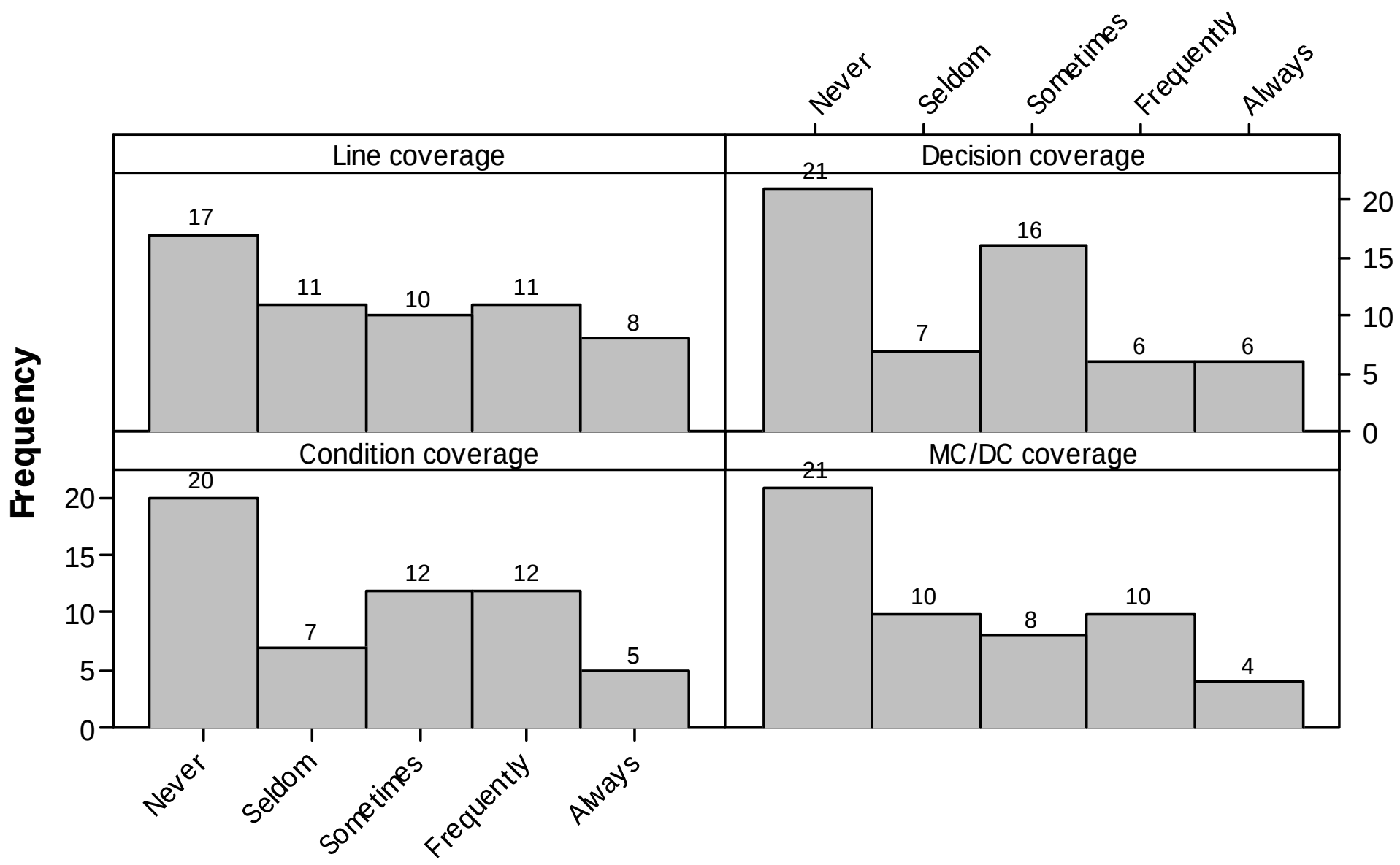

**Average values:**

| | |
|---|---|
| 2.68 | 2.44 |
| 2.55 | 2.36 |

**Median values:**

| | |
|---|---|
| 3 | 2,5 |
| 3 | 2 |

**Figure 36- Usage of code coverage metrics**

### 4.6.7 Other Quality-related Metrics (Q 31)

There are various test and quality metrics which are used by testers [24]. We also surveyed the usage frequency of a few of those metric and results are shown in Figure 37. Again, the histograms are generally skewed towards left, implying the generally low usage of these metrics.

- (Average) Number of defects per Line of Code (LOC)
- Testers defect (bug) detection productivity (bugs found per day by each tester)
- Total number of defects detected per day (week, or month)
- Number of tests cases executed within a period of time
- Number of passing user acceptance tests

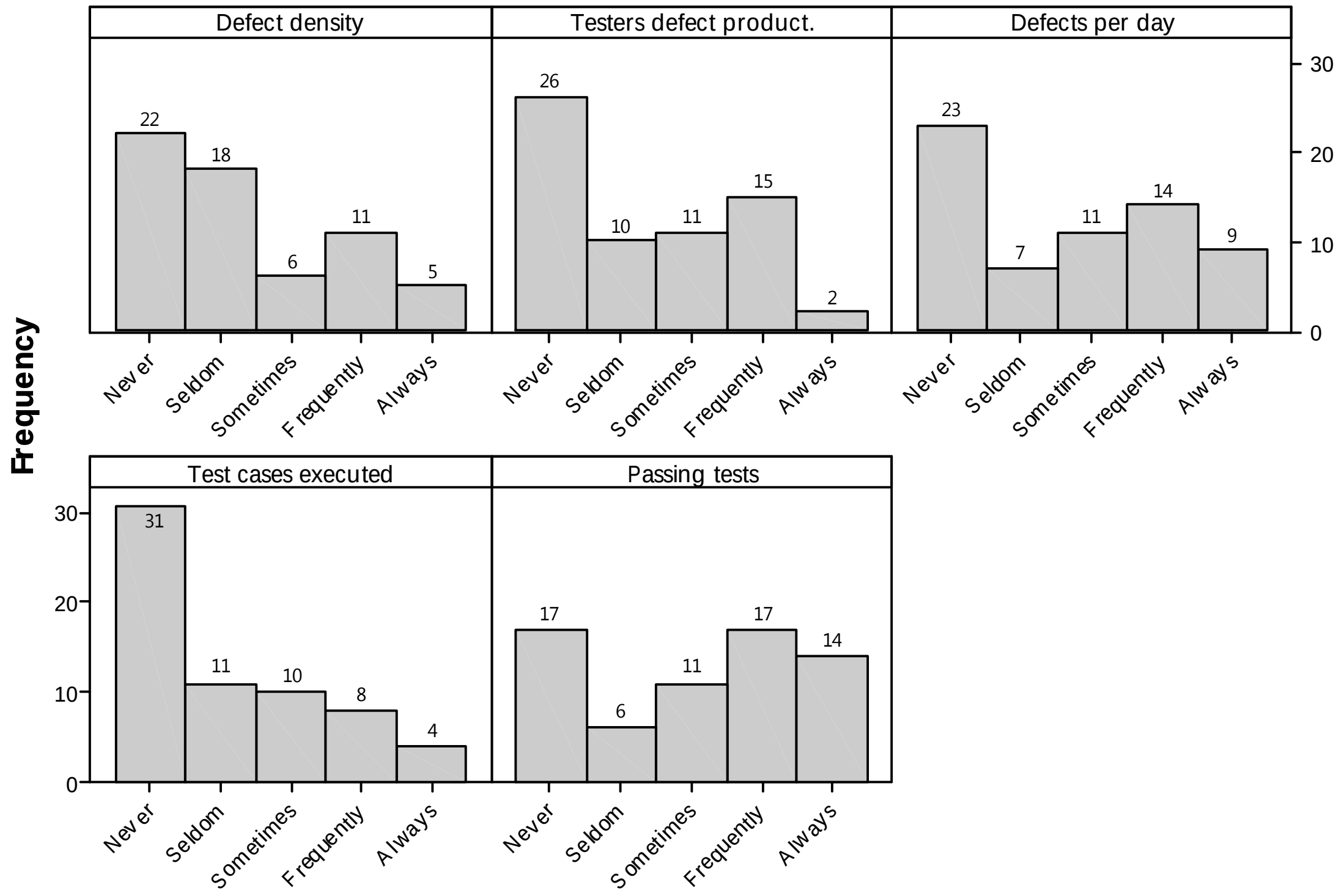

**Average values:**

| | | |
|---|---|---|
| 2,33 | 2,32 | 2,67 |
| 2,10 | 3,07 | 2,10 |

**Median values:**

| | | |
|---|---|---|
| 2 | 2 | 3 |
| 2 | 3 | |

**Figure 37- Test and quality metrics**

### 4.6.8 Ratio of Testers to Developers (Q 32)

This question was designed to find out the tester to developer ratio used in different teams. Results are shown in Figure 38. In most companies, testers are outnumbered by developers, with ratios ranging from 1:2 to 1:5 and higher. According to the figure, the frequency is decreasing as the (tester: developer) proportion grows from 1:5 to 1:2. Besides, a number of companies either make no distinction between these two roles or do not measure this metric.

In fact, the tester to developer ratio (tester: developer) has been an actively debated issue in the software industry lately. James Whittaker, a practitioner test engineering manager has a blog post [58] on the topic where he compares two leading software companies, in terms for their tester to developer ratio. He points out that in the two selected large leading technology companies; the ratio varies between 1:1 and 1:3. He finally concludes that: *"Test managers should be trying to find that sweet spot"* [58]. Iberle and Bartlett have an online article [59] on how to tackle this problem for a company, i.e., estimating the suitable tester to developer ratio. They present an interesting model and approach to come up with realistic tester to developer ratio for any given company/project. In our results, it seems that the most popular "sweet spot" chosen by managers have been from 1:2 to 1:5.

There are other studies on the issue of tester to developer ratio. Cusumano and Yoffie [60] report about the competition of Microsoft and Netscape in 1998 in browser development. They report that Netscape employed fewer testers and they then examined the implications of such a decision to other aspects of the development process, e.g., based on the analysis of data, they found that fewer testers would delay the product's release time).

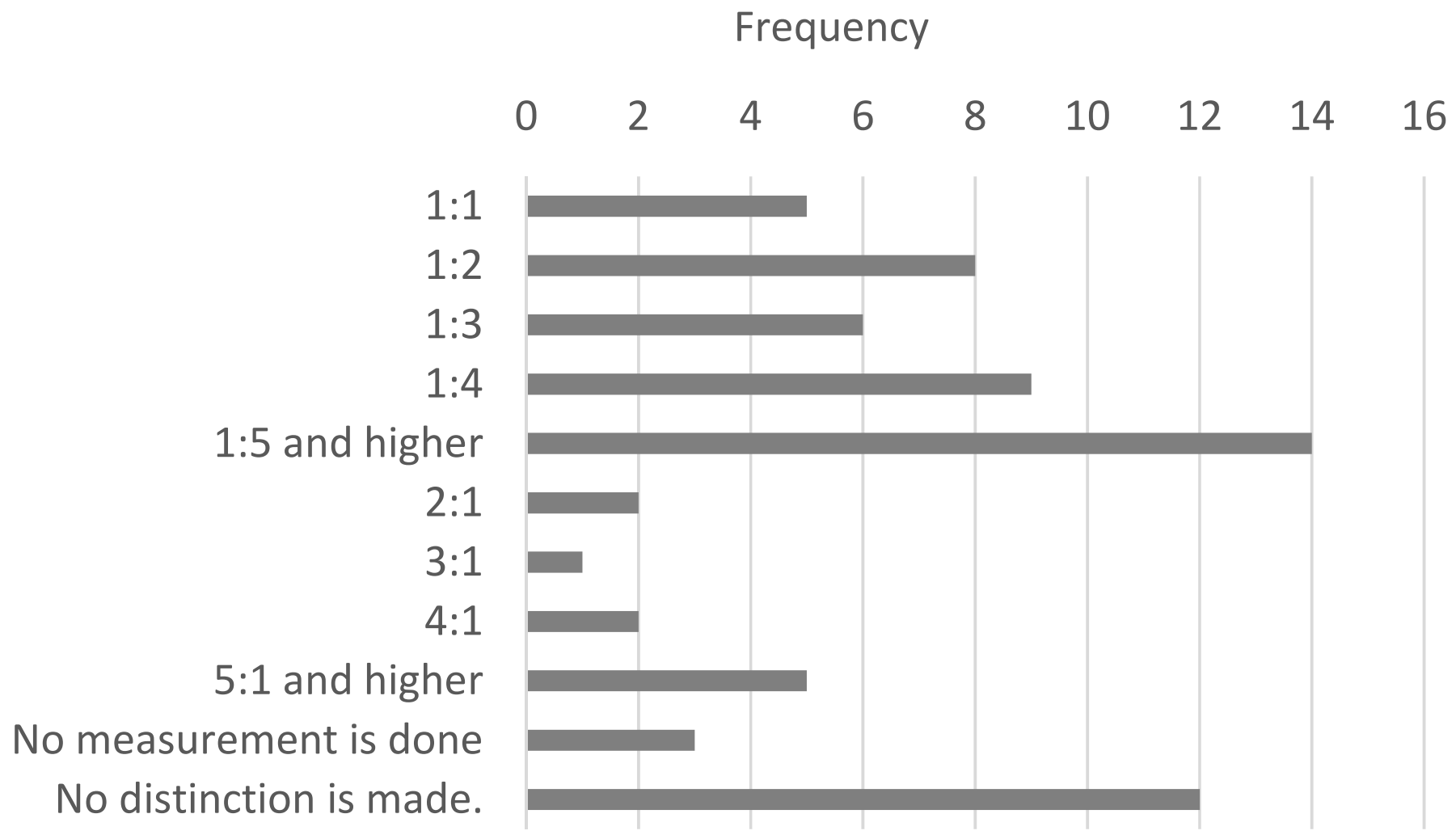

**Figure 38- Ratio of testers to developers (testers: developers)**

### 4.6.9 Criteria for Terminating Testing (Q 33)

In the software testing literature, many criteria have been proposed for ending testing. We had listed the following six criteria for this question and respondents could also add their own criteria.

- Code coverage analysis (e.g., when you reach a predetermined threshold of statement coverage, e.g., 90%)
- Fixed time duration
- Fixed budget
- Informal
- No bugs are found any more (reliability growth/saturation models)
- All test cases are executed without finding more defects (bugs)

The participants were asked to rate these criteria in the 5-point Likert scale. Figure 39 shows the results. Average values of the votes have also been shown in Figure 39. The last criterion seems to be the most widely used criterion for ending testing activities. Code coverage metrics seems to be not that popular, denoting their poor acceptance in the Turkish industry.

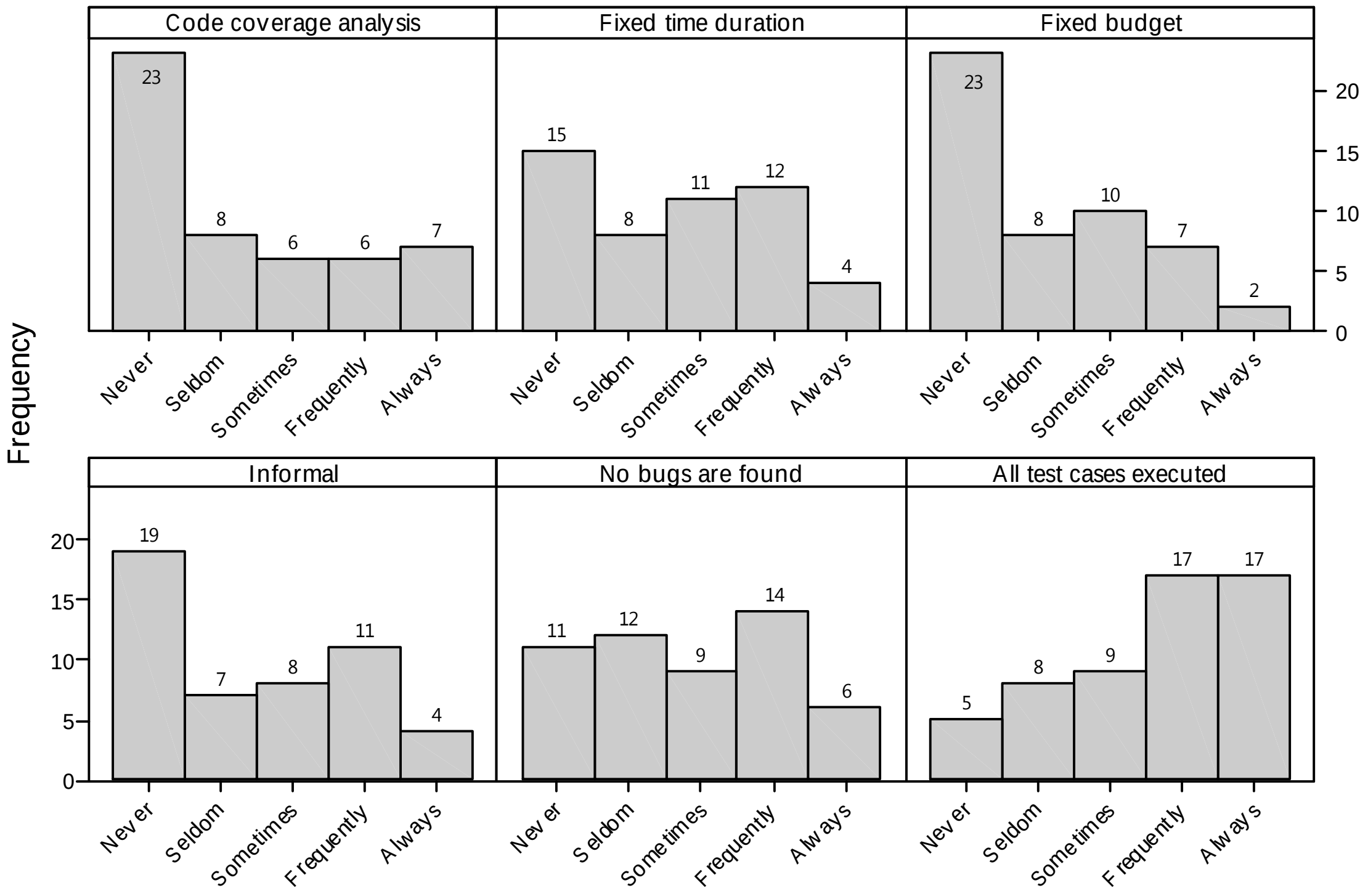

**Average values:**

| | | |
|---|---|---|
| 2,32 | 2,64 | 2,14 |
| 2,46 | 2,84 | 3,58 |

**Median values:**

| | | |
|---|---|---|
| 2 | 3 | 2 |
| 2 | 3 | 4 |

**Figure 39- Criteria for terminating testing**

The 2013 survey report by the Turkish Testing Board (TTB) [33] also had a question about the test-exit criteria. Two of that report's choices are the same as our choices in Figure 39, i.e., time and budget constraints. The measures in the TTB report were percentages, e.g., 11% of participants reported that their test exit criterion was budget constraints. To be able to compare our results with those of the TTB survey, we converted our Likert data to percentage values as follows. The average value for the constant budget data set in Figure 39 is 2.1, thus its percentage value would be (2.1-1)/5=22%. Similarly, the deadline series' percentage would be (2.6-1)/5=32%. Comparison of our results with those of the TTB survey is shown in Table 6. In both surveys, time constraints have been reported to be more the case than budget constraints.

**Table 6-Test termination criteria: comparison of our results with the results of the TTB survey [33].**

| | Our Survey | TTB survey [33, 61] |
|---|---|---|
| Time constraints/ deadline | 32% | 50% |
| Budget constraints | 22% | 11% |

### 4.7 SOFTWARE MAINTENANCE

Here we report the following results next:

- Types of maintenance (Q 34)
- Type of challenges during maintenance (Q 35 and 37)
- Description of maintenance challenges (Q 36)

#### 4.7.1 Types of Software Maintenance (Q 34)

This question asked the frequencies of the type of maintenance activities performed by software organizations. We provided the following types of maintenance in the survey.

- Corrective maintenance: Reactive modification of a software product performed after delivery to correct discovered problems.
- Adaptive maintenance: Modification of a software product performed after delivery to keep a software product usable in a changed or changing environment.
- Perfective maintenance: Modification of a software product after delivery to improve performance or maintainability.

Figure 40 displays the results and average values in a Likert scale of 1 (never) to 5 (always). Corrective maintenance is the most widely-used maintenance type whereas perfective maintenance is the least used.

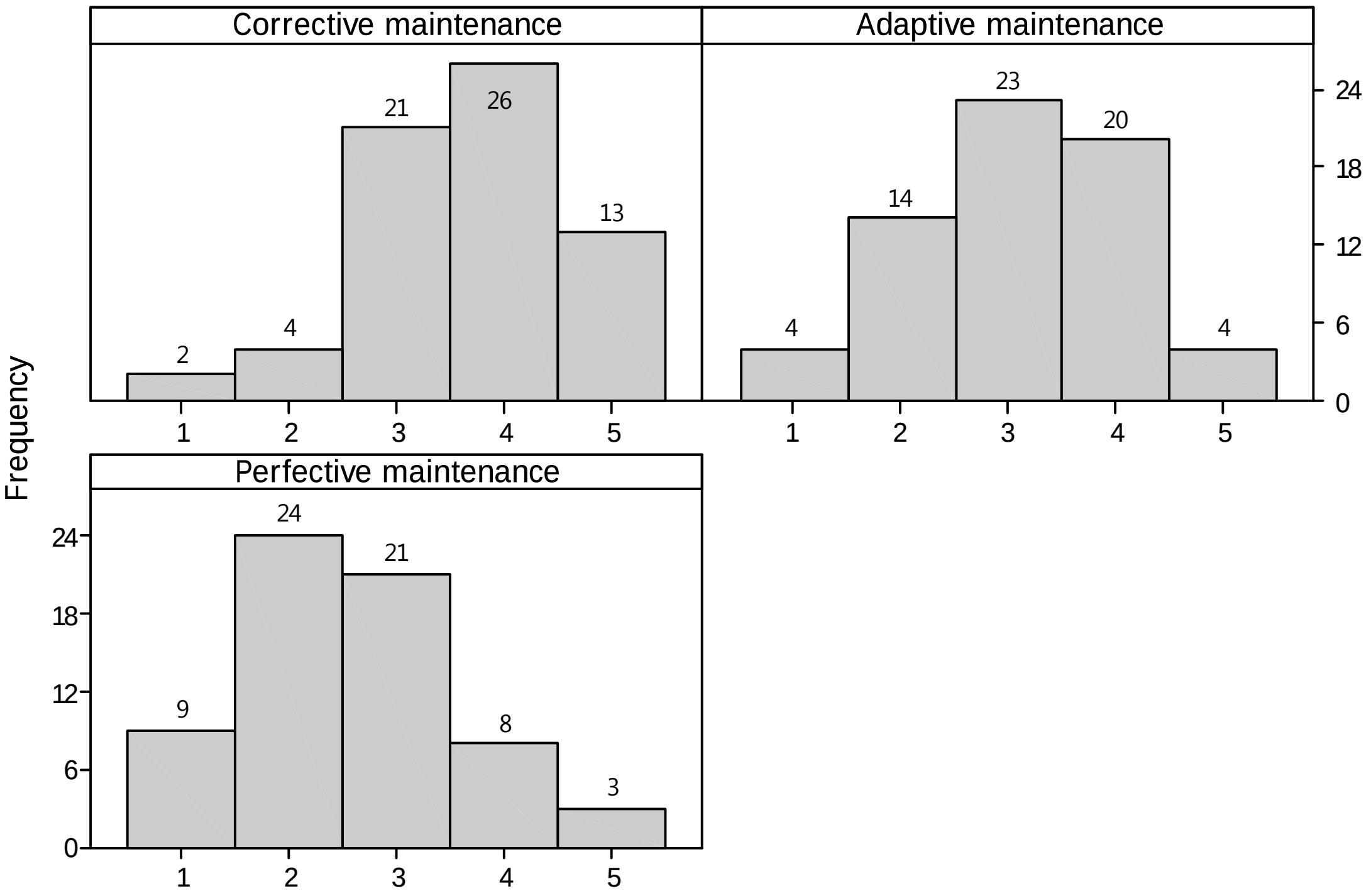

**Average values:**

| | |
|---|---|
| 3.67 | 3.09 |
| 2.57 | |

**Median values:**

| | |
|---|---|
| 4 | 3 |
| 2 | |

**Figure 40- Types of software maintenance activities used**

### 4.7.2 Level of Challenges during Maintenance (Q 35)

Participants were asked to rank the level of challenge they experience in each of the three maintenance types. Interestingly, most and least frequently performed maintenance types (as per the answers of Q 34) are also the ones software engineers experience challenges most and least respectively (as we reported in the previous question, Q 34).

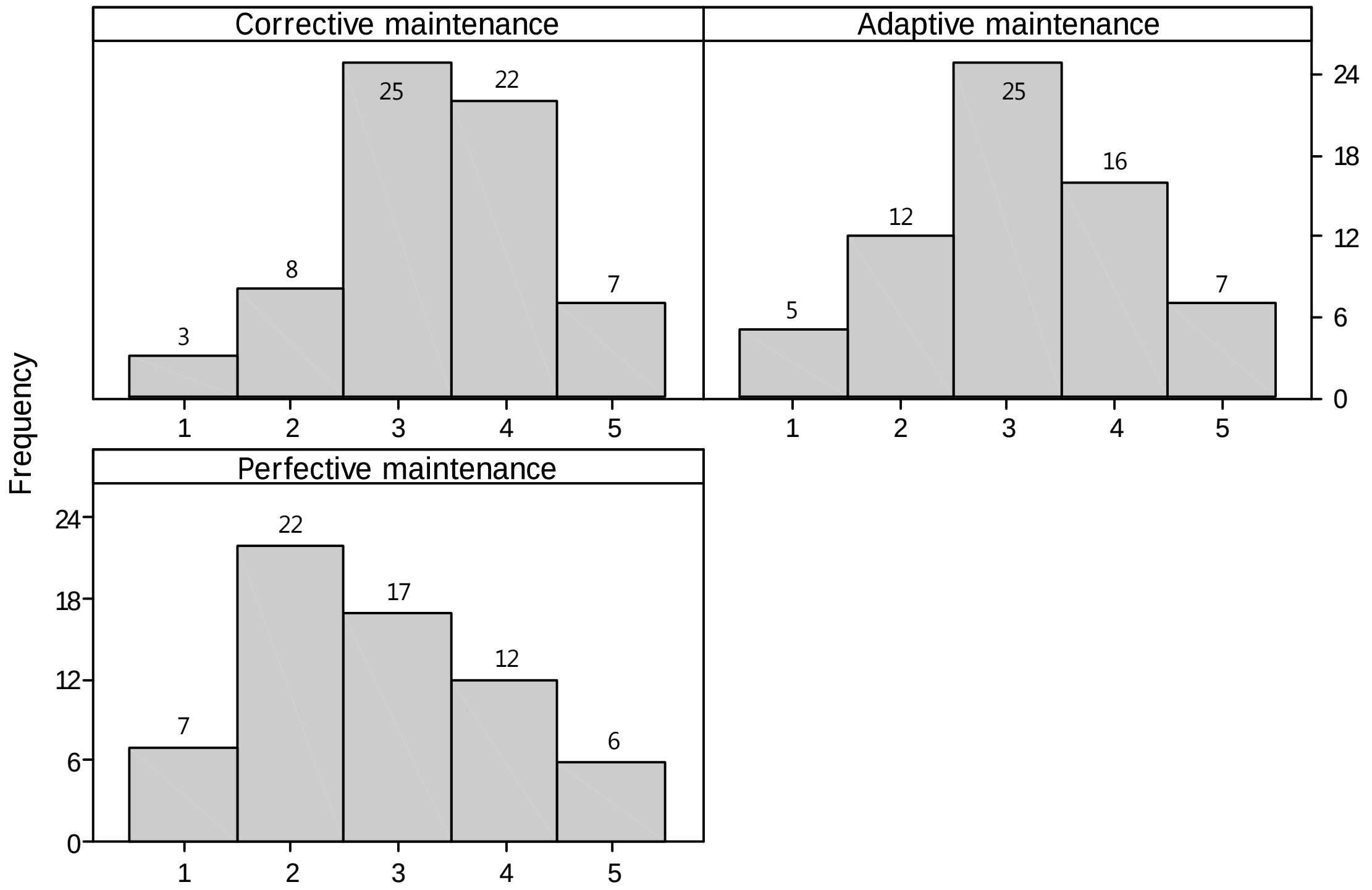

**Average values:**

| | |
|---|---|
| 3.34 | 3.12 |
| 2.81 | |

**Median values:**

| | |
|---|---|
| 3 | 3 |
| 3 | |

**Figure 41- Levels of challenges during maintenance activities**

### 4.7.3 Description of Maintenance Challenges (Q 36)

We were keen to find out the challenges of the software maintenance phase in the industry. Respondents were asked to summarize the challenges in their own words. Some blamed poor testing practices as the cause of high corrective maintenance effort. A few comments about this are:

- *"Defects emerge because of redundant test data"*
- *"Performance tests might not be performed in specified environments"*

Another respondent thinks managers are responsible for poor testing resulting in high corrective maintenance costs: *"Managers do not understand importance of testing, so we cannot persuade them to establish an independent test team"*.

Many others think change management and related practices are at the top of the list of causes of maintenance difficulties. One respondent pointed out that *"motivation and pressure of not breaking the working code"* is a challenge. Another agreed by stating a challenge is to *"detect the defects and resolve them without causing other defects"*. Two others commented in this issue by specifically mentioning an issue about regression tests as:

- *"We experience difficulties when regression tests are not done properly"*
- *"Finding out the module that is affected by the change is difficult"*

Other respondents stated the causes of maintenance difficulties include: poor maintenance support, architecture changes, personnel turnovers, undocumented code, time constraints, detecting defects in large projects, uncertain/volatile customer needs, and poor defect management.

### 4.7.4 Type of Challenges during Maintenance (Q 37)

As a follow-up question to Q 35 and 36, respondents were asked about the types of challenges experienced during maintenance from a given list. The following options were presented and respondent could select multiple choices.

- Different programming styles used by original developers
- Insufficient documentation
- Lack of tools in support of maintenance

Figure 42 presents the results. Insufficient documentation seems like a crucial issue for maintenance. Different programming styles is also an important type of challenge.

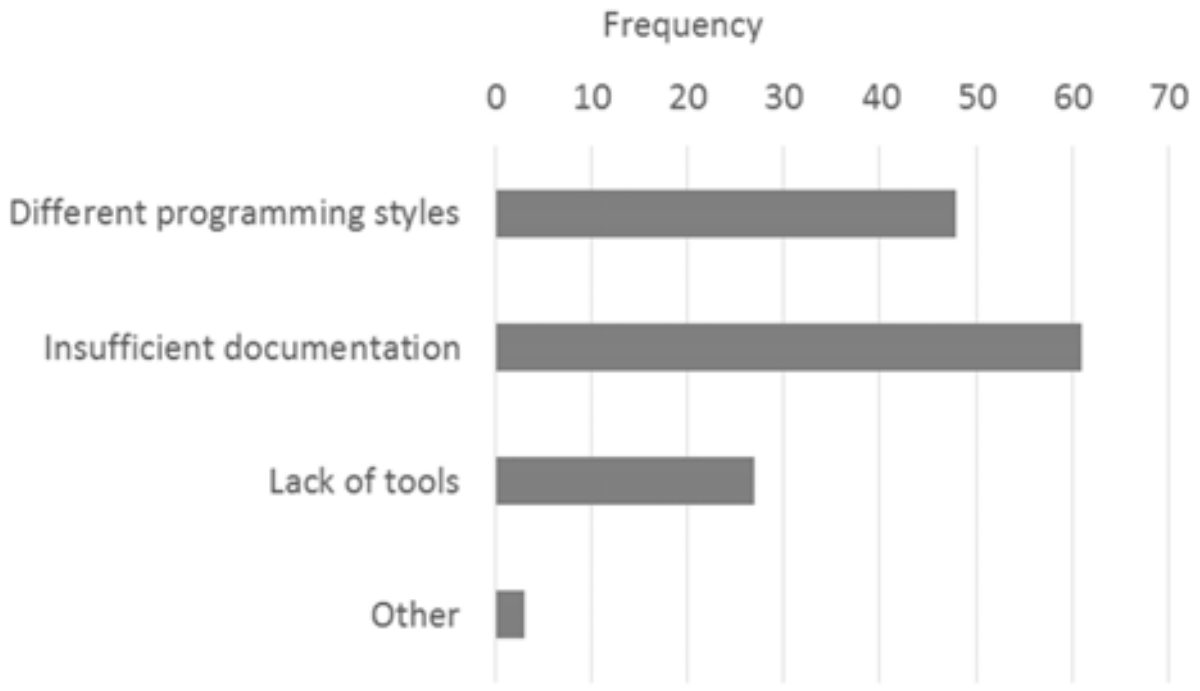

**Figure 42- Types of maintenance challenges**

## 4.8 SOFTWARE CONFIGURATION MANAGEMENT, RELEASE PLANNING AND SUPPORT

We report the following results next:

- Practices during product release and delivery (Q 38)
- Practices during product support (Q 39)

### 4.8.1 Practices during Product Release and Delivery (Q 38)

Q 38 asked participants to rank the following practices carried out by organizations during release and delivery of software projects, in a Likert scale. Histograms of participants' answers to each of the following options are presented in Figure 43.

- We have a formal release gate.
- Representatives from marketing are involved in release decisions.
- Representatives from development are involved in release decisions.
- Representatives from QA or test are involved in release decisions.
- Representatives from client support are involved in release decisions.
- We involve clients in the release process.
- QA alone makes the decision about product release.
- We release urgent patches without comprehensive testing.
- We collaborate closely with clients when we release software.
- We have a separate test and release environment.

Looking at the skewness and average values in Figure 43, most and least popular activities are: having representatives from QA and test in release decisions, and QA making the product release decisions alone, respectively. These are in line with Kirk and Tempero's study [17] stating that 51% of their respondents reported that the QA team is always involved in release decisions and 73% reported the QA team never decides alone for releases.

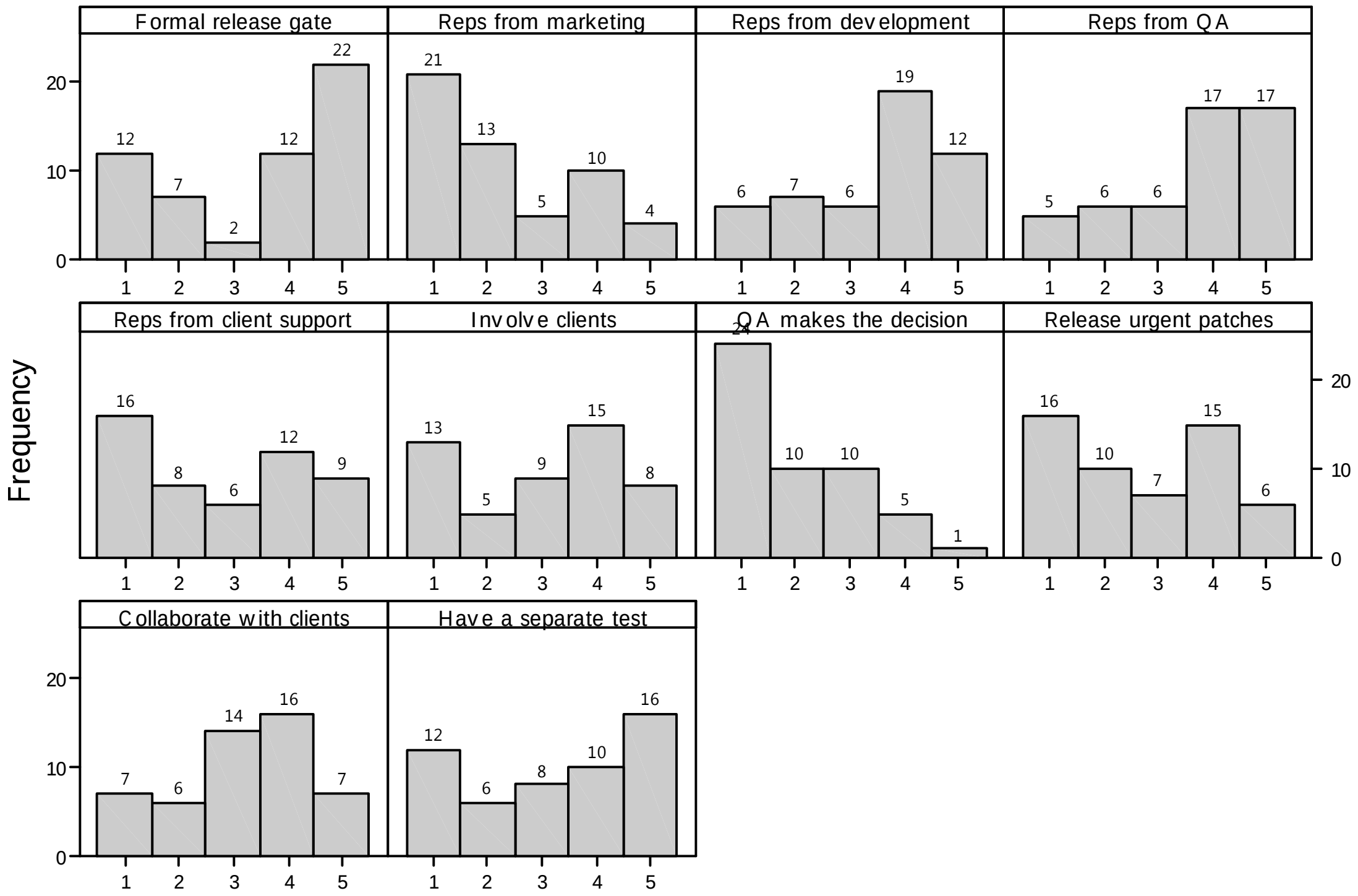

**Average values:**

| | | | |
|---|---|---|---|
| 3.45 | 2.30 | 3.48 | 3.69 |
| 2.80 | 3.00 | 1.98 | 2.72 |
| 3.20 | 3.23 | | |

**Median values:**

| | | | |
|---|---|---|---|
| 4 | 2 | 4 | 4 |
| 3 | 3 | 2 | 3 |
| 3 | 3,5 | | |

**Figure 43- Practices during product delivery**

### 4.8.2 Practices during Product Support (Q 39)

Q 39 was about practices carried out during the software support phase. The following practices were provided in the list and respondents answered them on a binary basis (yes or no), e.g., either their developers work with clients on-site or not.

- Developers work with clients on-site.
- A dedicated person is assigned to each major client.
- Our support team receives formal product training.
- Our client issue tool is separate from our defect tracking tool.
- We use a tool for client issue tracking.
- Our support team, developers and QA work closely together.
- We provide tiered support.
- We have a dedicated support team.

Figure 44 presents the results. The most popular practice is having a dedicated support team with 54% response rate. On the other hand, least popular practice in this category was to have a dedicated person assigned to each major client with 24% response rate.

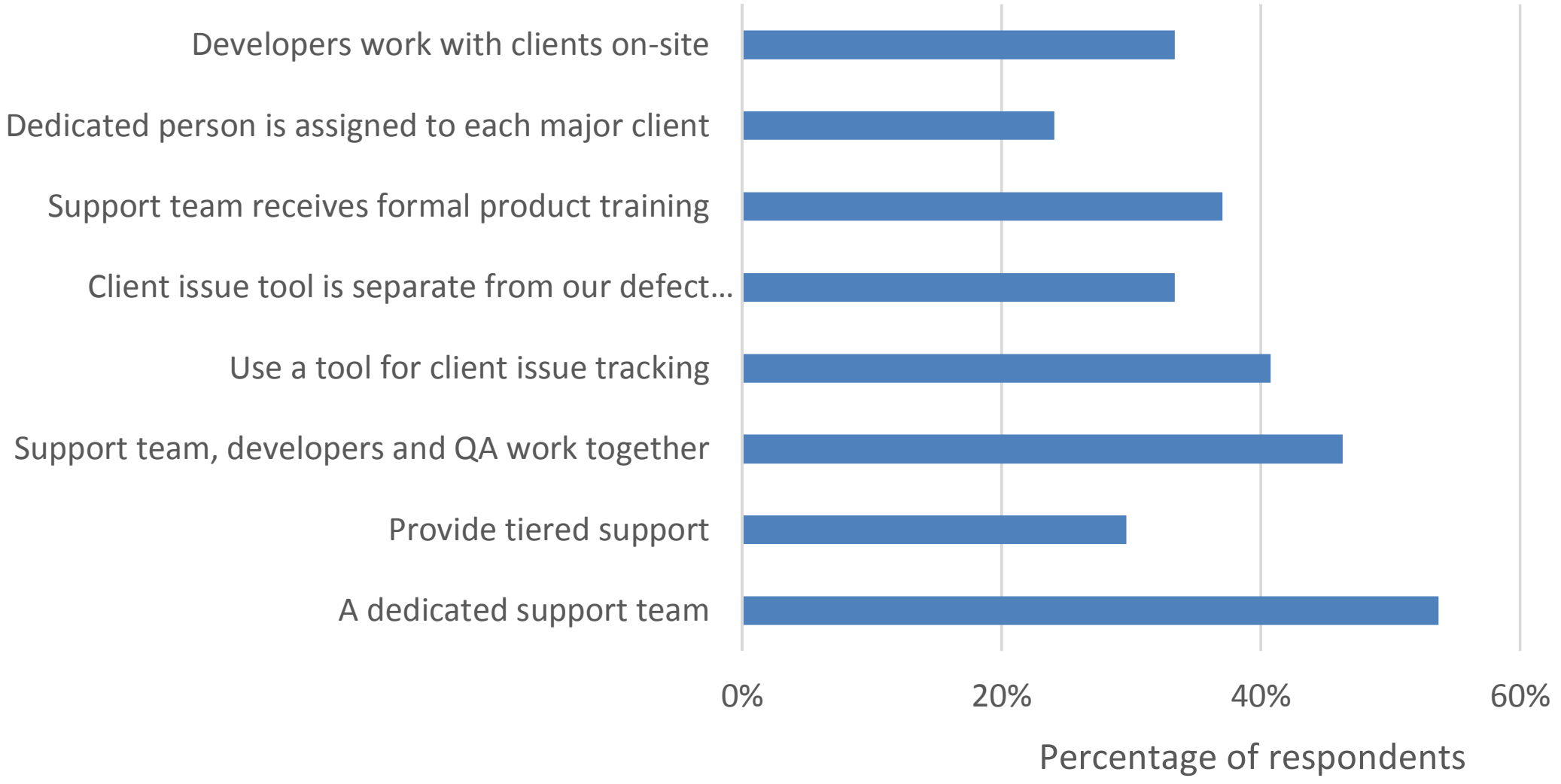

**Figure 44- Practices during product support**

## 4.9 Software Project Management

We report the following results next:

- Measurements and estimations in project management (Q 40)
- Planning and monitoring projects (Q 41)

### 4.9.1 Measurements and Estimations in Project Management (Q 40)

Q 40 asked to rank the use of measurements and estimations in project management practices again in the Likert scale. The following practices were available to respondents.

- Decisions taken by projects managers are based on systematic metrics and measurements.
- We perform systematic cost and effort estimations before each project.
- We utilize formal cost estimation approaches, e.g., Constructive Cost Model (COCOMO)

Figure 45 presents the results. Interestingly, while systematic cost and effort estimation is popular, formal cost estimation approaches are not, according to Figure 45. In other words, organizations perform estimations, but do not use formal approaches for estimation.

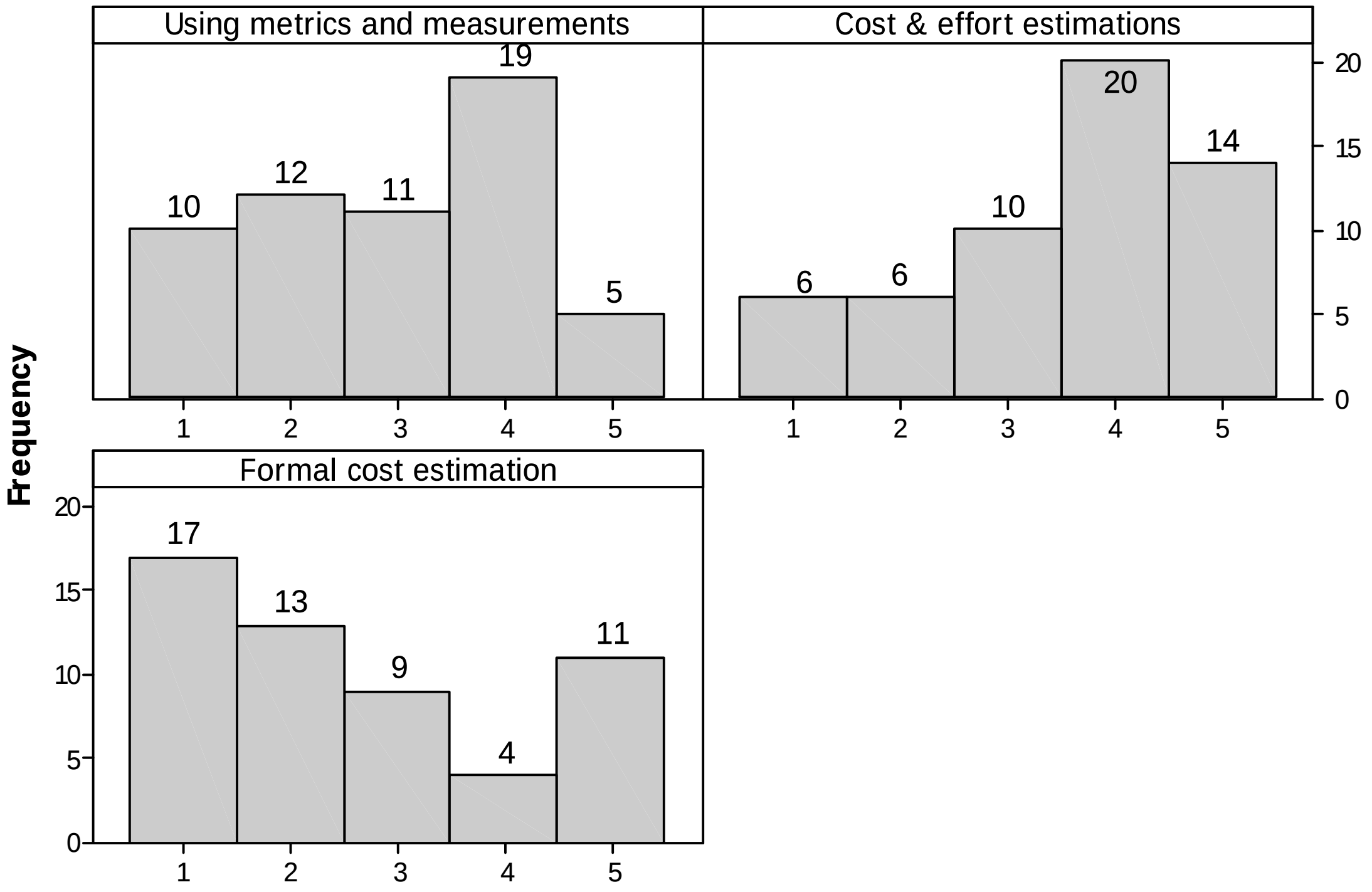

**Average values:**

| | |
|---|---|
| 2.95 | 3.54 |
| 2.61 | |

**Median values:**

| | |
|---|---|
| 3 | 4 |
| 2 | |

**Figure 45- Measurement and estimation related management practices**

### 4.9.2 Planning and Monitoring Projects (Q 41)

Q 41 was about planning and monitoring project management practices. The following five practices were presented to be ranked in a Likert scale.

- In our projects, we create weekly operational plans
- In our projects, we create plans for the entire project but we do not tailor defined standard processes for planning
- In our projects, we create plans for the entire project based on the defined standard process
- In our projects, the management team creates (weekly/monthly) progress reports
- In our project, we perform (weekly/monthly) progress meetings

Figure 46 presents the results. The most widely performed practice in this category is having periodic progress meetings.

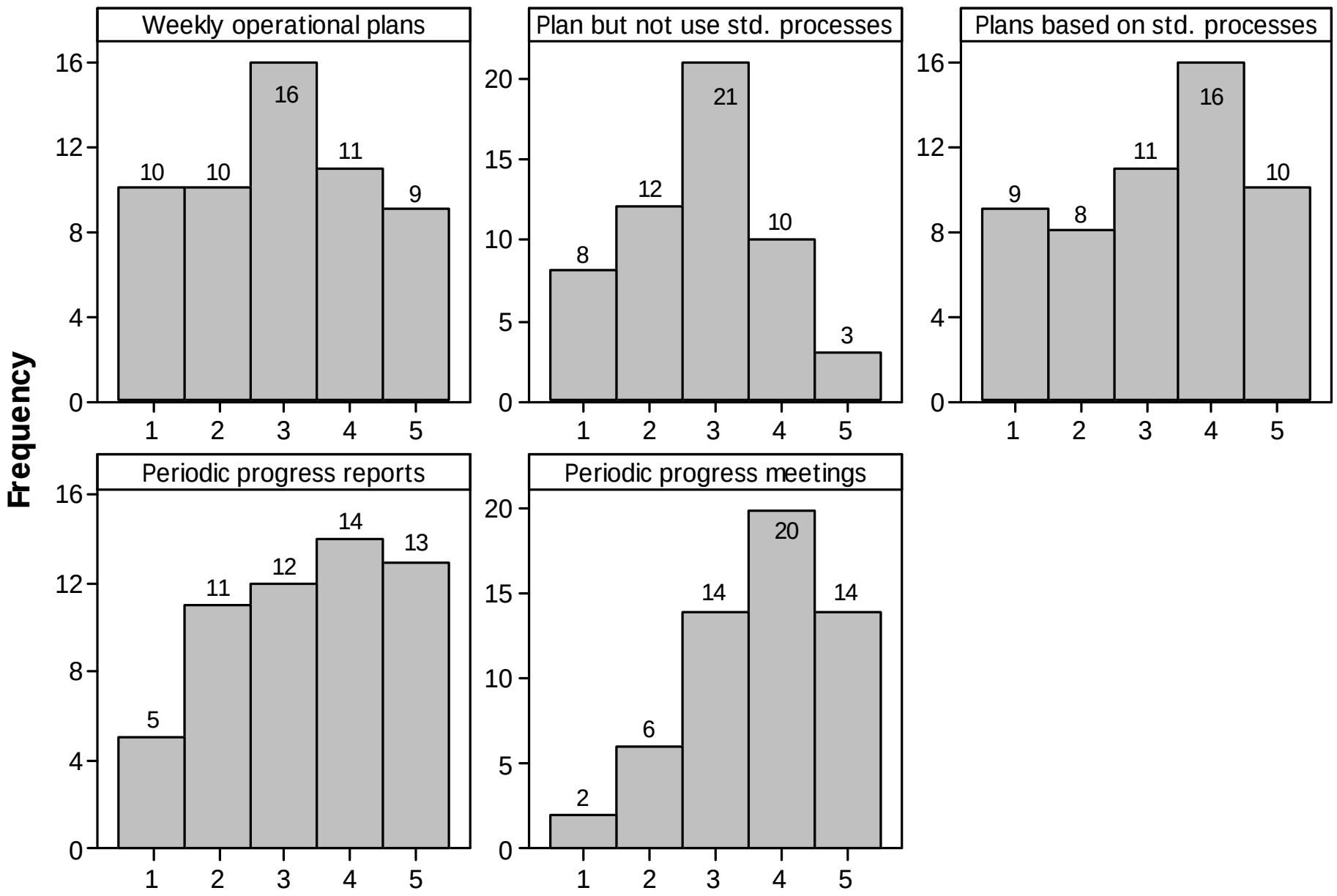

**Average values:**

| | | |
|---|---|---|
| 2.98 | 2.78 | 3.18 |
| 3.34 | 3.68 | |

**Median values:**

| | | |
|---|---|---|
| 3 | 3 | 3 |
| 3 | 4 | |

**Figure 46- Planning and monitoring practices**

## 4.10 Software Engineering Tools

The only question in this category was usage of SE tools (Q 42).

### 4.10.1 Usage of Software Engineering Tools (Q 42)

We also wanted to know the tools used in SE activities. Respondents were asked about how often they use the following types of tools.

- We use software requirements tools (e.g., for documenting requirements)
- We use software design tools (e.g., UML modeling tools)
- We use software testing tools (e.g., automating GUI testing)
- We use software maintenance tools (e.g., for comprehension or reverse engineering of design artifacts)
- We use SE process tools (e.g., for process modeling and process management)
- We use software quality tools (e.g., for static code analysis)
- We use software configuration management tools (e.g., for version management, release and build)
- We use software project management tools (e.g., for project planning, tracking, and risk management)

Figure 47 presents the results. The histogram for configuration management tools is significantly skewed towards right, which means most organizations use configuration management tools. Project management, design, and requirements tools are also popular. However, software maintenance tools are less widely used. Sökmen [15] also reported in 2010 that usage of project management (53%) and analysis and design (52%) tools is high, but configuration management tool usage (26%) is lower than what we see in our study.

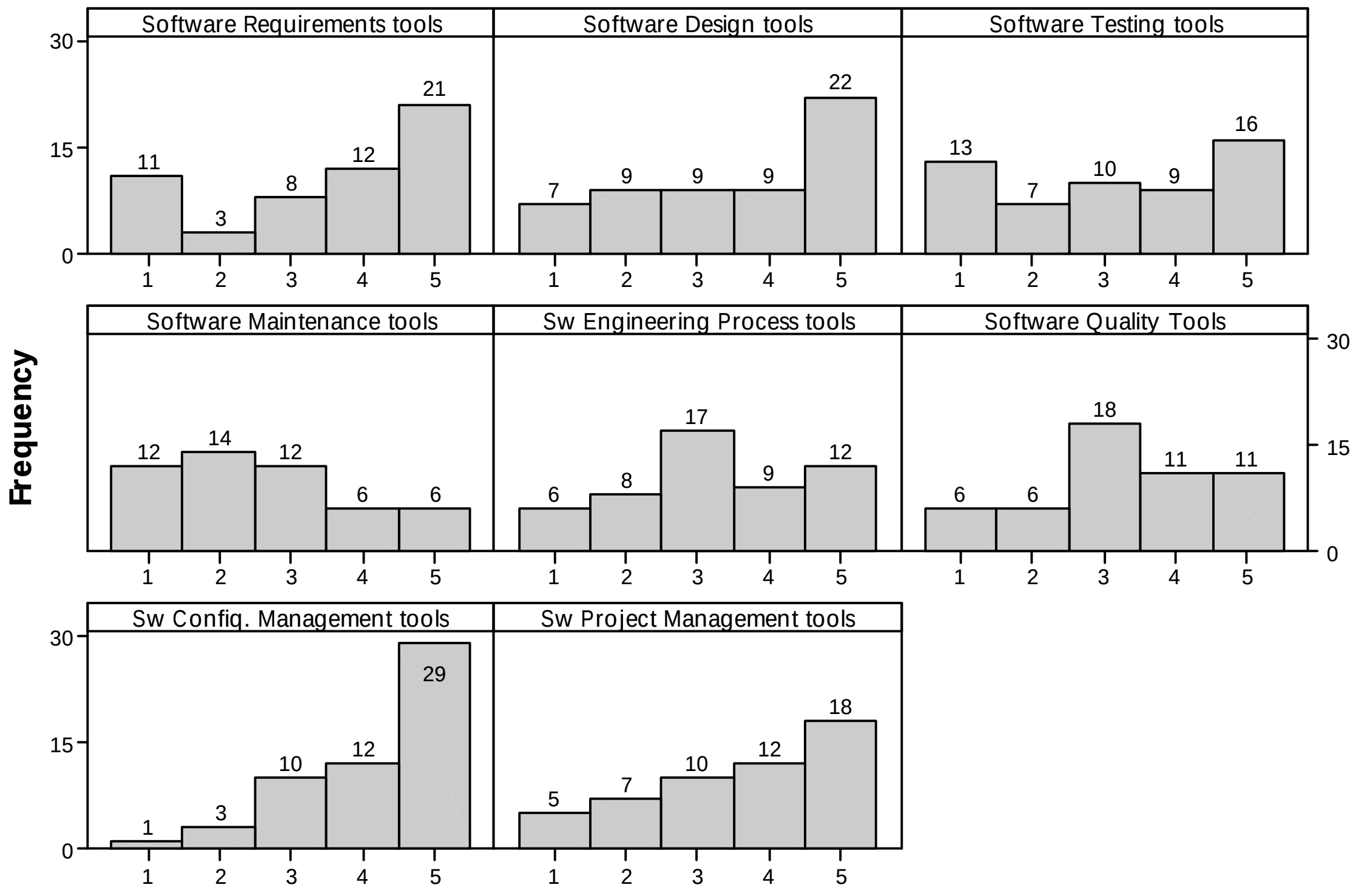

**Average values:**

| | | |
|---|---|---|
| 3.53 | 3.54 | 3.14 |
| 2.60 | 3.25 | 3.29 |
| 4.18 | 3.60 | |

**Median values:**

| | | |
|---|---|---|
| 4 | 4 | 3 |
| 2 | 3 | 3 |
| 5 | 4 | |

**Figure 47- Software engineering tools**

## 4.11 SOFTWARE QUALITY ASSURANCE

The only question in this category was software quality practices (Q 43).

### 4.11.1 Software Quality Practices (Q 43)

Our last consideration about phases was on quality assurance practices. The following practices were available for participants to rate in the Likert scale.

- Our firm follows industry-standard quality specifications in its projects, e.g., DO-178B (Software Considerations in Airborne Systems and Equipment Certification), ISO/IEC 9126 (Software engineering — Product quality), or IEEE 730-1998 (Standard for Software Quality Assurance Plans)
- We have well-defined verification and validation targets (e.g., defect density) in our projects
- We conduct systematic peer reviews and audits in our projects
- We have well-defined approaches for managing defects (e.g., defect characterization, bug triage guidelines)

Figure 48 presents the results. Peer reviews and audits are the most widely performed quality practice. It is a good sign that many companies "always" conduct all the four above practices.

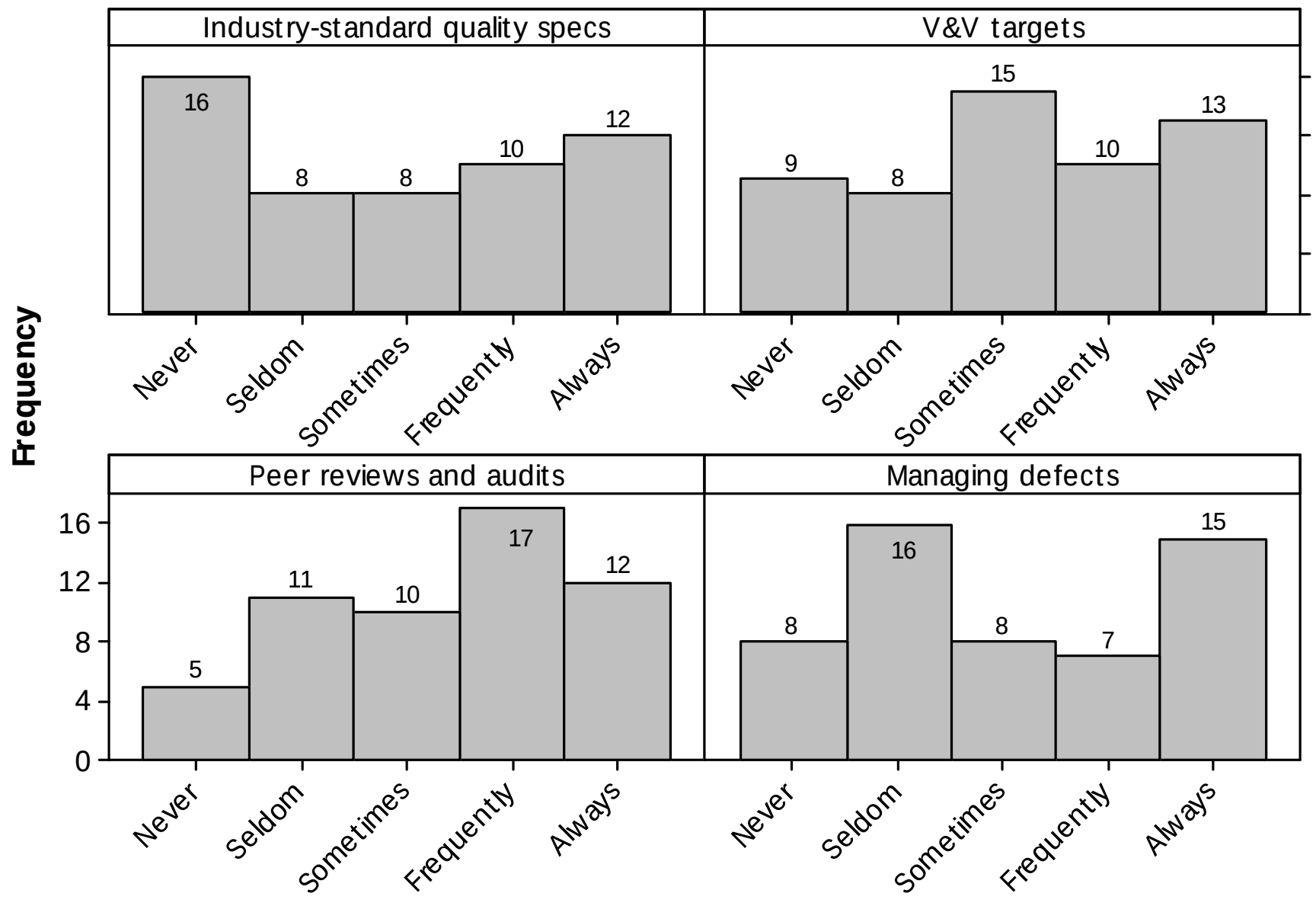

**Average values:**

| | |
|---|---|
| 2.89 | 3.18 |
| 3.36 | 3.09 |

**Median values:**

| | |
|---|---|
| 3 | 3 |
| 4 | 3 |

**Figure 48- Software quality practices**

## 4.12 RESEARCH AND INTERACTION WITH UNIVERSITY RESEARCHERS

In this category of question, we wanted to measure the extent of research activities in various firms and also the extent of interaction with university researchers. Here we report the following results:

- Having dedicated R&D departments/units (Q 44)
- Interacting with the SE university researchers (Q 45)
- Reading technical SE papers and articles (Q 46)

### 4.12.1 Having Dedicated R&D Departments/Units (Q 44)

We were curious if having a dedicated R&D department is common among SE firms. 36 respondents mentioned that their firms have R&D departments, while 24 respondents mentioned not having such departments. Roughly, 40% of the firms in our pool had R&D departments, specialized in SE research and aiming at developing new better ways to develop software. In the open-ended part of this question, participants mostly mentioned lack of management support for R&D units, and R&D units' focus being mostly on software process improvement.

### 4.12.2 Interacting with the Software Engineering University Researchers (Q 45)

To have a vibrant SE industry, it is important that the SE practitioners constantly interact with the university researchers. Different from the previous Likert-scale questions in this survey, the available options of this question were: (1) About once a month, (2) About once every 6 months, (3) About once a year, (4) Very seldom, and (5) Never.

As seen in Figure 49, the distribution is skewed towards right, indicating that interaction with university researchers is very low. This is rather disappointing as most software engineers in the Turkish industry very seldom or never interact with university researchers to discuss their SE success stories or their SE problems/challenges. We should state that, to encourage industry-academia collaboration, there have been recent new funding programs by the Turkish government, e.g., program #1505 (University-Industry Cooperation Support Program) by the Scientific and Technological Research Council of Turkey (acronym in Turkish: TÜBİTAK) and a R&D support program called "San-Tez" by the Ministry of Science, Industry and Technology.

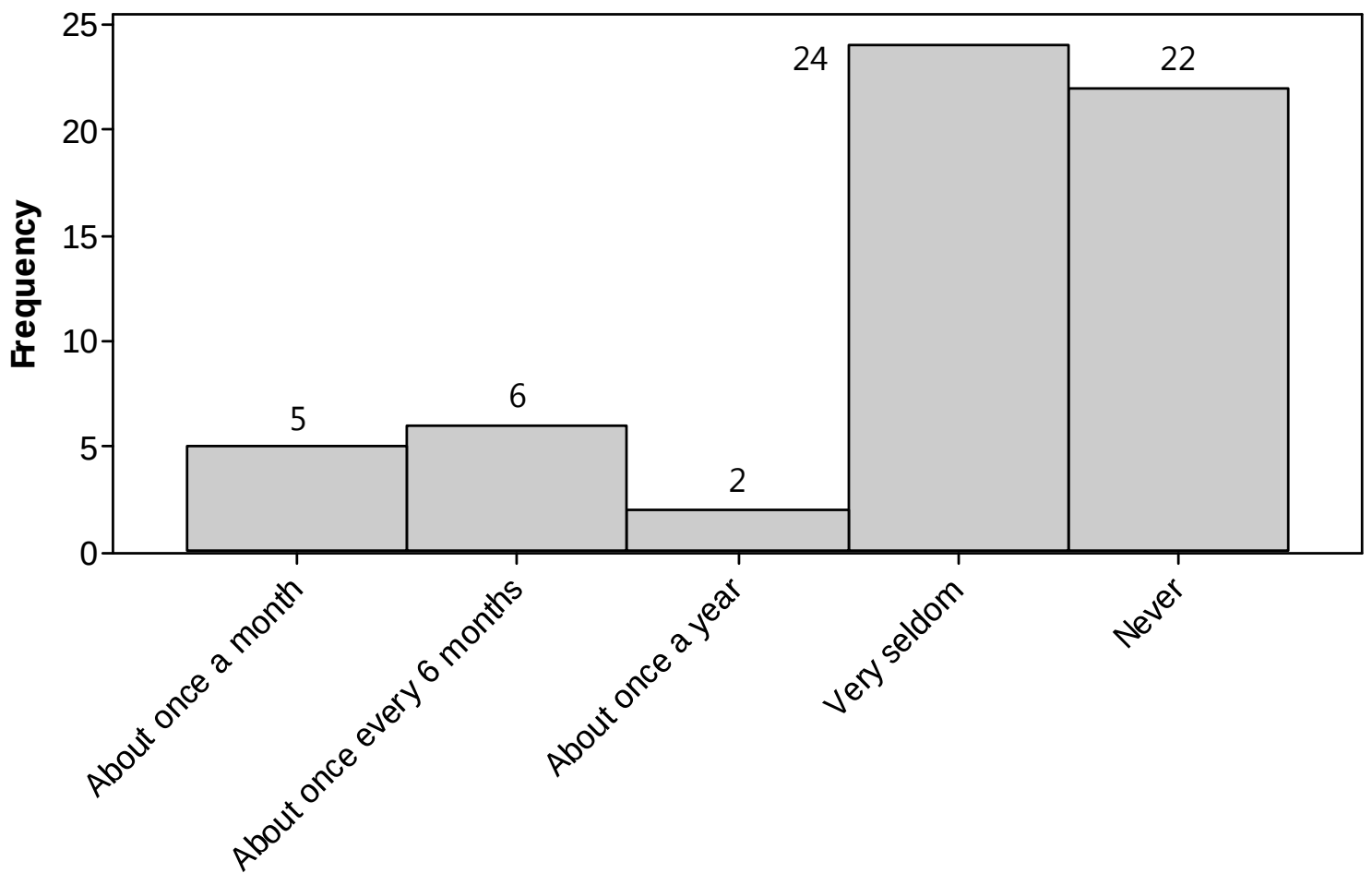

**Figure 49- Interacting with university researchers**

### 4.12.3 Reading Technical SE Papers and Articles (Q 46)

Finally, participants were asked how often they read technical papers or articles published in SE journals, magazines and conferences. The answer options were the same as the previous question: (1) About once a month, (2) About once every 6 months, (3) About once a year, (4) Very seldom, and (5) Never. Results are shown in Figure 50. As we can observe, most software engineers read materials about their professions either once a month or very seldom (less than once a year).

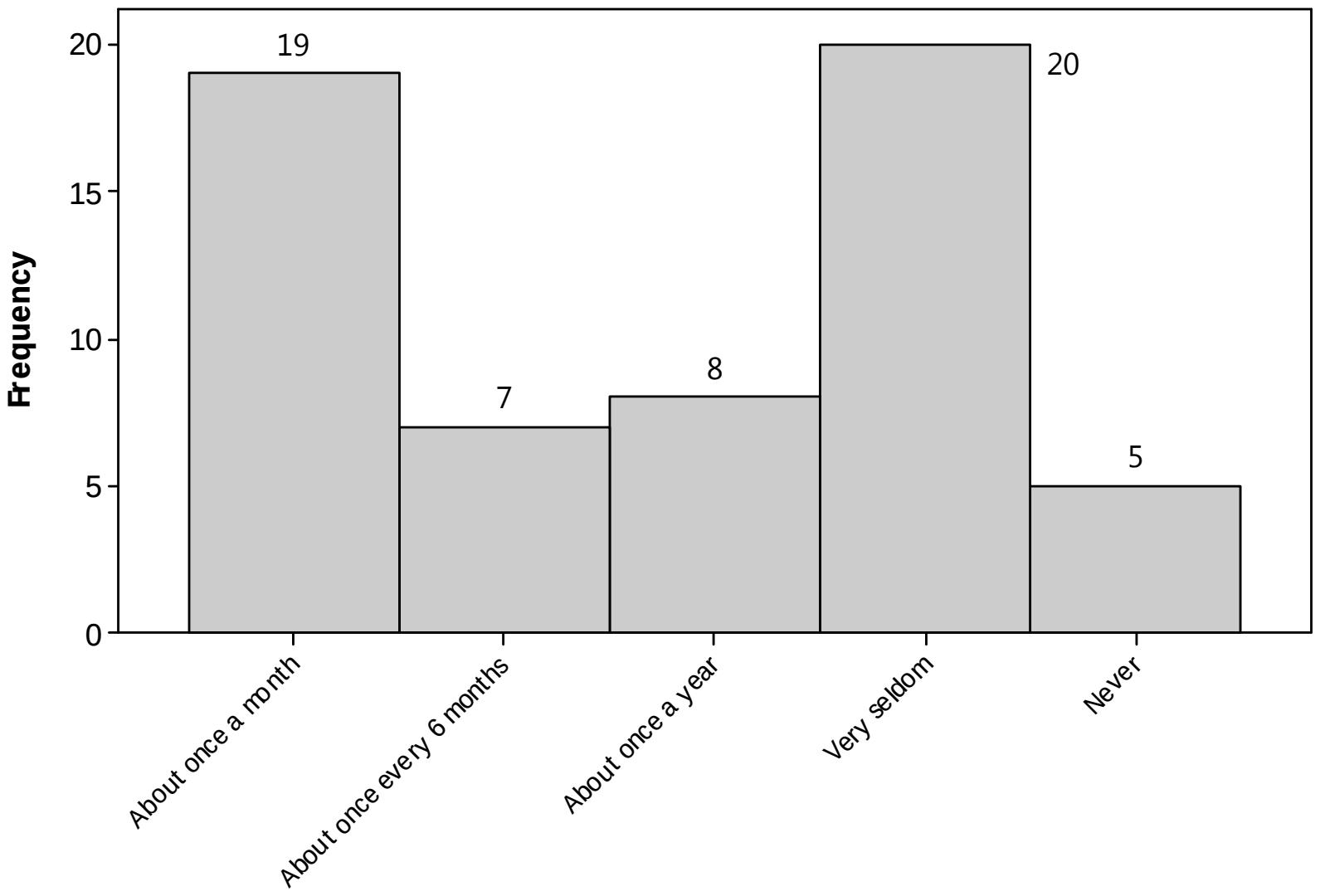

**Figure 50- Reading technical papers**

## 5 Discussions

Summary of our findings are discussed in Section 5.1. Section 5.2 discusses the lessons learned. Potential threats to the validity of our study and steps we have taken to minimize or mitigate them are discussed in Section 5.3.

### 5.1 Summary of Findings

Our survey gathered responses from 202 practitioner software engineers working in the Turkish companies. According to our demographic data, software firms from major Turkish cities (e.g., Ankara, Istanbul, Izmir, Bursa, Eskisehir, Gebze, Kocaeli, Afyon, Balikesir and Samsun) where software development establishments are located, participated in this survey.

The target industry type of products developed by firms varied from military and defense, to banking/finance, to government (excluding military and defense), to IT and telecommunication and to health. There was a good mix of different project and company profiles (Sections 4.1 and 4.2) in our survey data, and this helped our results to be unbiased from certain types of development firms.

The survey revealed interesting insights and trends in the Turkish software industry. A highlight of the results for the questions after the profiles and demographics section (Q1-Q11) is discussed in the following:

- Q 12-As expected, software engineers allocate most effort on implementation phase (31%), whereas effort spent for testing, requirements, design and maintenance are almost equal (14%, 12%, 12% and 11% respectively) across different respondents.
- Q 13-Our pool of respondents reported that they are involved mostly in implementation, requirements, design and testing phases.
- Q 14-Software engineers in Turkey experience most challenge in requirements followed by design, communication with management, testing and communication with end users.
- Q 15-Most commonly followed software development methodologies are waterfall life-cycle, incremental development and Agile/lean development with adoption rates of 53%, 38% and 34% respectively.
- Q 16-The ISO 9000 family of standards and Capability Maturity Model Integration (CMMI) are dominating process improvement models each having a usage rate of more than 60% of the software organizations where respondents work for.
- Q 17-Process related practices such as learning from previous projects, using defined processes, using tasks defined by project managers and monitoring and assessing processes are used to some degree with a rating around 3 out of 5.
- Q 18-Establishing traceability to other requirements and work products is the most frequently performed requirements engineering practice.
- Q 19-Natural language (61%) and use cases (54%) are widely used for specifying requirements. Whereas, formal notations (8%) are not as common. A surprising finding is that almost 13% of the respondents do not document their requirements at all.
- Q 20-About half of the time, software engineers review their designs and tend to be creative about design.
- Q 21-Class design and architecture design are the most popular design activities.
- Q 22-Among thirteen design related quality attributes, most of which are rated as important by participants, modularity and reliability are the most favored quality attributes.
- Q 23- Java (34%) is the most commonly used language, although other languages are also used commonly that we cannot assume dominance by a single language. Still, when compared to selected Canadian and Worldwide studies, popularity of Java is relatively higher in Turkey.
- Q 24-Our analyses on responses for development related practices revealed that building when a feature is completed, "developers being responsible for high level design and implementation" and "carrying out design and coding together" are frequently performed practices. An interesting finding is that pair programming, which is a highly praised popular practice within agile methods, is not performed as frequently in Turkey.
- Q 25-In terms of testing related practices, the most and least widely performed activities in this list are "Developers test the product before release", and "Unit tests are formally reviewed", respectively.
- Q 26-Test-last Development (TLD) approaches (i.e., testing after development) are still much more popular than Test-driven (first) Development (TDD).
- Q 27-The three most widely used test types are functional/system testing, user acceptance testing, and integration testing, with average ratings of 4, 4, and 3.7 out of 5, respectively. On the other hand, the least widely used approaches are load/stress testing and security testing.
- Q 28-About 49% of the Turkish testers reported not using any of the formal test-case design approaches, while the Canadian ratio is much better, about 17%. We should investigate in follow-up studies why formal approaches have low popularity.
- Q 29- The level of test automation range anywhere between "Never" and "Always" in Turkey, as it was similar in Canada. This denotes that respondents' decisions on selecting automated versus manual testing varies, i.e., some favor automated testing, while others prefer performing manual testing.
- Q 30-Generally the popularity of all four coverage metrics is low in the Turkish software industry.
- Q 31-Other quality-related metrics such as "number of defects per Line of Code", and "number of tests cases executed within a period of time" are not widely used in Turkey
- Q 32-Developers outnumber testers in most organizations with proportions (tester: developer) ranging from 1:2 to 1:5 and higher. Also, in some companies either no distinction is made between these two roles or this metric is not measured at all.
- Q 33-In terms of criteria used for terminating testing "All test cases are executed without finding more defects" seems to be the most widely used criterion by far.

- Q 34 and Q 35- Corrective maintenance is the most favored maintenance activity type whereas perfective maintenance is the least with ratings around 3.7 and 2.6 out of 5, respectively. These two maintenance types are also the ones software engineers experience challenges most and least respectively.
- Q 36-Poor testing, management, and change management are among the causes of maintenance challenges as stated by respondents.
- Q 37-Among the given list of possible challenges in maintenance, insufficient documentation and different programming styles used by original developers are at the top.
- Q 38-In terms of product release and delivery, most and least widely performed activities are "having representatives from QA and test in release decisions", and "QA making the product release decisions alone".
- Q 39-According to the responses, 54% of the participants have a dedicated support team.
- Q 40-Our analysis revealed that most practitioners in Turkey perform cost and effort estimations but do not use formal estimation methods.
- Q 41-Most popular project monitoring practice is performing periodic (weekly/monthly) progress meetings.
- Q 42-Most frequently used software engineering tools are configuration management, project management, design, and requirements tools in Turkish software industry.
- Q 43-Most frequently performed quality practices are peer reviews and audits.
- Q 44-About 40% of the organizations have a dedicated R&D department or unit.
- Q 45-Industry and academia collaboration is weak in Turkey that most software engineers almost never interact with university researchers as part of their work.
- Q 46-There is no common behavior among software engineers in terms of following technical papers and articles about their profession that some read such materials frequently (once a month) whereas some infrequently (less than once a year).

### 5.2 Lessons Learned

Our experience throughout this study revealed the importance of survey strategy once again. Authors' experiences in similar studies were helpful in both design and execution of the survey. Similar to the surveys in [23, 45], we categorized the questions as per SWEBOK (version 2004) [37]. Also in question design, we not only used our domain knowledge but also benefited from reviewing similar surveys and consulting industrial partners. These efforts aimed at ensuring use of a terminology that is familiar to the respondents and keeping the questions at a reasonable number without sacrificing domain coverage. In execution, we devised a progressive publicity plan. Each publicity task was initialized with a few days apart. This enabled us to monitor and control the survey population as we were able to observe exponential increases in total number of participants after each invitation sent.

Another lessons learned from this survey is that poor implementation of software engineering practices are closely related to economics and psychology rather than just technical issues. Performing state-of-the-art software engineering practices are usually put in a trade-off with organizations' culture, resources and cost considerations. So, we believe the researchers should avoid attributing poor practices in software engineering solely to lack of knowledge and skills of practitioners.

We devised an inevitably long survey with 46 questions, despite our efforts to keep it shorter in the very wide domain of software engineering knowledge. We received feedbacks from respondents enquiring if it was possible to save progress and continue answering later. Unfortunately this was not possible due to tool limitations, but we will keep it in mind in selecting surveying tool in future survey studies.

### 5.3 Threats to Validity

One threat to external validity lies in the demographic distribution of response samples. As reported in Section 3.2, our participants were mainly invited through the researchers' network of partners/contracts in Turkish software companies. Companies out of our contact network were not probably properly represented in the survey population. To mitigate this validity issue, we attempted to enlarge our email-list to include all major Turkish companies and technology development zones, placed messages in online Social and professional networks, e.g., LinkedIn, Yahoo! Groups and Twitter, and allocated two months of time for all invitees to finish the online questionnaire. Our sample size and geographic distribution of samples are quite reasonable to make a rough conclusion for Turkish software industry and clearly we do not have any intention to generalize our findings to other countries and regions.

In terms of construct validity, the issue relates to whether we are actually measuring the real-world software engineering practices in this survey. What we did was common to other survey studies—we counted the votes for each question and then made statistical inferences. It is believed that results based on such voting data can, to a certain extent, reflect the opinions of the majority of Turkish practitioners.

On the other hand, the validity of our conclusions could be of our concern. We attempted to conclude, qualitatively, that the SE practices have economics and psychology aspects as well as purely technical concerns. For our survey, we attempted to reduce the bias by seeking support from the statistical results of each software engineering aspect.

Finally, we did not intend to establish any causal relationship in this study, thus the discussion of internal validity is not applicable for our study.

## 6 Conclusions and Future Works

Thanks to the results of this survey, we have observed that the Turkish industry is vibrant, growing and competitive that many software teams are eager for adopting state-of-the-art software engineering practices and approaches. The authors are glad to see that awareness and knowledge about these practices and approaches are rising in the industry, that many organizations have been establishing them within their development operations. We believe that the encouraging findings reported in this study will expand in time as the young Turkish software industry matures.

However, encouraging figures should not overshadow some shortcomings. We have observed that there is still a long way to improve on using well-established requirements engineering and testing related practices. These two areas are where most displeasure is stated about. Also we observed that practitioners have a tendency to avoid formal or relatively complex best practices (e.g. TDD, formal languages, pair programming, and metric usage) in general, but especially in these two areas. We suggest organizations to invest in managerial support and training efforts for increasing the awareness and knowledge about software requirements and testing.

Software engineering managers and researchers, as guilty of intensifying their efforts on understanding and improving the technical aspects, tend to ignore the social aspects of software engineering. In this study, we have seen that social aspects are important as well that many practitioners have expressed their discontent and difficulties on issues like communication, collaboration, and team structure.

We plan to investigate the following future work directions:

- Further similar studies are needed in other regions and countries to be able to compare the latest trends in software engineering practices.
- It would also be a good idea to assess the maturity of SE practices in Turkey and elsewhere in future works by aligning them to standards such as the Capability Maturity Model Integration (CMMI).
- We found in question 19 of the survey that 13% of the participants did not report documenting the software requirements at all. This is an important finding that requires further analysis for causes (is it because of low quality awareness or emergence of Agile methods or something else?). Also, it would be important to analyze the potential effects of not documenting requirements (e.g., how is the end product affected by not documenting requirements?). We plan to empirically investigate these questions.
- We are intrigued by the observation that too many respondents have stated their dissatisfaction about social aspects of software engineering in various ways in questions Q44 and Q14. It would be interesting to elaborately investigate these aspects in software organizations from both their vertical and horizontal structure perspectives.
- We have noticed that there may be "cross-factor" correlations among difference practices. For example, there may be a correlation between the target industry sector of a company and use of size measurement approaches or compliance with the CMMI in that firm. We plan to conduct detailed hypotheses-based cross-factor correlation analysis in future works.
- As another future work direction, we aim at identifying the open problems of the Turkish software industry more systematically so that the research community could work on. There have been articles targeting this issue, e.g., [62-64]. For example, in a recent 2014 paper [62], Begel and Zimmermann conducted a survey among the Microsoft software engineers and proposed 145 questions for software engineering scientists. Panfilis and Berre discussed in [63] the open issues and concerns on component-based software engineering. Last but not least, in a book titled "Software Creativity 2.0" [64], Robert Glass discusses the gap between the software academia and industry and calls on both sides to work on several concrete topics that he proposes.

## Acknowledgements

Vahid Garousi-Yusifoğlu was supported by Atilim University and the Visiting Scientist Fellowship Program (#2221) of the Scientific and Technological Research Council of Turkey (TÜBİTAK). We would like to sincerely thank all the software engineers from across Turkey who anonymously participated in the survey. Last but not least, we are grateful to our colleagues and friends (Ozden Ozcan Top, Elif Aydin, Burak Coşkun, and Ergin Topcu and others) who helped in the publicity of our survey.

## REFERENCES

[1] P. Naur and B. Randell, "Software Engineering: Report of a conference sponsored by the NATO Science Committee," *Scientific Affairs Division, NATO, Brussels,* 1969.

[2] L. L. Beck and T. E. Perkins, "A Survey of Software Engineering Practice: Tools, Methods, and Results," *IEEE Transactions on Software Engineering,* vol. SE-9, pp. 541-561, 1983.

[3] M. Zelkowitz, R. Yeh, R. Hamlet, J. Gannon, and V. Basili, "Software Engineering Practices in the US and Japan," *IEEE Computer,* vol. 00, pp. 57-66, 1984.

[4] B. Curtis, H. Krasner, and N. Iscoe, "A field study of the software design process for large systems," *Communications of the ACM,* vol. 31, pp. 1268-1287, 1988.

[5] M. Cusumano and C. Kemerer, "A quantitative analysis of US and Japanese practice and performance in software development," *Management Science,* vol. 36, pp. 1384-1406, 1990.

[6] J. D. Blackburn, G. D. Scudder, and L. N. Van Wassenhove, "Improving speed and productivity of software development: a global survey of software developers," *IEEE Transactions on Software Engineering,* vol. 22, pp. 875-885, 1996.

[7] J. Singer, T. Lethbridge, N. Vinson, and N. Anquetil, "An examination of software engineering work practices," in *Proceedings of Conference of the Centre for Advanced Studies on Collaborative Research*, 1997.

[8] J. Holt, "Current practice in software engineering: a survey," *Computing & Control Engineering Journal,* vol. 8, pp. 167 - 172, 1997.

[9] S. Dutta, M. Lee, and L. V. Wassenhove, "Software Engineering in Europe: A study of best practices," *Software, IEEE,* pp. 82-90, 1999.

[10] L. Groves, R. Nickson, G. Reeve, S. Reeves, and M. Utting, "A Survey of Software Development Practices in the New Zealand Software Industry," in *Australian Software Engineering Conference*, 2000, pp. 189-201.

[11] M. Cusumano, A. MacCormack, C. F. Kemerer, and W. Crandall, "A global survey of software development practices," in *Center for eBusiness@ MIT*, 2003, pp. 1-17.

[12] T. Aytaç, S. Ikiz, and M. Aykol, "A SPICE-Oriented, SWEBOK-Based Software Process Assessment on a National Scale: Turkish Software Sector Survey," in *International Conference on Process Improvement and Capability dEtermination (SPICE) Conference*, 2003.

[13] C. Denger and M. H. Raimund L. Feldmann, Christin Lindholm, Forrest Shull, "A snapshot of the state of practice in software development for medical devices," *First International Symposium on Empirical Software Engineering and Measurement,* pp. 485-487, 2007.

[14] M. M. Aykol, "Software engineering and software management practices in Turkey (in Turkish: Türkiye'de yazılım mühendisliği ve yazılım yönetimi uygulamaları)," MSc thesis, Bahçeşehir University, Turkey, 2009.

[15] N. Sökmen, "Competency level of the software industry in Turkey and guidelines for enhancement of companies and the sector (in Turkish: Türkiye'de Yazılım Üreticilerinin Yetkinlik Düzeyi Firmaların ve Sektörün Gelişimi)," 2010.

[16] E. Egorova, M. Torchiano, and M. Morisio, "Evaluating the Perceived Effect of Software Engineering Practices in the Italian Industry," *Lecture Notes in Computer Science* pp. 100-111, 2009.

[17] D. Kirk and E. Tempero, "Software Development Practices in New Zealand," in *Public report*, 2012.

[18] F. Vonken, J. Brunekreef, A. Zaidman, and F. Peeters, "Software Engineering in the Netherlands: The State of the Practice," *Technical Report TUD-SERG-2012-022, Delft University of Technology, Software Engineering Research Group,* 2012.

[19] T. C. Lethbridge, J. Diaz-Herrera, R. J. J. LeBlanc, and J. B. Thompson, "Improving software practice through education: Challenges and future trends," in *Proc. of International Conference on Future of Software Engineering (FOSE)*, 2007, pp. 12-28.

[20] G. Tirpançeker, "Turkish software sector and value added by this sector (in Turkish: Türkiye Yazılım Sektörü ve Yazılımın Yarattığı Katma Değerler)," http://www.sde.org.tr/userfiles/file/Gulara_Tirpanceker_SDE_2011Aral%C4%B1k-2.pdf, *Turkish Software Industry Association (YASAD)*, 2011.

[21] The Standish Group, "Extreme CHAOS," http://www.standishgroup.com/sample_research/showfile.php?File=extreme_chaos.pdf, 2001 [cited: Oct. 2009].

[22] The Standish Group, "CHAOS Manifesto," *https://secure.standishgroup.com/newsroom/chaos_manifesto.php*, Oct. 2009 [cited: Oct. 2009].

[23] V. Garousi and T. Varma, "A Replicated Survey of Software Testing Practices in the Canadian Province of Alberta: What has Changed from 2004 to 2009?," *Journal of Systems and Software,* vol. 83, pp. 2251-2262, 2010.

[24] V. Garousi and J. Zhi, "A Survey of Software Testing Practices in Canada," *Journal of Systems and Software,* vol. 86, pp. 1354–1376, May 2013.

[25] V. Garousi, A. Coşkunçay, A. B. Can, and O. Demirörs, "A Survey of Software Testing Practices in Turkey," in *Turkish National Software Engineering Symposium (Ulusal Yazılım Mühendisliği Sempozyumu, UYMS)*, 2013.

[26] V. Garousi, "Recent Trends in Software Testing: Opportunities for Industry-Academia Collaborations," *Invited speaker, YouTube Corporation, San Bruno, California,* June 30, 2010.

[27] V. Garousi, "Better Software Testing through University/Industry Collaborations," *Invited talk, Calgary Software Quality Discussion Group (SQDG),* 2011.

[28] V. Garousi, "A Systematic Approach to Software Test Automation and How to Increase its ROI," *Invited Talk, TestIstanbul industry conference, Istanbul, Turkey,* May 23-24, 2013.

[29] V. Garousi, "Success stories in Systematic Software Testing: A Canadian-Turkish Perspective," *Invited Talk, "Ankara Testing Days" industry conference, Ankara, Turkey,* May 2014.

[30] Turkish Software Industry Association (YASAD), "Software: the new strength of the economy (in Turkish: Yazilim: ekonominin yeni kalkinma gücü)," http://www.yasad.org.tr/UserFiles/File/yasad_rapor.pdf, *Turkish Software Industry Association (YASAD),* 2009.

[31] D. U. Güneş, "Software industry in Turkey," *Turkish Software Industry Association (YASAD),* http://yasad.org.tr/UserFiles/File/YASAD_Presentation_ENG.pdf, 2010.

[32] M. U. Akkaya, Z. Baktır, M. Canlı, A. Çekiç, H. R. Çetin, M. Duran, *et al.*, "The software sector in Turkey (Turkish: Türkiye'de Yazilim Sektörü)," http://www.sde.org.tr/userfiles/file/TURKIYEDE_YAZILIM_%20SEKTORU.pdf, 2012.

[33] Turkish Testing Board, "Turkey Software Quality Report 2013," in http://testistanbul.org/TSQR.html, 2013.

[34] Turkish Testing Board, "Turkey Software Quality Report 2012," in http://testistanbul.org/TSQR.html, 2012.

[35] Turkish Testing Board, "Turkey Software Quality Report 2011," in http://testistanbul.org/TSQR.html, 2011.

[36] Turkish Testing Board, "Turkey Software Quality Report 2014," in http://testistanbul.org/TSQR.html, 2014.

[37] P. Bourque and R. E. Fairley, "Guide to the Software Engineering Body of Knowledge, Version 2004," *IEEE Computer Society,* http://www.computer.org/portal/web/swebok/2004guide, 2004.

[38] CMMI Institute, "Published CMMI® Appraisal Results," *https://sas.cmmiinstitute.com/pars,* Last accessed: July 2014.

[39] Turkish Standards Institute, "SPICE-certified organizations (the page is in Turkish: SPICE Belgelendirilen Kuruluşlar)," http://bilisim.tse.org.tr/-b-standardlar-b-/spice/-b-belgelendi-ri-len-kurulu%C5%9Flar-b-, Last accessed: July 2014.

[40] E. Egorova, M. Torchiano, and M. Morisio, "Actual vs. perceived effect of software engineering practices in the Italian industry," *Journal of Systems and Software,* vol. 83, pp. 1907–1916, 2010.

[41] M. Lubars, C. Potts, and C. Richter, "A review of the state of the practice in requirements modeling," in *Proceedings of the IEEE International Symposium on Requirements Engineering*, 1992, pp. 2-14.

[42] U. Nikula, J. Sajaniemi, and H. Kälviäinen, "A State-of-the-practice Survey on Requirements Engineering in Small- and Medium-sized Enterprises," 2000.

[43] C. J. Neill and P. A. Laplante, "Requirements engineering: The state of the practice," *IEEE Software,* vol. 20, pp. 40-45, 2003.

[44] P. Rodríguez, J. Markkula, M. Oivo, and K. Turula, "Survey on agile and lean usage in finnish software industry," in *Proceedings of the ACM-IEEE international symposium on Empirical software engineering and measurement*, ed, 2012, p. 139.

[45] A. M. Geras, M. R. Smith, and J. Miller, "A Survey of Software Testing Practices in Alberta," *Canadian Journal of Electrical and Computer Engineering,* vol. 29, pp. 183-191, 2004.

[46] V. R. Basili, "Software modeling and measurement: the Goal/Question/Metric paradigm," Technical Report, University of Maryland at College Park1992.

[47] A. Coşkunçay, A. Betin-Can, O. Demirörs, and V. Garousi, "Full questions for the Survey of Software Engineering Practices in Turkey," *https://drive.google.com/file/d/0B6dKdxaNjBENZVBaMDNKM0lzTjQ/edit?usp=sharing,* Last Accessed: Dec. 2013.

[48] D. A. Garvin, *Managing quality: the strategic and competitive edge*: Free Press, 1988.

[49] Various StackExchange user members, "Why is there such limited support for Design by Contract in most modern programming languages?," *http://programmers.stackexchange.com/questions/128717/why-is-there-such-limited-support-for-design-by-contract-in-most-modern-programm*, Last accessed: July 2014.

[50] Various Slashdot user members, "Why is "Design by Contract" not more popular?," *http://ask.slashdot.org/story/07/03/10/009237/why-is-design-by-contract-not-more-popular*, Last accessed: July 2014.

[51] TIOBE Group, "TIOBE Index for ranking the popularity of Programming languages " *http://www.tiobe.com/index.php/content/paperinfo/tpci/index.html*, Last accessed: Dec. 2013.

[52] E. M. Maximilien and L. Williams, "Assessing test-driven development at IBM," in *Proceedings of International Conference on Software Engineering*, 2003, pp. 564-569.

[53] T. Bhat and N. Nagappan, "Evaluating the efficacy of test-driven development: industrial case studies," in *Proceedings of the International Symposium on Empirical Software Engineering*, 2006, pp. 356-363.

[54] S. A. Jolly, V. Garousi, and M. M. Eskandar, "Automated Unit Testing of a SCADA Control Software: An Industrial Case Study based on Action Research," in *IEEE International Conference on Software Testing, Verification and Validation (ICST)*, 2012, pp. 400-409.

[55] C. Pinheiro, V. Garousi, F. Maurer, and J. Sillito, "Introducing Automated Environment Configuration Testing in an Industrial Setting," in *Proceedings of the International Conference on Software Engineering and Knowledge Engineering, Workshop on Software Test Automation, Practice, and Standardization*, 2010, pp. 186-191.

[56] Z. Sahaf, V. Garousi, D. Pfahl, R. Irving, and Y. Amannejad, "When to Automate Software Testing? Decision Support based on System Dynamics – An Industrial Case Study," in *Proc. of International Conference on Software and Systems Process, In Press*, 2014.

[57] Y. Amannejad, V. Garousi, R. Irving, and Z. Sahaf, "A Search-based Approach for Cost-Effective Software Test Automation Decision Support and an Industrial Case Study," in *Proc. of International Workshop on Regression Testing, co-located with the Sixth IEEE International Conference on Software Testing, Verification, and Validation, In Press*, 2014, pp. 302-311.

[58] J. Whittaker, "Google vs. Microsoft, and the Dev:Test Ratio Debate," *http://blogs.msdn.com/james_whittaker/archive/2008/12/09/google-v-microsoft-and-the-dev-test-ratio-debate.aspx*, Dec. 2008 [cited: Oct. 2009].

[59] K. Iberle and S. Bartlett, "Estimating Tester to Developer Ratios (or Not)," in *Pacific Northwest Software Quality Conference, http://www.stickyminds.com/s.asp?F=S6174_ART_2*, 2001.

[60] M. A. Cusumano and D. B. Yoffie, "Software Development in Internet Time," *IEEE Computer,* vol. 32, pp. 60-69, 1999.

[61] Turkish Testing Board (TTB), "Turkish Software Quality Reoprt 2013-2014," *http://www.testistanbul.org/TurkeySoftwareQualityReport_2013_2014.pdf*, 2013.

[62] A. Begel and T. Zimmermann, "Analyze This! 145 Questions for Data Scientists in Software Engineering," in *Proceedings of the International Conference on Software Engineering*, 2014.

[63] S. D. Panfilis and A. J. Berre, "Open issues and concerns on Component Based Software Engineering," in *International Workshop on Component-Oriented Programming*, 2004.

[64] R. L. Glass, *Software Creativity 2.0*: Publisher: developer.* Books, 2006.